\documentclass[11pt]{article}
\usepackage{jcappub}
\usepackage{bm}
\usepackage{graphicx}
\usepackage{graphbox}
\usepackage{caption}
\usepackage{subcaption}

\newcommand*\dif{\mathop{}\!\mathrm{d}}
\newcommand{\fnl}{f_{\mathrm{NL}}}
\newcommand{\lh}{\left(}
\newcommand{\rh}{\right)}

\title{The bispectra of galactic CMB foregrounds and their impact on
primordial non-Gaussianity estimation}
\author[a,d]{Gabriel Jung,}
\author[b,c]{Benjamin Racine,}
\author[a]{and Bartjan van Tent}

\affiliation[a]{Laboratoire de Physique Th\'eorique (UMR 8627), CNRS, Univ.\ Paris-Sud,
  Universit\'e Paris-Saclay, 91405 Orsay, France}
\affiliation[b]{Institute of Theoretical Astrophysics, University of Oslo, P.O. Box 1029 Blindern, NO-0315 Oslo, Norway}
\affiliation[c]{Harvard-Smithsonian Center for Astrophysics, Cambridge, MA 02138, USA}
\affiliation[d]{Department of Theoretical Physics, University of the Basque Country UPV/EHU, 48080 Bilbao, Spain}

\emailAdd{gabriel.jung@protonmail.com}
\emailAdd{racinebenjamin@gmail.com}
\emailAdd{bartjan.van-tent@th.u-psud.fr}

\abstract{We use the binned bispectrum estimator to determine the bispectra of
the dust, free-free, synchrotron, and AME galactic foregrounds using
maps produced by the \texttt{Commander} component separation method from Planck
2015 data. We find that all of these peak in the squeezed configuration,
allowing for potential confusion with in particular the local primordial
shape. Applying an additional functionality implemented in the binned
bispectrum estimator code, we then use these galactic bispectra as
templates in an $f_\mathrm{NL}$ analysis of other maps. After testing
and validating the method and code with simulations, we show that
we detect the dust in the raw 143~GHz map with the expected amplitude
(the other galactic foregrounds are too weak at 143~GHz to be detected)
and that no galactic residuals are detected in the cleaned CMB map.
We also investigate the effect of the mask on the templates and the effect
of the choice of binning on a joint dust-primordial $f_\mathrm{NL}$
analysis.}


\arxivnumber{}

\begin{document}

\maketitle
\flushbottom

\section{Introduction}
\label{sec:introduction}

The exploration of the CMB as a source of high-precision information on
cosmology and as the best (if somewhat opaque) window on
the primordial universe started in earnest with the first WMAP release in 2003 \cite{Spergel:2003cb}.  
The Planck satellite with its three releases in 2013, 2015, and 2018 \cite{Ade:2013sjv, Adam:2015rua, Akrami:2018vks} raised
the game to unprecedented levels of precision. Still, the amount of information
we have about the primordial universe, and in particular on the period of
inflation, remains very limited. Apart from looking for possible new
observables, it is also very important to work as much as possible on the
observables that we do have, from both ends: from the observational side to
improve estimators and data cleaning to get as precise a value as we can, and
from the theoretical side to improve (inflationary) predictions so that we can
draw the theoretical consequences from the observations.
Microwave observations are contaminated by astrophysical
foregrounds, which can be extra-galactic or galactic in origin. In order to
improve the quality of the data used for the CMB analyses, these foregrounds
are first removed as much as possible by component separation methods, which
produce so-called cleaned CMB maps, although these still contain foreground
residuals at some level.

Some of the most important inflationary observables are the non-Gaussianity
parameters $f_\mathrm{NL}$. Non-Gaussianity means that not all information
about the CMB is contained in its two-point correlation function / power 
spectrum, as would be the case for a Gaussian distribution. The lowest-order 
deviation from Gaussianity will lead to a non-zero bispectrum, the Fourier 
transform (or spherical harmonic transform on the celestial sphere) of the 
three-point correlation function. Standard single-field slow-roll
inflation produces an unobservably small non-Gaussian signal \cite{Maldacena:2002vr, Acquaviva:2002ud}, but other
inflation models predict larger amounts. Moreover, different types of models
predict differently shaped bispectra,
and we can look for the presence of any of them. The amplitude of each is
parametrized by its own $f_\mathrm{NL}$ parameter. Some of the most important
bispectrum templates are the local shape \cite{Gangui:1993tt} (typically produced by multiple-field
inflation models) and the equilateral \cite{Creminelli:2005hu} and orthogonal \cite{Senatore:2009gt} shapes (typically produced
by single-field models with non-standard kinetic terms), see \cite{Chen:2010xka} for a review. So far there is no
detection of a primordial $f_\mathrm{NL}$ value inconsistent with zero, but this
null detection with precise error bars has led to the exclusion of inflation
models that predict too much non-Gaussianity~\cite{Ade:2013ydc, Ade:2015ava}.

In order to extract any information about primordial bispectral non-Gaussianity
from the CMB data, given that this information is
primarily parametrized in the form of the $f_\mathrm{NL}$ amplitude
parameters of the different bispectrum shapes, we need an estimator for
$f_\mathrm{NL}$. This estimator should be unbiased as well as optimal (or
effectively optimal given the accuracy of the experiment), which
means it has the smallest variance theoretically possible, to extract the
primordial $f_\mathrm{NL}$ from real data contaminated by astrophysical
foreground residuals and experimental effects like noise. In addition the
estimator implementation should be fast enough to make data analysis
possible in practice. Three such estimators were developed and used for the
official Planck analysis: the KSW estimator \cite{Komatsu:2003iq, Yadav:2007rk, Yadav:2007ny} (this was the only one used for
the official WMAP analysis as well, as the other ones did not yet exist at
that time), the binned estimator~\cite{Bucher:2009nm, Bucher:2015ura}, and
the modal estimator \cite{Fergusson:2009nv, Fergusson:2010dm, Fergusson:2014gea}. All three are based
on the same theoretical exact $f_\mathrm{NL}$ estimator, but differ in the
approximations made in their implementations to make them fast enough for
practical use. In addition to being $f_\mathrm{NL}$ estimators, the binned
and modal estimators also allow for the determination of the full bispectrum
of the data. 

In this paper we will use the binned bispectrum estimator to determine the
bispectra of various galactic foregrounds (dust, free-free, anomalous microwave
emission (AME), synchrotron), and then use those bispectra as templates to 
determine the corresponding $f_\mathrm{NL}$ parameters in other
maps. The aim of the paper is threefold. In the first place it is a proof
of concept. In fact, the ability to determine bispectra from maps and then
use them as templates was one of the original motivations for developing the
binned bispectrum estimator, but so far this potential ability had not yet
been put to the test in practice. Secondly, it is interesting to study the
bispectra of these galactic foregrounds as an aim in itself, and see how they
correlate with the primordial templates.\footnote{It should be noted that for 
the purpose of studying the non-Gaussianity of a galactic foreground in itself, 
the bispectrum would probably not be the best tool. Due to their localized 
(non-isotropic) nature, an approach in pixel space instead of harmonic space 
would seem more logical. Minkowski functionals, for example, have been used to study galactic synchrotron radiation in the context of 21-cm line studies \cite{2018MNRAS.481..970R}. However, in this paper we are primarily interested in 
seeing how much these foregrounds contaminate a determination of primordial 
non-Gaussianity, which is in general isotropic and for which the bispectrum is 
then an optimal tool.} 
Finally, in the third place we want to test if any detectable 
galactic non-Gaussianity remains in the cleaned Planck CMB maps.
The quality of these maps has been tested in many different ways, mostly
using the power spectrum, but also by seeing if primordial $f_\mathrm{NL}$
measurements remained optimal, and they passed these tests. Still, it is good 
to also test for the presence of non-Gaussian galactic residuals directly.

The fact that we restrict ourselves to galactic foregrounds is because for
the most important extra-galactic foregrounds templates already exist in
analytic form \cite{Komatsu:2001rj, Lacasa:2013yya} (theoretically or heuristically determined), and can for example be used to compute biases \cite{2018arXiv180707324H}. 
But no such templates exist for galactic foregrounds. One of the advantages of the
binned bispectrum estimator is that it does not necessarily require templates
in analytic form, but can also deal with a numerical binned template.
We restrict ourselves in this paper to temperature maps only. A preliminary
exploration of the polarization maps and simulations of the 2015 Planck
release showed that these were not yet sufficiently accurate to make
a similar analysis in polarization meaningful.

The paper is organized as follows. In section \ref{sec:binn-bisp-estim}, we
review the binned bispectrum estimator, or at least those aspects of it that
we will need. One of the early questions we had, was if we should include
a linear correction term in the foreground bispectrum templates. To answer
that, we had to look at the theoretical derivation of the linear term and
its assumptions, and found that it makes no sense to add a linear correction
for highly non-Gaussian bispectra. We include
a full derivation in appendix~\ref{sec:variance-appendix}.
In sections~\ref{sec:foregrounds} and~\ref{sec:analyses}, our data
analysis results using the binned bispectrum estimator on data from
the 2015 Planck release are presented. In
section~\ref{sec:foregrounds}, several galactic foregrounds are
studied at the bispectral level, with some additional results in
appendix~\ref{ap:weights-bisp-shap}. The newly determined templates from
these foregrounds are then applied to several CMB maps (Gaussian
simulations and real data) in section~\ref{sec:analyses}. We conclude
in section~\ref{sec:conclusion}.

\section{The binned bispectrum estimator}
\label{sec:binn-bisp-estim}

In this section, we recall the main aspects of the binned bispectrum estimator. The reader interested in more details or in a more general formulation containing also E-polarization is invited to look at \cite{Bucher:2015ura}.

\subsection{Bispectrum}
\label{sec:bispectrum}

A convenient description of maps on the celestial sphere $S^2$ is in terms of
their expansion in spherical harmonics $Y_{\ell m}$, as a function of the
position on the sky $\hat\Omega = (\theta,\varphi)$:
\begin{equation}
  M(\hat\Omega)=\sum\limits_{\ell=2}^{\ell_\mathrm{max}}\sum\limits_{m=-\ell}^{\ell}
  a_{\ell m}Y_{\ell m}(\hat\Omega).
\end{equation}
Here the mode coefficients $a_{\ell m}$ contain the information in the map,
which in our case will be maps of temperature anisotropies of the microwave sky,
$M(\hat\Omega) = (T(\hat\Omega) - T_0)/T_0$ with $T_0$ the temperature monopole.
The lower limit $\ell=2$ is imposed by the fact that the monopole $\ell=0$ does
not contain any information about anisotropies by definition, and the dipole
$\ell=1$ is too contaminated by our motion with respect to the rest frame of
the CMB to be useful in practice. The upper limit $\ell_\mathrm{max}$ is imposed
by the resolution and the noise level of the experiment.

From these harmonic coefficients $a_{\ell m}$ we can define the power spectrum,
which is the Fourier transform of the two-point correlation function,
\begin{equation}
  \label{eq:power-spectrum-and-bispectrum}
  \mathcal{C}_{\ell m, \ell' m'}= \langle a_{\ell m}^{\;} a_{\ell' m'}^*\rangle.
\end{equation}
If the fluctuations were completely Gaussian, this function would fully
describe the probability distribution function of the temperature fluctuations
in the CMB. In the case of rotational invariance (which in practice is broken
by masking and by anisotropic noise, see the next subsection), there is no
$m$-dependence and the power spectrum is diagonal in multipole space
\begin{equation}
  \label{eq:power-spectrum}
 \langle a_{\ell m}^{\;} a_{\ell' m'}^*\rangle = C_{\ell}\,\delta_{\ell\ell'}\delta_{m m'}.
\end{equation}

In this paper, we are mainly interested in deviations from Gaussianity.
In that case the power spectrum does no longer contain all information about the
map and higher-order correlation functions must also be considered. In this
paper we are only interested in the bispectrum, which is the Fourier transform
of the three-point correlation function, and the function in which generic
primordial non-Gaussianity can most easily be detected. In the case of
rotational invariance the $m$-dependence is again trivial and we can use the
angle-averaged bispectrum:
\begin{equation}
  \label{eq:bispectrum}
  B_{\ell_1 \ell_2 \ell_3} \equiv \left\langle\int_{S^2}\dif \hat\Omega \,
  M_{\ell_1}(\hat\Omega) M_{\ell_2}(\hat\Omega) M_{\ell_3}(\hat\Omega)\right\rangle,
\end{equation}
where $M_{\ell}$ is a map of the temperature anisotropies filtered at the
multipole $\ell$:
\begin{equation}
  \label{eq:map}
  M_\ell(\hat\Omega)=\sum\limits_{m=-\ell}^{\ell}a_{\ell m}Y_{\ell m}(\hat\Omega).
\end{equation}
Using the Gaunt integral,
\begin{equation}
  \label{eq:gaunt-integral}
  \int\dif\hat\Omega\,Y_{\ell_1 m_1} Y_{\ell_2 m_2} Y_{\ell_3 m_3} =h_{\ell_1 \ell_2 \ell_3}\begin{pmatrix}
\ell _1 &\ell _2 &\ell _3 \\ m_1 & m_2 & m_3\\
\end{pmatrix},
\end{equation}
where the object with the parentheses is a Wigner $3j$-symbol and the $h_{\ell_1 \ell_2 \ell_3}$ are defined as
\begin{equation}
  \label{eq:numtriangles}
  h_{\ell_1 \ell_2 \ell_3}
  \equiv
  \sqrt{\frac{(2\ell_1+1)(2\ell_2+1)(2\ell_3+1)}{4\pi}}
  \begin{pmatrix}
    \ell _1 &\ell _2 &\ell _3\cr 0 & 0 & 0\cr
  \end{pmatrix} ,
\end{equation} 
the bispectrum~\eqref{eq:bispectrum} can be rewritten in terms of the harmonic
coefficients $a_{\ell m}$:
\begin{equation}
  \label{eq:bispectrum-alm}
  B_{\ell_1 \ell_2 \ell_3}  =
  h_{\ell_1 \ell_2 \ell_3}
  \sum_{m_1,m_2,m_3} \begin{pmatrix}
    \ell _1 &\ell _2 &\ell _3\cr
    m_1 & m_2 & m_3\cr
  \end{pmatrix}
  \langle a_{\ell_1 m_1} a_{\ell_2 m_2} a_{\ell_3 m_3} \rangle.
\end{equation}
The quantity $B_{\ell_1 \ell_2 \ell_3}/ h_{\ell_1 \ell_2 \ell_3}^2$ is generally called
the reduced bispectrum in the literature. Because of the Wigner 3$j$-symbols,
$B_{\ell_1 \ell_2 \ell_3}$ vanishes if either the parity condition
($\ell_1 + \ell_2 + \ell_3$ even) or the triangle condition
($|\ell_1-\ell_2|\leq \ell_3 \leq \ell_1 + \ell_2$) (or both) are not
satisfied.

\subsection{Linear correction and variance}
\label{sec:variance}

While in the previous section we considered idealized data, in this section we take into account several effects present in real experimental data. 

We model the angular response of the instrument by a beam window function, which is often well approximated by a Gaussian beam:
\begin{equation}
  \label{eq:gaussian-beam}  b_\ell=\exp\left[-\frac{1}{2}\frac{\ell(\ell+1)\theta_\mathrm{FWHM}^2}{8\ln2}\right],
\end{equation}
which is entirely characterized by the full width at half maximum $\theta_\mathrm{FWHM}$ (in radians). Moreover, the instrument noise can be described by the noise power spectrum $N_\ell$. This means that one has to substitute the CMB power spectrum $C_\ell$ by the one measured by a real experiment:
\begin{equation}
  \label{eq:experimental-power-spectrum}
  C_\ell \longrightarrow b_\ell^2 C_\ell + N_\ell.
\end{equation}
When comparing the observed bispectrum to a theoretical template (see \cite{Bucher:2015ura} for an overview), we also need to take the beam into account:
\begin{equation}
  \label{eq:experimental-bispectrum}
  B_{\ell_1 \ell_2 \ell_3}^{\mathrm{th}} \longrightarrow b_{\ell_1}b_{\ell_2}b_{\ell_3} B_{\ell_1 \ell_2 \ell_3}^{\mathrm{th}}.
\end{equation}

While the equations from the previous section assume rotational invariance, real data have two major sources of anisotropy: the mask and the anisotropic noise. 
Bright objects in the microwave domain, of galactic or extragalactic origins, prevent observations of the CMB over the whole sky. While component separation helps to mitigate these foregrounds (see \cite{Adam:2015tpy, Adam:2015wua} and references therein), it is often necessary to mask the galactic plane and the strongest point sources in the remaining sky. Besides the anisotropy induced by the mask, different effects have then to be taken into account. First, there is a multiplicative bias that needs to be included in the power spectrum and the bispectrum using the factor $f_{\mathrm{sky}}$, where $f_{\mathrm{sky}}$ is the unmasked fraction of the sky:
\begin{equation}
  \label{eq:fsky-approximation}
  C_\ell^{\mathrm{masked}} \simeq f_{\mathrm{sky}}C_\ell^{\mathrm{unmasked}}, \qquad B_{\ell_1 \ell_2 \ell_3 }^{\mathrm{masked}} \simeq
   f_{\mathrm{sky}} B_{\ell_1 \ell_2 \ell_3 }^{\mathrm{unmasked}} .
\end{equation}
It is called the $f_{\mathrm{sky}}$ approximation \cite{Komatsu:2001wu} and it is valid for small enough masks. In addition to that, one has to be careful of different edge effects induced by the mask (see section 4 of \cite{Bucher:2015ura} for a discussion of these issues). One effect that is easy to see is the fact that the mask acts as a step function in real space, so it has an influence over a large range of multipoles in harmonic space. Here, we will use the standard technique of inpainting to deal with these issues: the masked region is filled in before being filtered in multipole space. For a complete description of the inpainting approach and a comparison to other possibilities, see \cite{Gruetjen:2015sta}. The simplest method to fill in the masked regions of the map is diffusive inpainting \cite{Bucher:2015ura} and it was used for the Planck non-Gaussianity results \cite{Ade:2013ydc, Ade:2015ava}. After filling in the masked part by the average value of the map, diffusive inpainting simply consists in giving to each masked pixel the average value of its neighbouring pixels and iterating this procedure (2000 times in the case of Planck).

Another source of anisotropy comes from the scanning pattern of the satellite which makes the noise anisotropic, because some parts of the sky are observed more often than others. This makes the rotationally invariant bispectrum \eqref{sec:bispectrum} unsuited to describe the non-Gaussianity of the CMB temperature fluctuations and leads to large error bars. However, in the case of weak non-Gaussianity this issue can be dealt with by adding a simple linear correction to the cubic expression of the bispectrum \cite{Creminelli:2005hu}
\begin{equation}
  \label{eq:linear-correction}
  B^\mathrm{obs}_{\ell_1 \ell_2 \ell_3} \equiv \frac{1}{f_{\mathrm{sky}}}\int_{S^2 \setminus \mathcal{M}}\dif \hat\Omega \, \left[M_{\ell_1} M_{\ell_2} M_{\ell_3} - M_{\ell_1}\langle M_{\ell_2} M_{\ell_3}\rangle - M_{\ell_2}\langle M_{\ell_1} M_{\ell_3}\rangle - M_{\ell_3}\langle M_{\ell_1} M_{\ell_2}\rangle\right],
\end{equation}
where the integration is performed on the celestial sphere with the masked part denoted by $\mathcal{M}$ excluded. This is in fact the third-order Wick product of the maps, and it is a known result that for a Gaussian variable, it is the cubic statistic with the smallest variance \cite{Donzelli:2012ts}. Then, one can show that the variance of the bispectrum is given by (see appendix \ref{sec:variance-appendix} for the details)
\begin{equation}
  \label{eq:variance}
  V_{\ell_1 \ell_2 \ell_3} =
  g_{\ell_1 \ell_2 \ell_3} \frac{h_{\ell_1 \ell_2 \ell_3}^2}{f_\mathrm{sky}}
  (b_{\ell_1}^2 C_{\ell_1} + N_{\ell_1}) (b_{\ell_2}^2 C_{\ell_2} + N_{\ell_2}) (b_{\ell_3}^2 C_{\ell_3} + N_{\ell_3}),
\end{equation}
which only depends on parameters previously introduced, and on the combinatorial factor $g_{\ell_1 \ell_2 \ell_3}$ which takes the value 6, 2 or 1 depending on whether 3, 2, or no $\ell$'s are equal, respectively.

\subsection{Binning}
\label{sec:binning}

One of the main issues with the different equations above when used with WMAP or Planck temperature maps is the very high number of valid triplets (up to $\mathcal{O}(10^9)$ in the case of Planck). Different techniques have been developed to deal with the enormous amount of computations required, like the KSW \cite{Komatsu:2003iq, Yadav:2007rk, Yadav:2007ny} and skew-$C_\ell$ \cite{Munshi:2009ik} estimators using factorisable templates or the modal estimator \cite{Fergusson:2009nv, Fergusson:2010dm, Fergusson:2014gea} decomposing the bispectrum into a sum of uncorrelated templates. Here we adopt the binned bispectrum estimator, originally introduced in \cite{Bucher:2009nm} and further developed in \cite{Bucher:2015ura}. It relies on the simple idea that the variation with $\ell$ of many physically motivated bispectral templates is smooth. In that case, instead of having ($\ell_\mathrm{max} - \ell_\mathrm{min} + 1$) maximally filtered maps $M_{\ell}$ with $\ell$ in $[\ell_\mathrm{min}, \ell_\mathrm{max}]$, it is possible to use broader filters:
\begin{equation}
  \label{eq:binned-map}
  M_i(\hat\Omega) = \sum\limits_{\ell \in \Delta_i} M_{\ell}(\hat\Omega),
\end{equation}
where $\Delta_i = [\ell_i,~\ell_{i+1} - 1]$ with $i=0,...,(N_\mathrm{bins} -1)$ and $\ell_{N_\mathrm{bins}} = \ell_\mathrm{max} + 1$ define the binning of the multipole space.

Similar to \eqref{eq:linear-correction}, the observed binned bispectrum is
\begin{equation}
  \label{eq:binned-bispectrum}
    B^\mathrm{obs}_{i_1 i_2 i_3} \equiv \frac{1}{f_{\mathrm{sky}}\Xi_{i_1 i_2 i_3}}\int_{S^2 \setminus \mathcal{M}}\dif \hat\Omega \, \left[M_{i_1} M_{i_2} M_{i_3} - M_{i_1}\langle M_{i_2} M_{i_3}\rangle - M_{i_2}\langle M_{i_1} M_{i_3}\rangle - M_{i_3}\langle M_{i_1} M_{i_2}\rangle\right],
\end{equation}
where the normalization factor $\Xi_{i_1 i_2 i_3}$ is the number of $\ell$-triplets within the ($i_1$, $i_2$, $i_3$) bin-triplet satisfying the two selection rules (triangle inequality \cite{Luo:1993xx} and parity condition). When it comes to binning the multipole space, it was shown in \cite{Bucher:2015ura} that another rule of selection is necessary to avoid numerical inaccuracies when computing the observed bispectrum due to the pixelization of the celestial sphere. We use here the same selection criterion that the ratio of valid $\ell$-triplets to the ones satisfying only the parity condition (but not the triangle inequality) in a bin triplet should be at least 1~$\%$.
It has been shown (see \cite{Bucher:2015ura}) that the variance has the same form as \eqref{eq:variance}, with an extra factor $1/(\Xi_{i_1 i_2 i_3})^2$:
\begin{equation}
  \label{eq:binned-variance}
  V_{i_1 i_2 i_3}=\frac{1}{f_{\mathrm{sky}}}\frac{g_{i_1 i_2 i_3}}{(\Xi_{i_1 i_2 i_3})^2} \sum\limits_{\ell_1 \in \Delta_1}\sum\limits_{\ell_2 \in \Delta_2}\sum\limits_{\ell_3 \in \Delta_3} h_{\ell_1 \ell_2 \ell_3}^2 (b_{\ell_1}^2C_{\ell_1}+N_{\ell_1}) (b_{\ell_2}^2C_{\ell_2}+N_{\ell_2}) (b_{\ell_3}^2C_{\ell_3}+N_{\ell_3}).
\end{equation}

Determining an optimal binning is not an easy task, however as we will see later, there exists a simple criterion to verify how good a given binning is.

\subsection{$\fnl$ estimator}
\label{sec:estimator}

The main use of the binned bispectrum estimator is to determinate the dimensionless parameter $\fnl$ corresponding to the amplitude of non-Gaussianity. It is based on the well-known result of \cite{Komatsu:2001rj} and takes the form
\begin{equation}
  \label{eq:fnl-estimator}
  \hat{f}_\mathrm{NL} =  \frac{\langle B^\mathrm{th, exp}, B^\mathrm{obs} \rangle} {\langle B^\mathrm{th, exp}, B^\mathrm{th, exp} \rangle},
\end{equation}
where the bispectral inner product is defined by
\begin{equation}
  \label{eq:inner-product}
  \langle B^A, B^B \rangle = \sum\limits_{i_1 \leq i_2 \leq i_3} \frac{B_{i_1 i_2 i_3}^A B_{i_1 i_2 i_3}^B}{V_{i_1 i_2 i_3}}.
\end{equation}
In these expressions, $B^\mathrm{th, exp}$ describes a theoretical bispectral template as it would be observed through the experiment beam, $B^\mathrm{th, exp}_{\ell_1 \ell_2 \ell_3} = b_{\ell_1} b_{\ell_2}  b_{\ell_3} B^\mathrm{th,\fnl=1}_{\ell_1 \ell_2 \ell_3}$. The observed bispectrum $B^\mathrm{obs}$ is given by \eqref{eq:binned-bispectrum}, which contains the linear correction term discussed previously. Note that the error on $\hat{f}_\mathrm{NL}$ is then given by $\mathrm{Var}(\hat{f}_\mathrm{NL}) = 1/\langle B^\mathrm{th, exp}, B^\mathrm{th, exp} \rangle$.
The variance $V_{i_1 i_2 i_3}$ is given by \eqref{eq:binned-variance}, hence it is evaluated from the theoretical power spectrum taking into account the characteristics of the experiment, such as its beam and noise. It is the main interest of \eqref{eq:inner-product}, because the product of the two bispectra is weighted by the variance, meaning that for example a very noisy bin (inducing a large variance) cannot play an important role.

We also use the shorthand notation of the bispectral inner product in several other important quantities. We need it to define a simple criterion to verify how optimal the choice of binning is. As it takes the same form for the binned bispectrum ($i$-triplets) and for the full one ($\ell$-triplets), it is easy to compare ideal and binned estimators using the relation
\begin{equation}
  \label{eq:ratio-ideal-binned}
  R \equiv \frac{\mathrm{Var}(\hat{f}_\mathrm{NL}^\mathrm{ideal})}{\mathrm{Var}(\hat{f}_\mathrm{NL}^\mathrm{binned})} = \frac{\langle B^\mathrm{th}, B^\mathrm{th} \rangle^\mathrm{binned}}{\langle B^\mathrm{th}, B^\mathrm{th} \rangle^\mathrm{no~binning}}.
\end{equation}
This ratio shows by how much binning the multipole space has increased the variance. It has been proven in \cite{Bucher:2015ura} that $R$ is a number between 0 and 1. Obviously, $R$ is closer to 1 if the binning better describes the considered theoretical shape.

We have at our disposal several types of theoretical bispectra to which the binned bispectrum of a map can be compared and which we will refer to as the standard shapes. Some are predictions of early Universe theories \cite{Babich:2004gb, Fergusson:2008ra} like the local \cite{Gangui:1993tt}, equilateral \cite{Creminelli:2005hu} and orthogonal \cite{Senatore:2009gt} shapes. Others have a much later origin, for example the different foregrounds contaminating the CMB signal. Like the primordial shapes mentioned above, some foregrounds of extra-galactic origin (CIB \cite{Lacasa:2013yya}, extra-galactic radio point sources \cite{Komatsu:2001rj}, lensing-ISW \cite{Goldberg:1999xm, Smith:2006ud,Lewis:2011fk}, etc.) can be described analytically and used to compute biases on primordial non-Gaussianity \cite{2018arXiv180707324H}, but galactic foregrounds cannot and are the focus of this paper (see section \ref{sec:foregrounds}).

One issue with these theoretical shapes is that they are not necessarily independent from each other. Hence it is important to evaluate if one can distinguish them in an experiment. A natural definition for their correlator is 
\begin{equation}
  \label{eq:shape-correlator}
  C_{IJ} = \frac{\langle B^I, B^J\rangle}{\sqrt{\langle B^I, B^I\rangle \langle B^J, B^J\rangle}}.
\end{equation}
The indices $I$ and $J$ label the theoretical shapes. $C_{IJ}$ is between -1 and 1, a value close to 0 indicates that the theoretical bispectra $I$ and $J$ are almost uncorrelated. Note that because the bispectral inner product depends on the characteristics of the experiment, the shape correlators do too. In this paper, we use the characteristics of the Planck mission corresponding to its 2015 data release. In table \ref{tab:corr_coeff_2015}, we recall the correlation coefficients between the standard shapes as determined in \cite{Bucher:2015ura}.

\begin{table}
  \begin{center}
    \begin{tabular}{l|cccccc}
      \hline
      & Local & Equil & Ortho & LensISW & UnclustPS & CIB \\
      \hline
      Local & 1 & 0.21 & -0.44 & 0.28 & 0.002 & 0.006\\
      Equilateral && 1 & -0.05 & 0.003 & 0.008 & 0.03\\
      Orthogonal &&& 1 & -0.15 & -0.003 & -0.001\\
      Lensing-ISW &&&& 1 & -0.005 & -0.03\\
      Unclustered point sources &&&&& 1 & 0.93\\
      CIB point sources &&&&&& 1\\
      \hline
    \end{tabular}
  \end{center}
  \caption{Correlation coefficients from \cite{Bucher:2015ura} between the standard bispectral shapes computed using the characteristics of the Planck experiment (temperature). We see a correlation between local and orthogonal and between local and lensing-ISW. Equilateral and orthogonal are mostly uncorrelated, and the correlation between the point source templates and the primordial ones is negligible.}
  \label{tab:corr_coeff_2015}
\end{table}

The formula \eqref{eq:fnl-estimator} for the estimator $\hat{f}_\mathrm{NL}$ is used for each shape independently (giving each time a parameter $\fnl^I$ describing a unique shape). However, if multiple shapes are present in the data, it is important to take into account their correlation. We can then define the joint estimator
\begin{equation}
  \label{eq:joint-estimator}
  \hat{f}_\mathrm{NL}^I = \sum\limits_J (F^{-1})_{IJ} \langle B^J, B^\mathrm{obs}\rangle,
 \end{equation}
where $F_{IJ}$ is the Fisher matrix,
 \begin{equation}
  \label{eq:Fisher-matrix}
  F_{IJ} = \langle B^I, B^J\rangle.
 \end{equation}
The variance of $\hat{f}_\mathrm{NL}^I$ goes from $1/F_{II}$ for a fully independent estimation to $(F^{-1})_{II}$ in the case of a joint analysis. One can show that the variance increases if there are several correlated templates. However, it is also possible that one of the $\fnl^I$ parameters (or more) is already known, for example the lensing-ISW template including its amplitude (i.e.\ $\fnl$ parameter) is entirely determined by theory. It is then possible to avoid the increase of variance due to the joint analysis, treating such a shape as a known bias. In the case of two shapes, subtracting the known $\fnl^{(2)}$ from the unknown $\fnl^{(1)}$ is as simple as
 \begin{equation}
   \label{eq:bias}
   \fnl^{(1)} = \frac{1}{F_{11}}\langle B^{(1)}, B^\mathrm{obs}\rangle - \frac{F_{12}}{F_{11}}\fnl^{(2)}.
 \end{equation}
 The demonstration is straightforward, it is sufficient to use \eqref{eq:joint-estimator} for the shapes $(1)$ and $(2)$ to substitute $\langle B^{(2)}, B^\mathrm{obs}\rangle$ by $\langle B^{(1)}, B^\mathrm{obs}\rangle$ and $\fnl^{(2)}$.

The binned bispectrum estimator was implemented numerically in Python and C by Van Tent, Bucher and Racine (see section 5 of \cite{Bucher:2015ura} for a complete description). This code was used for the official 2013 and 2015 Planck releases \cite{Ade:2013ydc, Ade:2015ava}. Here we use it for the different data analyses presented in this paper (see sections \ref{sec:foregrounds} and \ref{sec:analyses}), focusing on a new functionality  we added for the work presented in this paper, which is the ability to use the binned bispectrum of a given map (computed by the observational part) as a numerical template for the analysis of another map. To be more precise, we will use maps of galactic foregrounds and determine their binned bispectrum (for which there is no analytical expression) in the next section. Then, we will apply these templates to different maps in section \ref{sec:analyses}.

\section{Galactic foregrounds}
\label{sec:foregrounds}

In this section, we study several galactic foregrounds with the binned bispectrum estimator. As discussed previously, when studying non-Gaussianity in CMB data maps, the usual method is to compare the observed bispectrum to different theoretical shapes using the inner product \eqref{eq:inner-product}. The determined parameters $\fnl$ simply indicate to what extent these shapes are present in the data. Usually this method is applied to several shapes which have analytical expressions and it includes primordial non-Gaussianity (generated during inflation) but also late-time bispectra (generated after recombination) like extra-galactic foregrounds.
However, when observing the CMB, the main source of contamination is our own galaxy and there is no equivalent theoretical expression to describe the non-Gaussianity of galactic foregrounds yet. There are many techniques to clean the maps from the presence of different galactic foregrounds (see \cite{Adam:2015tpy, Adam:2015wua, Ade:2015qkp} for a review) and CMB analyses at the bispectral level are generally performed on these clean maps. In this section, we use the fact that an analytical formulation of theoretical shapes is not mandatory for use with the binned bispectrum estimator, allowing us to examine these foregrounds too. Indeed, to use the inner product \eqref{eq:inner-product}, one only needs the numerical binned theoretical bispectrum. This means that in principle, the binned bispectrum of any map determined numerically could be used as theoretical template for the analysis of another map under the condition that the binning is the same. In this way we determine templates using the maps of different galactic foregrounds from the 2015 Planck release obtained by the \texttt{Commander} component separation technique \cite{Eriksen:2004ss, Eriksen:2007mx}.\footnote{\url{https://pla.esac.esa.int}}

In section \ref{sec:analyses} we will use these new numerical templates on the CMB cleaned maps studied in \cite{Ade:2015ava}. To be more precise, we will use the \texttt{SMICA} \cite{Cardoso:2008qt} CMB map from the 2015 Planck release. We will also study the raw 143 GHz map, which is the dominant frequency channel in the \texttt{SMICA} map (see figure D.1 of \cite{Adam:2015tpy}). While it is the best channel to observe the CMB (best combination of a low noise level and a good resolution), that is not the case for the different foregrounds (at least if the goal was to study the physics of these foregrounds). Nevertheless, here we only need to estimate their eventual contamination to the CMB signal. At that frequency, the CMB dominates the sky after masking the brightest parts (galactic plane and strong point sources). For this, we use the temperature common mask of the Planck 2015 release, which is a combination of the masks of the different component separation methods \cite{Adam:2015tpy}. In section \ref{sec:noise-masks}, we will discuss the influence of the mask on the different foregrounds by using a smaller one (\texttt{Commander} mask). Finally, another important choice is the binning which was determined using the ratio $R$ defined in \eqref{eq:ratio-ideal-binned} to be optimal for the primordial shapes. It is true that this criterion has nothing to do with the galactic foregrounds, but our ultimate goal is to determine the primordial shapes optimally, not the galactic ones. To illustrate the method, we start by studying the case of thermal dust.

\subsection{Thermal dust}
\label{sec:dust}

Above 100 GHz, the strongest contamination from galactic foregrounds is due to small dust grains ($\sim$ 1 $\mu$m or smaller) present in the interstellar medium. This dust plays an important role in galactic evolution (chemistry of interstellar gas, etc., see the textbook \cite{draine2010physics} for example), but it also has a large influence on astrophysical observations. Indeed dust grains are heated by the UV starlight they absorb, so they emit a thermal radiation (infrared) in the frequency range of CMB experiments. This emission is  well described by a modified blackbody model also called greybody (see \cite{Ade:2013crn, Abergel:2013jba, Adam:2015wua})
\begin{equation}
  \label{eq:greybody-dust}
  I(\nu) = A \nu^{\beta_d}B_\nu(T_d), 
\end{equation}
where $B_\nu$ describes Planck's law, $T_d \sim 20 K$ is the mean temperature and $\beta_d \sim 1.5$ is the free emissivity spectral index.

\begin{figure}
  \centering
   \includegraphics[width=0.49\linewidth]{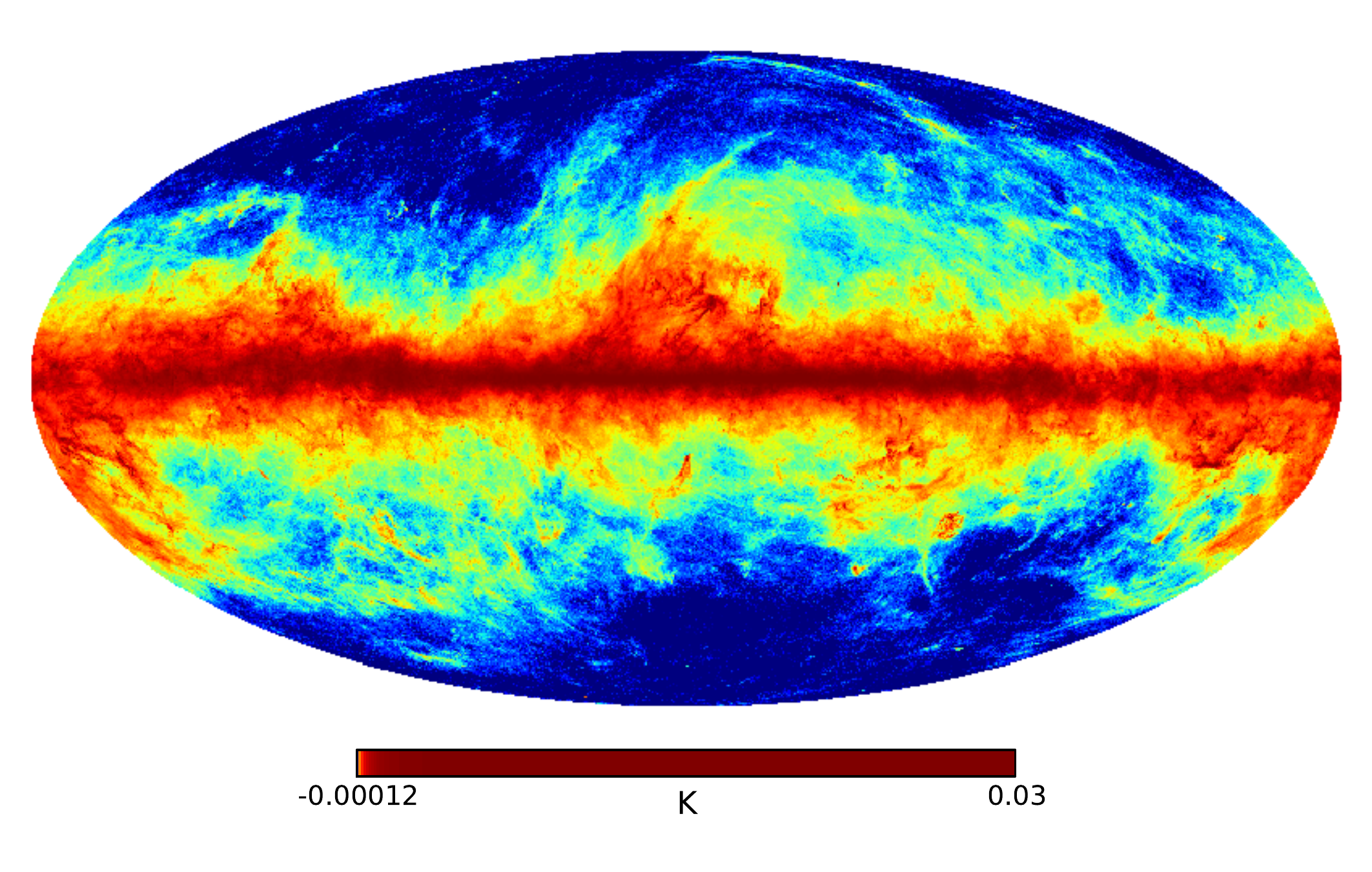}
 \includegraphics[width=0.49\linewidth]{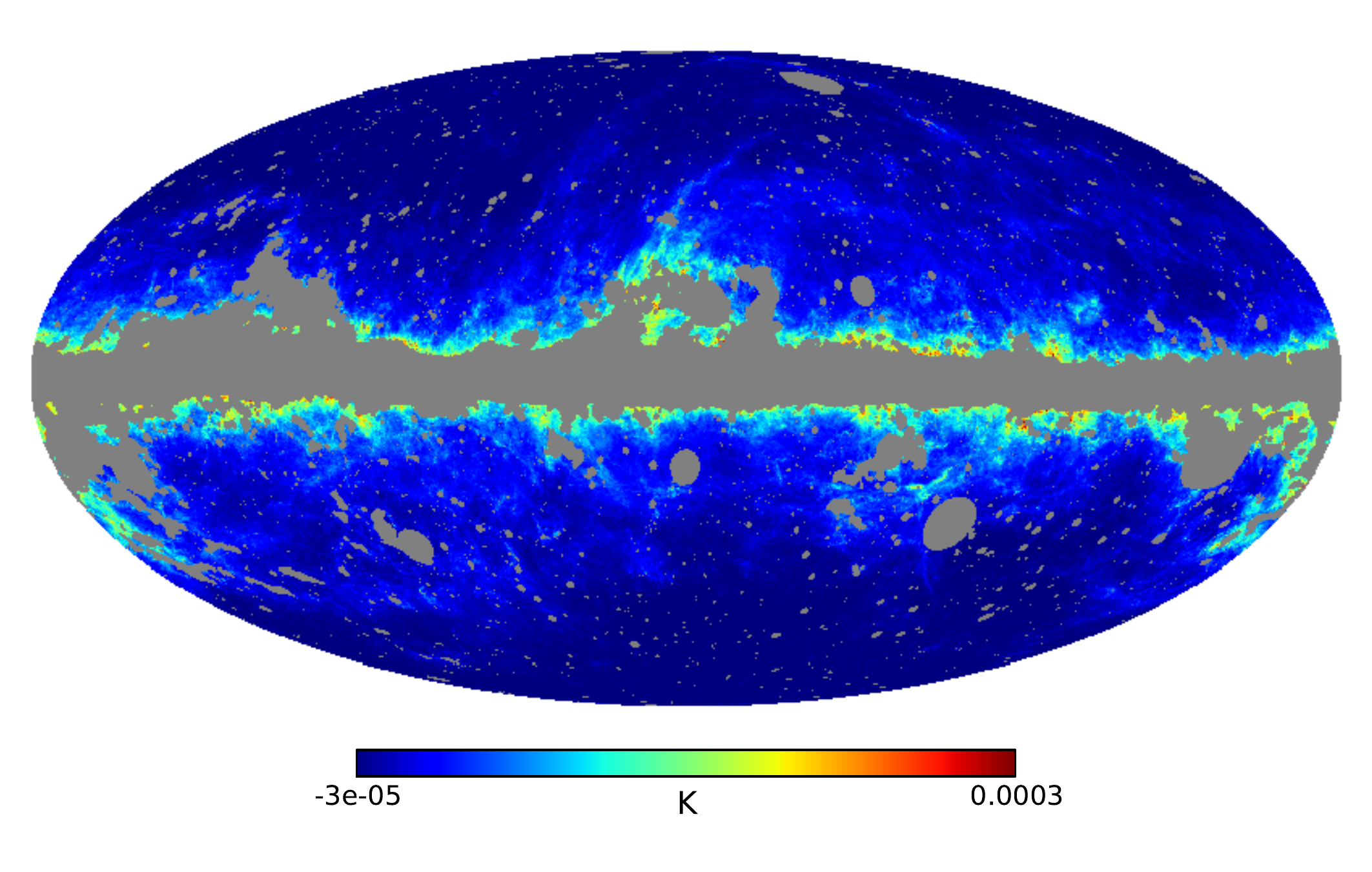}
  \caption{Unmasked (left) and masked (right) maps of thermal dust at 143 GHz from \texttt{Commander} (Planck 2015), using the common mask. On the left, the map is normalized using histogram equalization to highlight all the regions containing dust. On the right, the scale is linear, this shows that most of the signal is coming from the galactic plane near the mask. The range of the scale is also different.}
  \label{fig:dust-map}  
\end{figure}

Figure \ref{fig:dust-map} shows the map of the galactic thermal dust at 143 GHz, before and after applying the common mask. As mentioned before, we are interested in the contribution of the foregrounds in a CMB analysis (where a mask is always used to hide the galactic plane). Hence, the map on the right is the most important here because it is the actual contribution of dust that could be seen in a CMB analysis. In the following, we will be interested in the power spectrum and the bispectrum of this map. As expected, most of the signal comes from the galactic plane, and it is strongest close to the mask. Because of the dust localization, this emission is very non-Gaussian \cite{MivilleDeschenes:2007ya} and anisotropic (and this is also the case for the other galactic foregrounds studied in section \ref{sec:other-foregrounds}). The bispectrum is not the best tool to describe such a localized non-Gaussianity (an estimator in pixel space would be better). However, we are only interested in the impact of this galactic foreground on the primordial shapes. This requires us to be careful with the different expressions of section \ref{sec:binn-bisp-estim}, mostly derived using the weak non-Gaussianity approximation (see appendix \ref{sec:variance-appendix}). Concerning the observed bispectrum of the dust map, which is exactly what we need to make a dust template, it is still defined by \eqref{eq:binned-bispectrum}, but without the linear correction terms which are not justified here.

\begin{figure}
  \centering
  \includegraphics[width=0.66\linewidth]{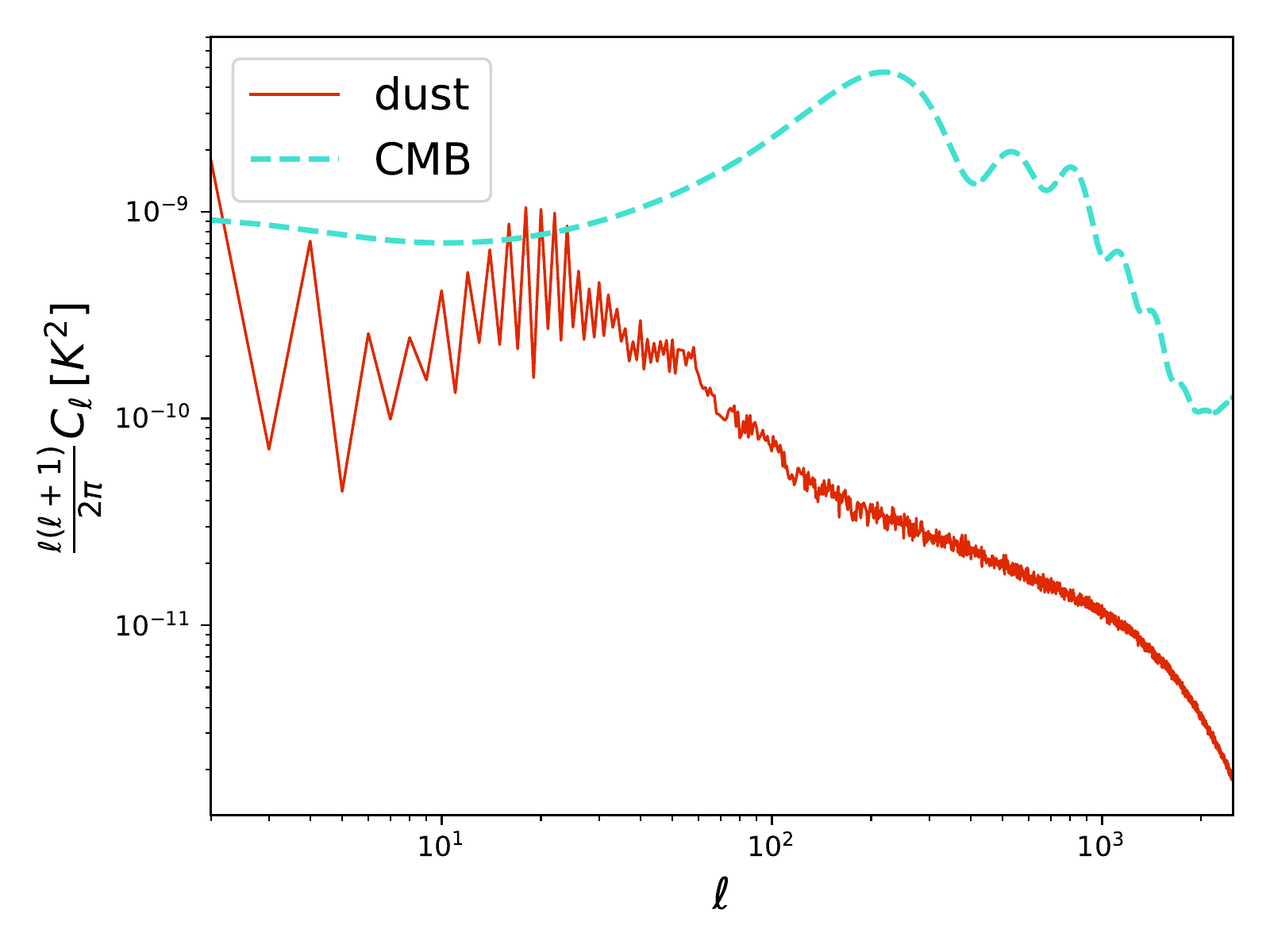}
  \caption{This figure shows the dust (at 143 GHz) and the CMB (Planck 2015 best fit) temperature power spectra (including the beam and the noise for both) as a function of the multipole $\ell$.}
  \label{fig:dust-power-spectrum}
\end{figure}

Before describing the dust bispectrum, it is interesting to examine the power spectra of the dust and CMB maps shown in figure \ref{fig:dust-power-spectrum}. It is clear that at 143 GHz, the CMB dominates except for the largest scales (smallest $\ell$) where the dust power spectrum has a sawtooth pattern. We can see that it is smaller (up to an order of magnitude) for each odd $\ell$ up to $\ell \sim \mathcal{O}(20)$. This is in fact due to the symmetry of the masked map in figure \ref{fig:dust-map} around the galactic plane when viewed on the largest scales. Because of this symmetry, the temperature is an even function of the angle $\theta$ (with the usual $\hat{\Omega}=(\theta,\varphi)$, where $\theta$ describes the latitude position), using the simple approximation that the mask can be seen as a band with all the dust signal on the border. The spherical harmonics $Y_{\ell 0}$ also have a similar symmetry around the galactic plane so they are the main contribution when decomposing in harmonic space. However, the $Y_{\ell 0}$ are even in $\theta$ only for $\ell$ even and they are odd for $\ell$ odd, so the odd terms have to be small. For the same reasons similar effects are expected in the dust bispectrum as far as large scales are concerned.

\begin{figure}
  \centering
\includegraphics[width=0.49\linewidth]{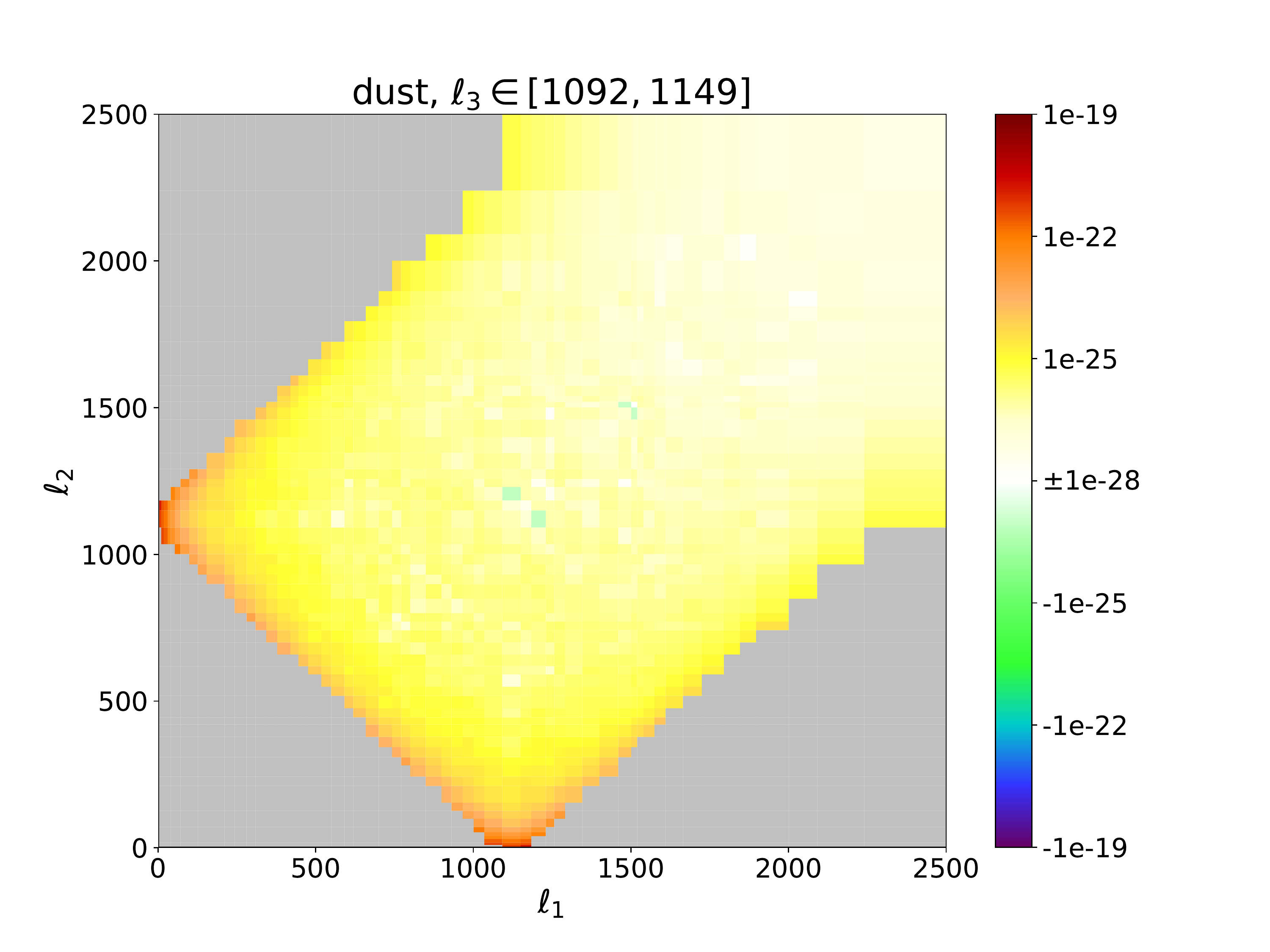}
  \includegraphics[width=0.49\linewidth]{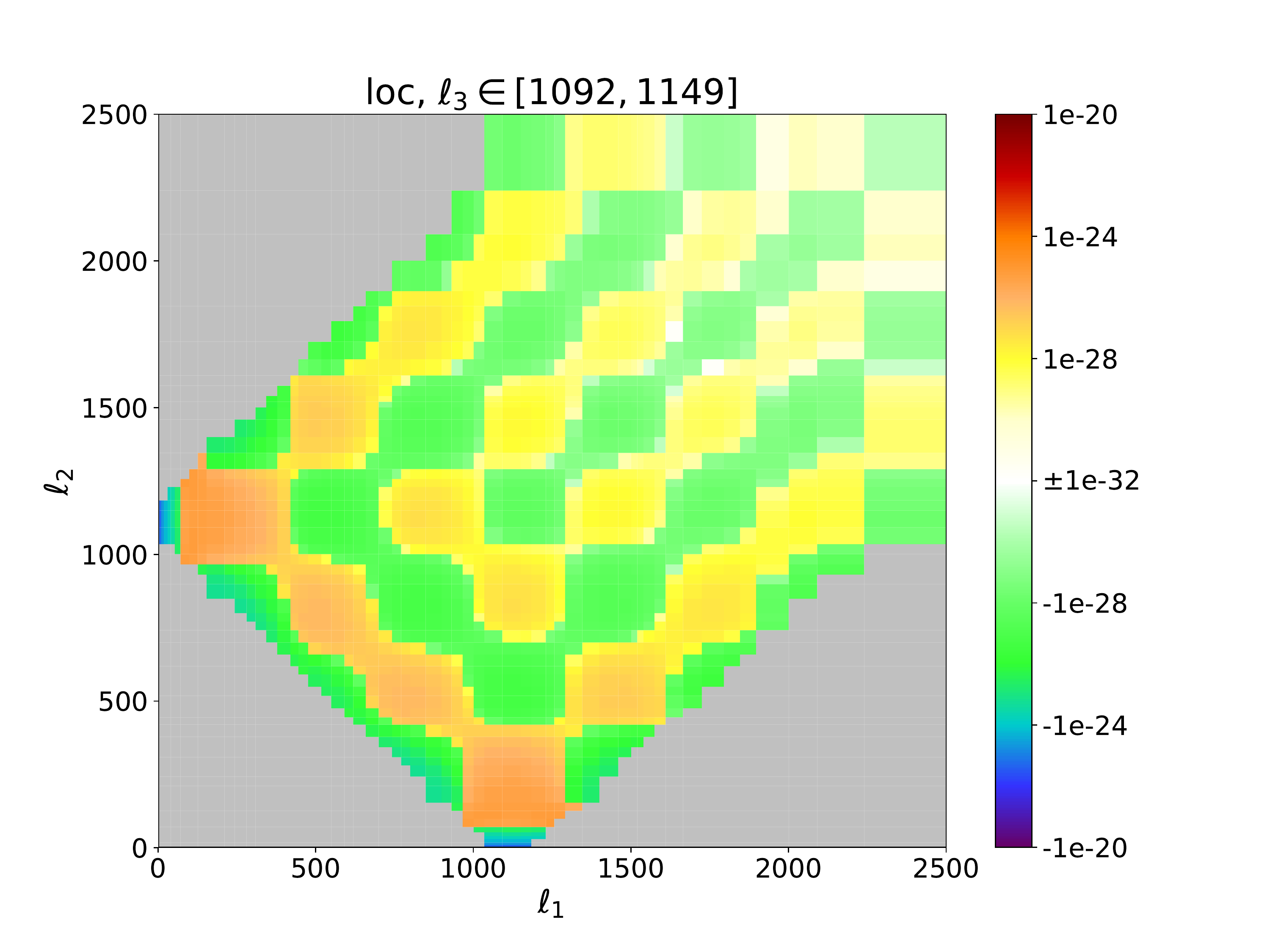}
  \caption{Left: observed binned bispectrum of the thermal dust as a function of the multipoles $\ell_1$ and $\ell_2$ for $\ell_3 \in [1092,1149]$.
  Right: theoretical bispectrum for the local shape in the case $\fnl^\mathrm{local}=1$ for the same bin of $\ell_3$. Note the difference of colour scale (both for maximal and minimal values).}
  \label{fig:dust-template}
\end{figure}

Moving on to the dust bispectrum, we use 2D-slices where the multipoles $\ell_1$ and $\ell_2$ go from 2 to 2500 but $\ell_3$ is in a chosen bin, in order to make it easy to visualize. Figure \ref{fig:dust-template} shows a slice ($\ell_3 \in [1092,1149]$) of the binned dust bispectrum compared to the local shape in the case $\fnl^\mathrm{local}=1$.\footnote{$\fnl^\mathrm{local}=1$ is still well within the observational bounds, but is very large compared to the predictions of standard slow-roll single-field inflation $\fnl^\mathrm{local}\sim\mathcal{O}(10^{-2})$.} If we compare the bispectrum amplitudes, it is clear that the dust is several orders of magnitude larger than the local shape. Moreover, as expected, acoustic oscillations present in both the CMB power spectrum and the local theoretical bispectrum are not there in the case of thermal dust.

\begin{figure}
  \centering
  \includegraphics[width=0.49\linewidth]{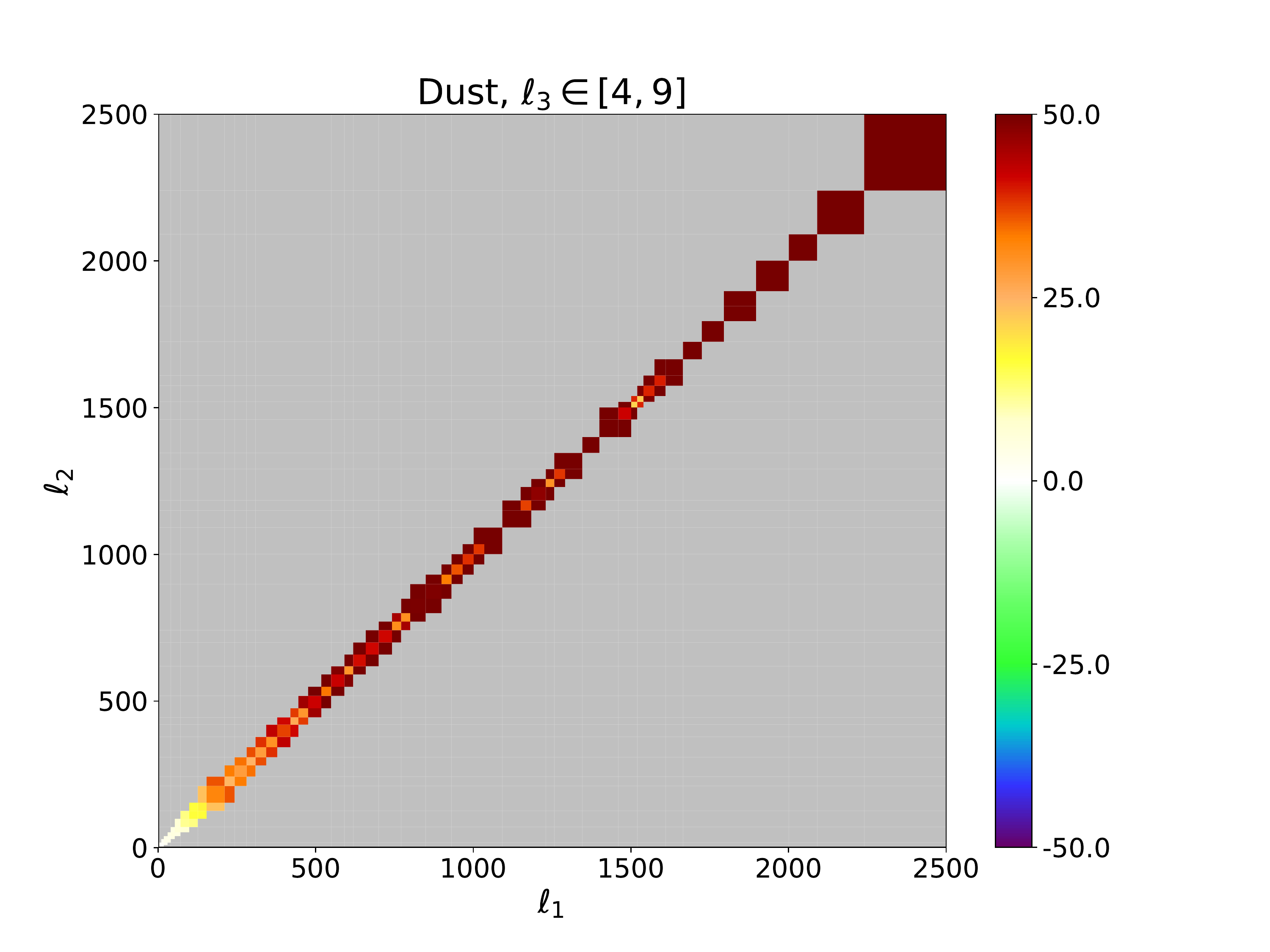}
  \includegraphics[width=0.49\linewidth]{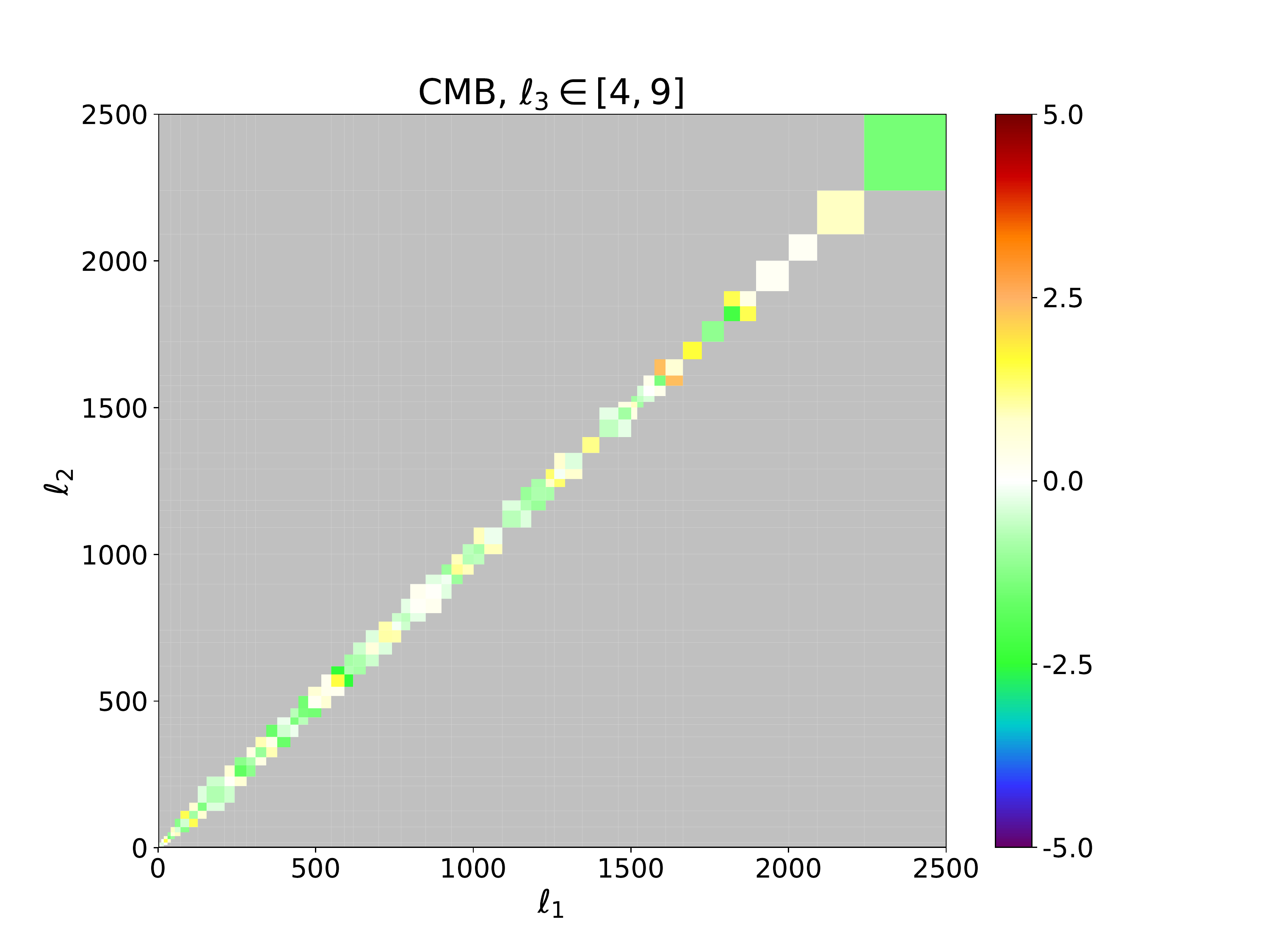}
  \includegraphics[width=0.49\linewidth]{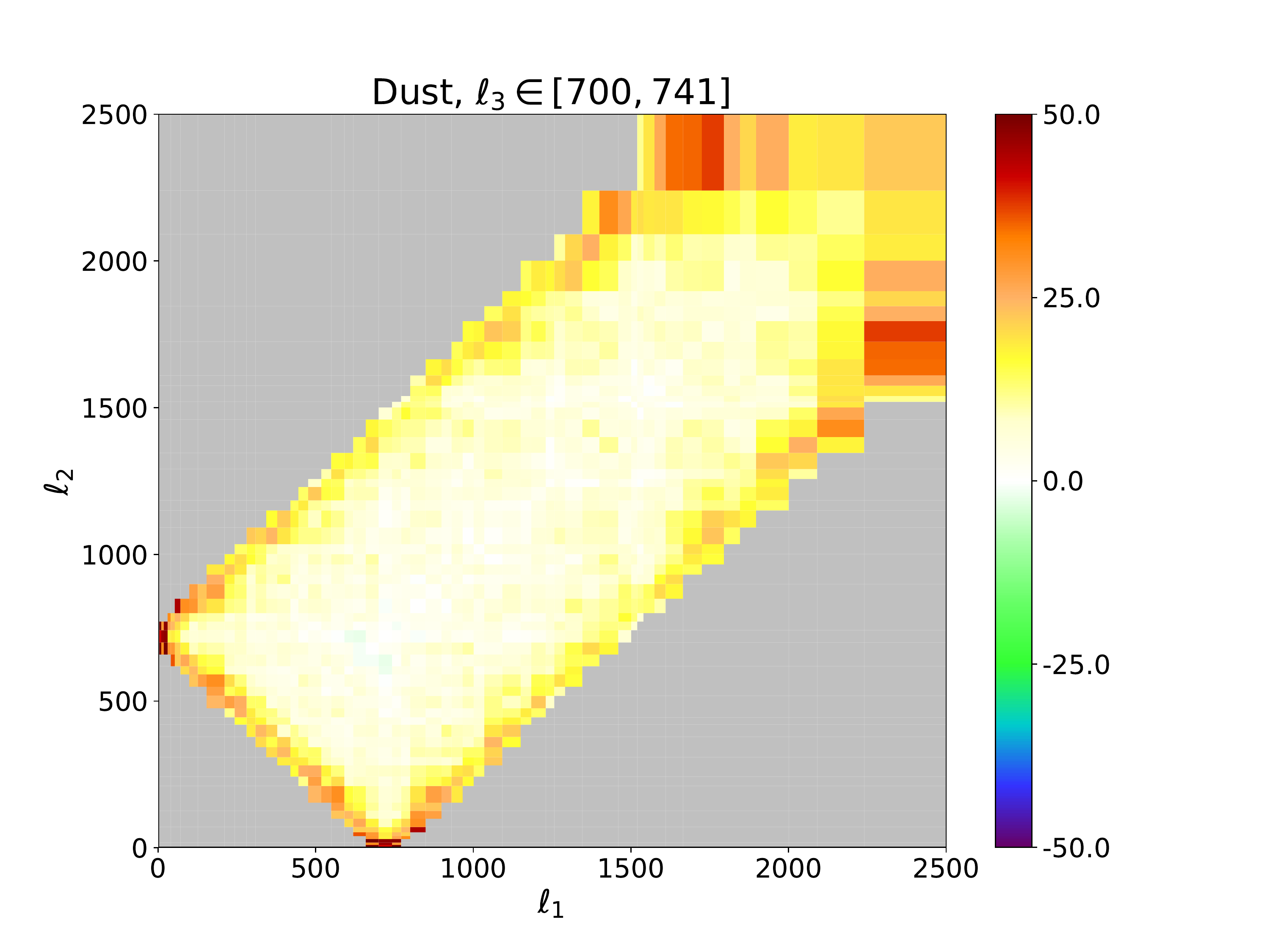}
  \includegraphics[width=0.49\linewidth]{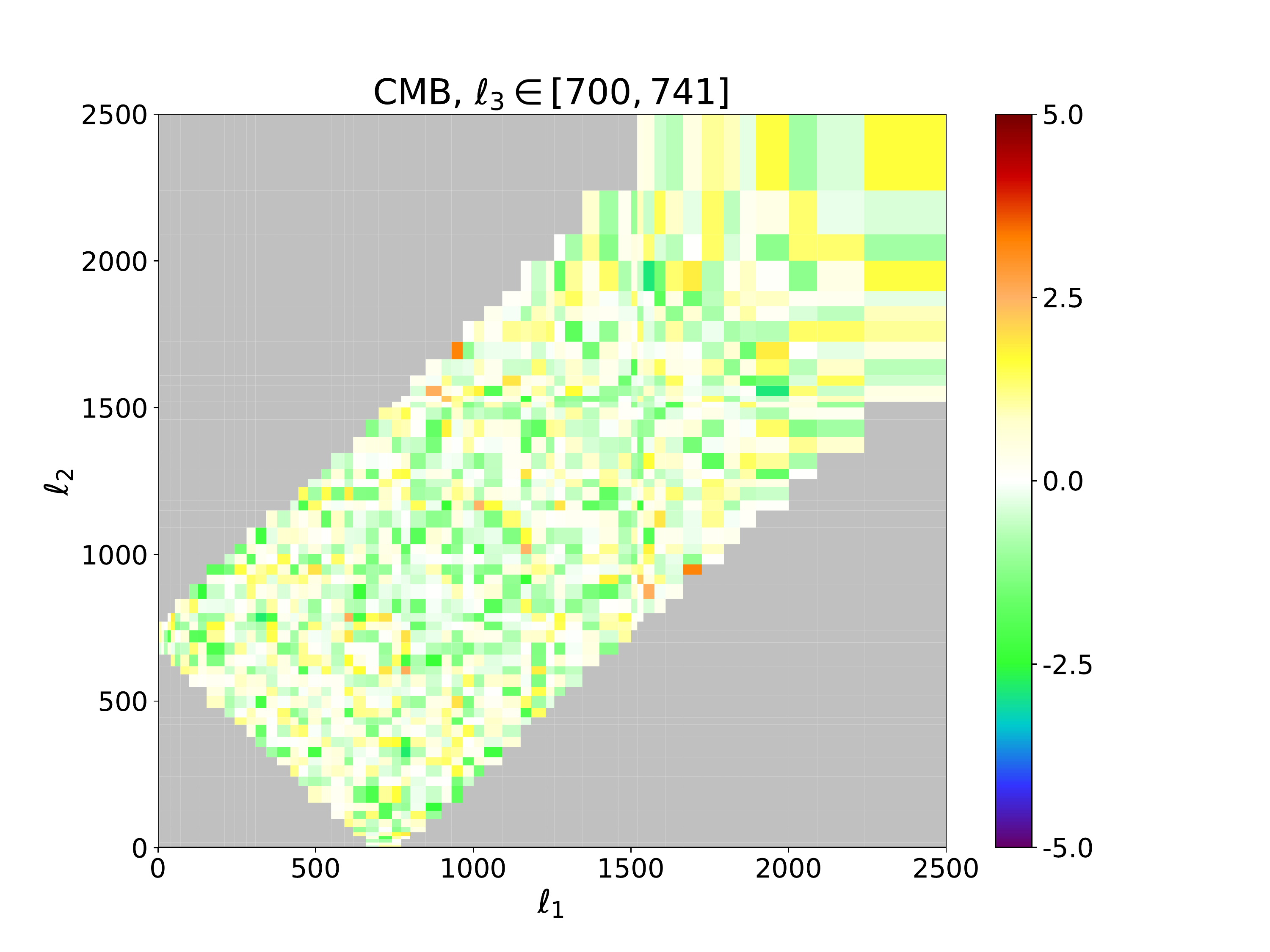}
  \includegraphics[width=0.49\linewidth]{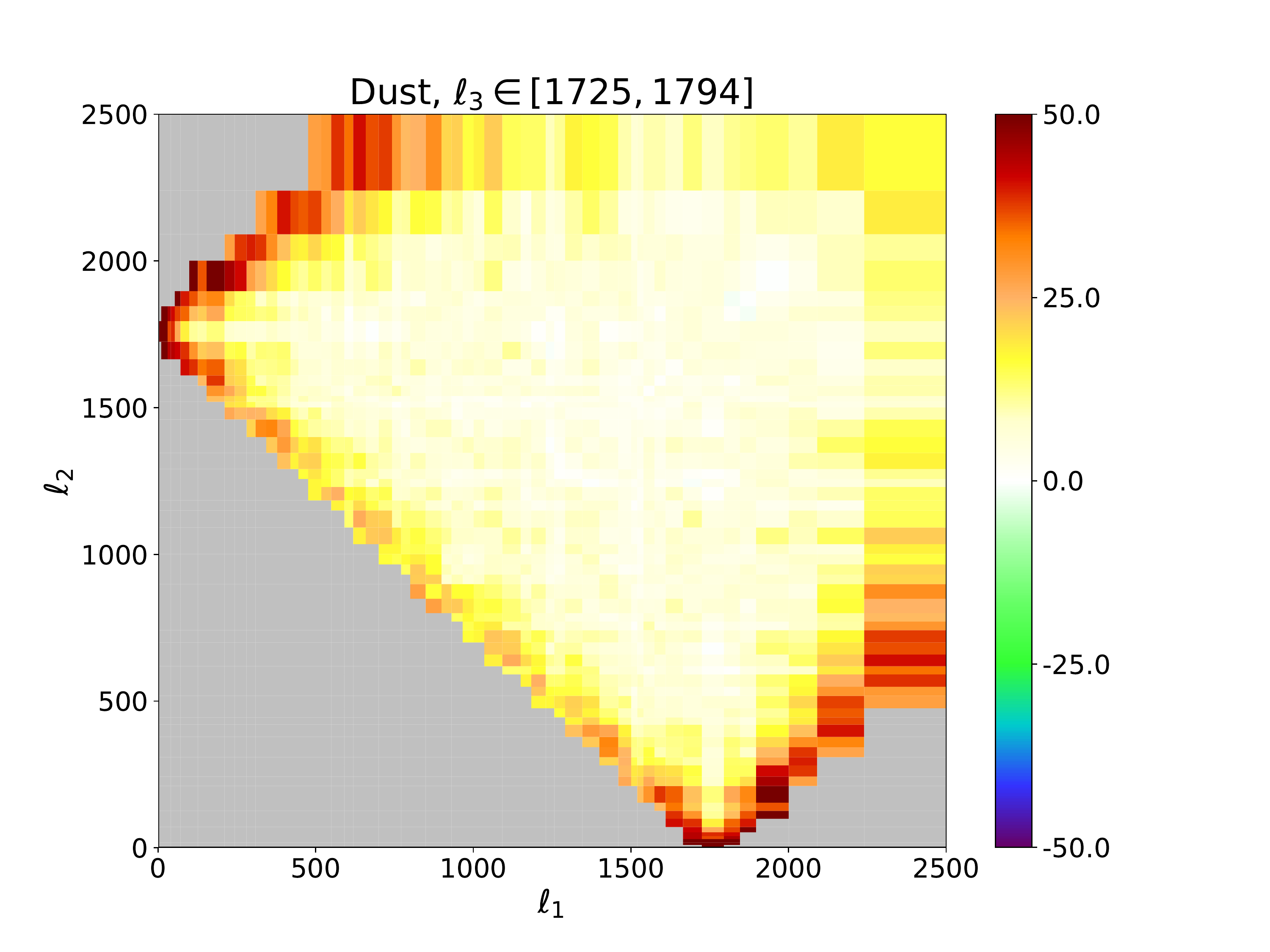}
  \includegraphics[width=0.49\linewidth]{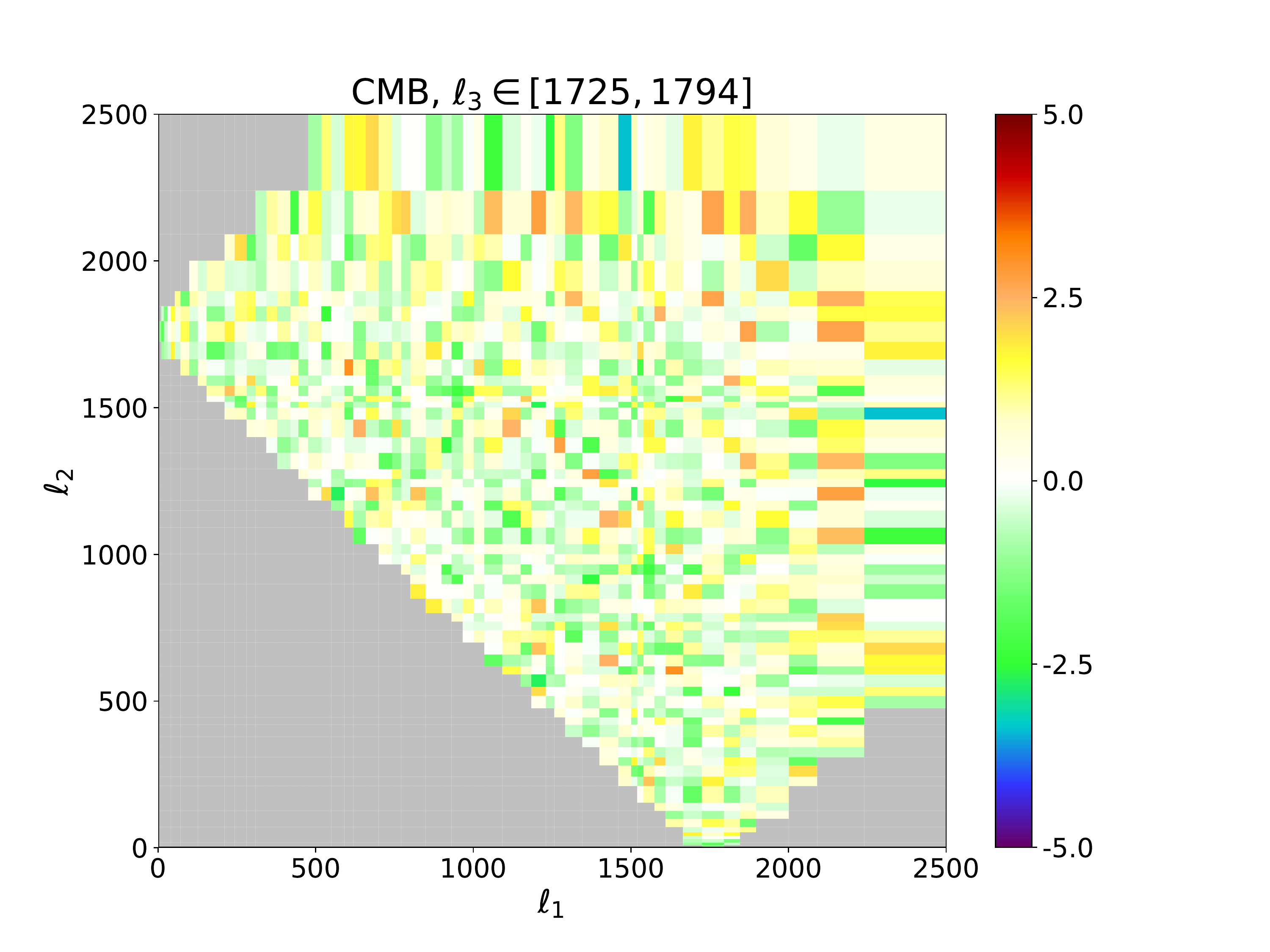}
  \caption{Left: bispectral signal-to-noise of the thermal dust as a function of the multipoles $\ell_1$ and $\ell_2$ for three different bins of $\ell_3$.
  Right: same for the CMB map studied later in section \ref{sec:cmb-analyses} for the same bins of $\ell_3$. Note the factor 10 difference in colour scale.}
  \label{fig:dust-bispectrum}
\end{figure}

However, the plots of figure \ref{fig:dust-template} are not well suited to describe quantitatively the non-Gaussian nature of these shapes. As in the case of the power spectrum, which peaks at low $\ell$ if we do not multiply by the factor $\ell(\ell+1)$, the CMB bispectrum as well as the dust bispectrum have a strong $\ell$ dependence. This means that we should use an adapted function of $\ell$ to highlight the true nature of a bispectral signal. A good choice is to use signal-to-noise plots \cite{Bucher:2015ura} as shown in figure \ref{fig:dust-bispectrum}: the bispectrum is divided by the square root of the variance of the map computed using the power spectrum, see \eqref{eq:variance}. It is important to note that this is different from the correlation coefficients \eqref{eq:shape-correlator} that we discuss later in this section where the variance of the cleaned CMB map is used. In this kind of plots, non-Gaussianity is simply represented by values large compared to $\mathcal{O}(1)$.

Figure \ref{fig:dust-bispectrum} shows the bispectral signal-to-noise ratio for three different slices of the dust template (on the left), but also for a cleaned CMB map which we will study in detail in section~\ref{sec:analyses} (on the right). It is now obvious that the dust map is very non-Gaussian and that indeed its bispectrum peaks in the squeezed configuration. This effect can be seen in the top plot (low $\ell_3$) but also on the left (low $\ell_1$) and on the bottom (low $\ell_2$) of the other plots. A squeezed configuration is expected when there are correlations between small-scale and large-scale effects. There is a simple physical explanation for the origin of these correlations. The large clouds of dust (i.e.\ large-scale fluctuations) have the highest intensity where they are the thickest along the line of sight. Moreover, the brightest parts have stronger fluctuations (small-scale), see \cite{MivilleDeschenes:2007ya} for a discussion, so the small-scale fluctuations are modulated by the large-scale ones which corresponds to a squeezed bispectrum.

The squeezed signal present in both the dust and the local shapes is a good indication that they are correlated. This can be verified in table \ref{tab:corr_coeff_dust} which gives the correlation coefficients between the dust and the standard shapes computed using \eqref{eq:shape-correlator} in the context of a CMB analysis (more details in section \ref{sec:analyses}), so the denominator of the inner product is the CMB bispectrum variance. There is an anti-correlation between the dust and local shapes (60~$\%$) because they have opposite signs (this anti-correlation was pointed out in \cite{Yadav:2007yy}). The local shape is itself correlated to the other primordial shapes (see table~\ref{tab:corr_coeff_2015}). However, this does not mean that the dust template has to be correlated to them too. And indeed, the dust and equilateral shapes are uncorrelated because the latter does not peak in the squeezed configuration. The correlations between local and dust (squeezed) do not come from the same multipole triplets as the correlations between local and equilateral (acoustic peaks). However, the orthogonal and dust shapes are a little correlated (around 15~$\%$), because the orthogonal bispectrum in the squeezed limit is large. The dust bispectrum template is very weakly correlated to extra-galactic foreground templates like unclustered point sources and CIB, but anti-correlated to lensing-ISW (which is known to be highly correlated to the local shape). An alternative representation of the bispectra of the different shapes, which shows in which regions of multipole space they dominate, is given in appendix~\ref{ap:weights-bisp-shap}.

\begin{table}
  \begin{center}
    \begin{tabular}{l|cccccc}
      \hline
      & Local & Equilateral & Orthogonal & Lensing-ISW & Point sources & CIB \\
      \hline
      Dust & -0.6 & 0.004 & 0.15 & -0.34 & 0.054 & 0.083\\
      \hline
    \end{tabular}
  \end{center}
  \caption{Correlation coefficients between the standard theoretical templates and the observed dust bispectrum computed using the characteristics of the Planck experiment (temperature).}
  \label{tab:corr_coeff_dust}
\end{table}

\subsection{Other foregrounds}
\label{sec:other-foregrounds}

Apart from dust, there exist other foregrounds which have a greater effect at low frequencies, of which we will study three here. In this section we use maps produced by the \texttt{Commander} method to separate foregrounds, but this time in addition to the Planck data, observations from WMAP between 23 and 94 GHz \cite{bennett2013nine} and a 408 MHz survey map \cite{haslam1982408} were also used to determine them. They have a lower resolution ($n_{\mathrm{side}}=256$) and a larger beam (60' FWHM Gaussian beam). For the sake of comparison of these foregrounds with the dust we discussed in the previous section, we will also use here a dust map with the same characteristics.

In the case the dust grains rotate rapidly (in addition to their thermal vibrations), they can produce a microwave emission which probably corresponds to the anomalous microwave emission (AME) \cite{Leitch:1997dx, Draine:1998gq}, large at low frequencies.

Dust is not the only component responsible for the contamination of the CMB signal; some interactions of electrons with the interstellar medium can also generate emissions. On the one hand, ultra-relativistic electrons (cosmic rays) spiraling in the galactic magnetic fields radiate. This synchrotron emission can be described by a power law $\nu^\beta$ with $\beta \simeq -3$ indicating that indeed, this radiation is significant at low frequencies \cite{haslam1982408}. On the other hand, electrons can be slowed down by scattering off ions. This generates the free-free emission \cite{Dickinson:2003vp}, also called bremsstrahlung.

The frequency dependence of the foregrounds and the CMB signal can be seen in figure 51 of \cite{Adam:2015wua}. As discussed, the synchrotron, the free-free and the spinning dust (AME) emissions dominate at low frequencies. The dust thermal emission is the main contribution at high frequencies and is of the same order as the CMB at 143 GHz (this of course depends on the choice of mask). 

\begin{figure}
  \centering  \includegraphics[width=0.49\linewidth]{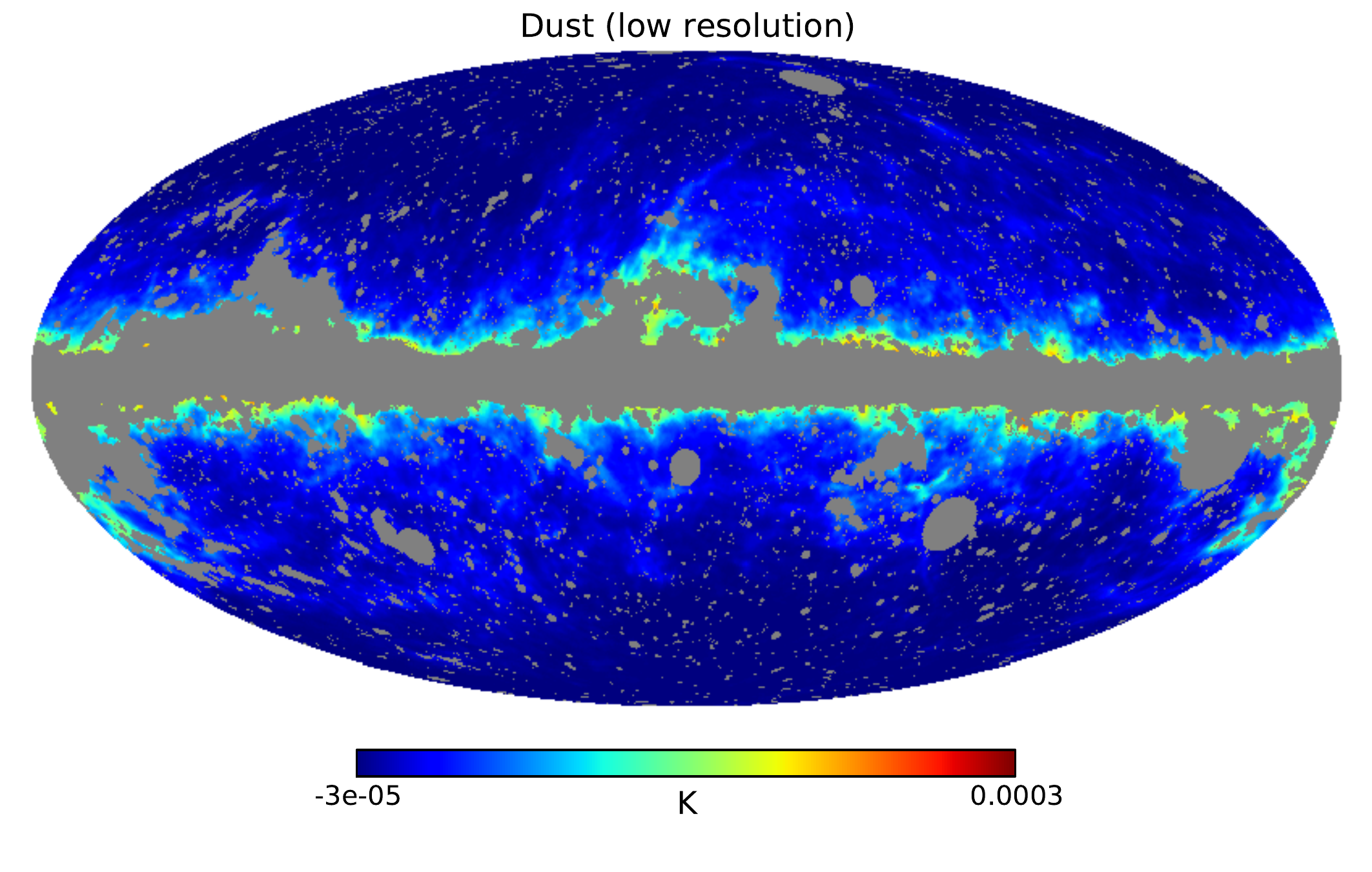}
\includegraphics[width=0.49\linewidth]{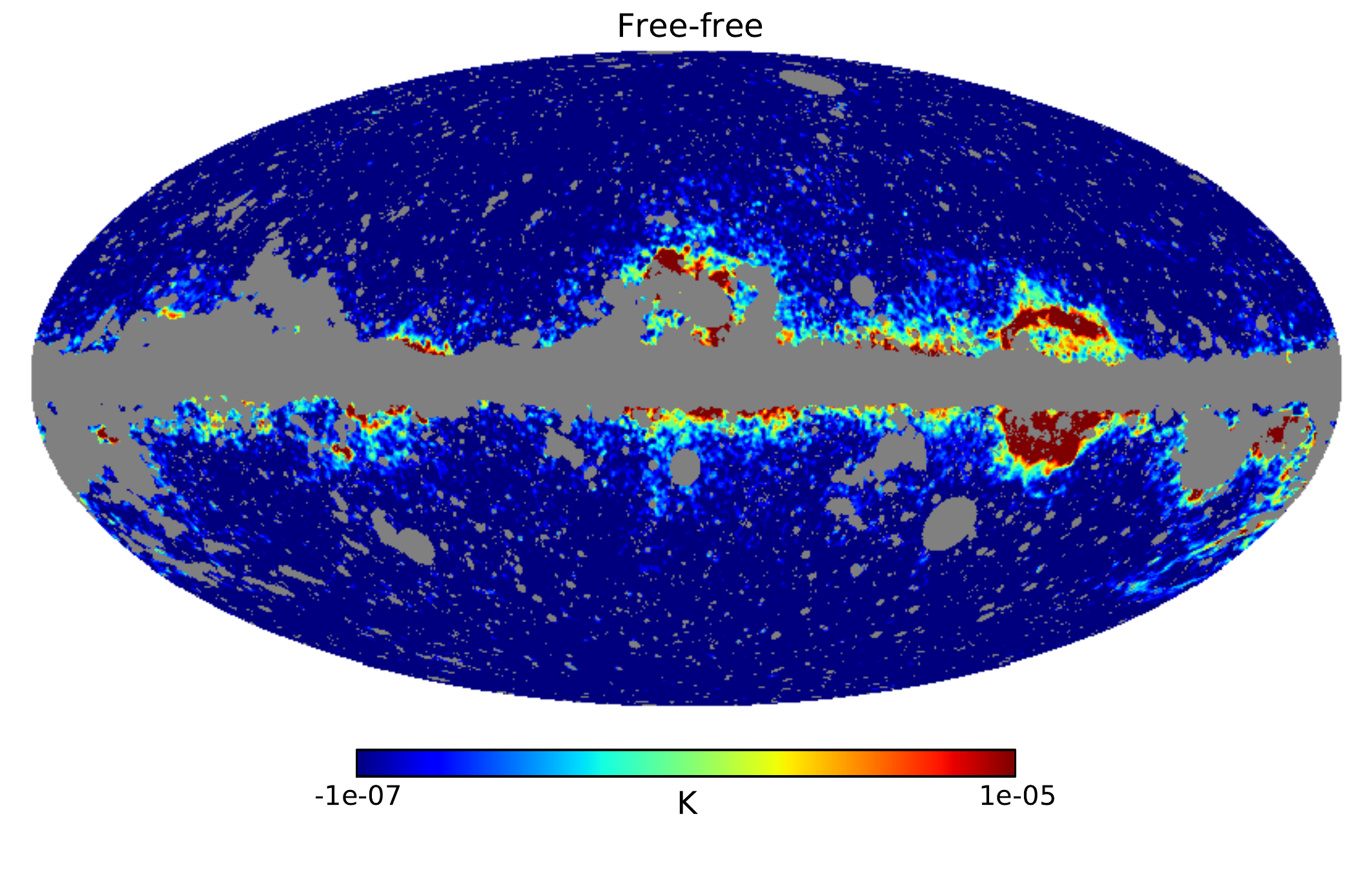}
\includegraphics[width=0.49\linewidth]{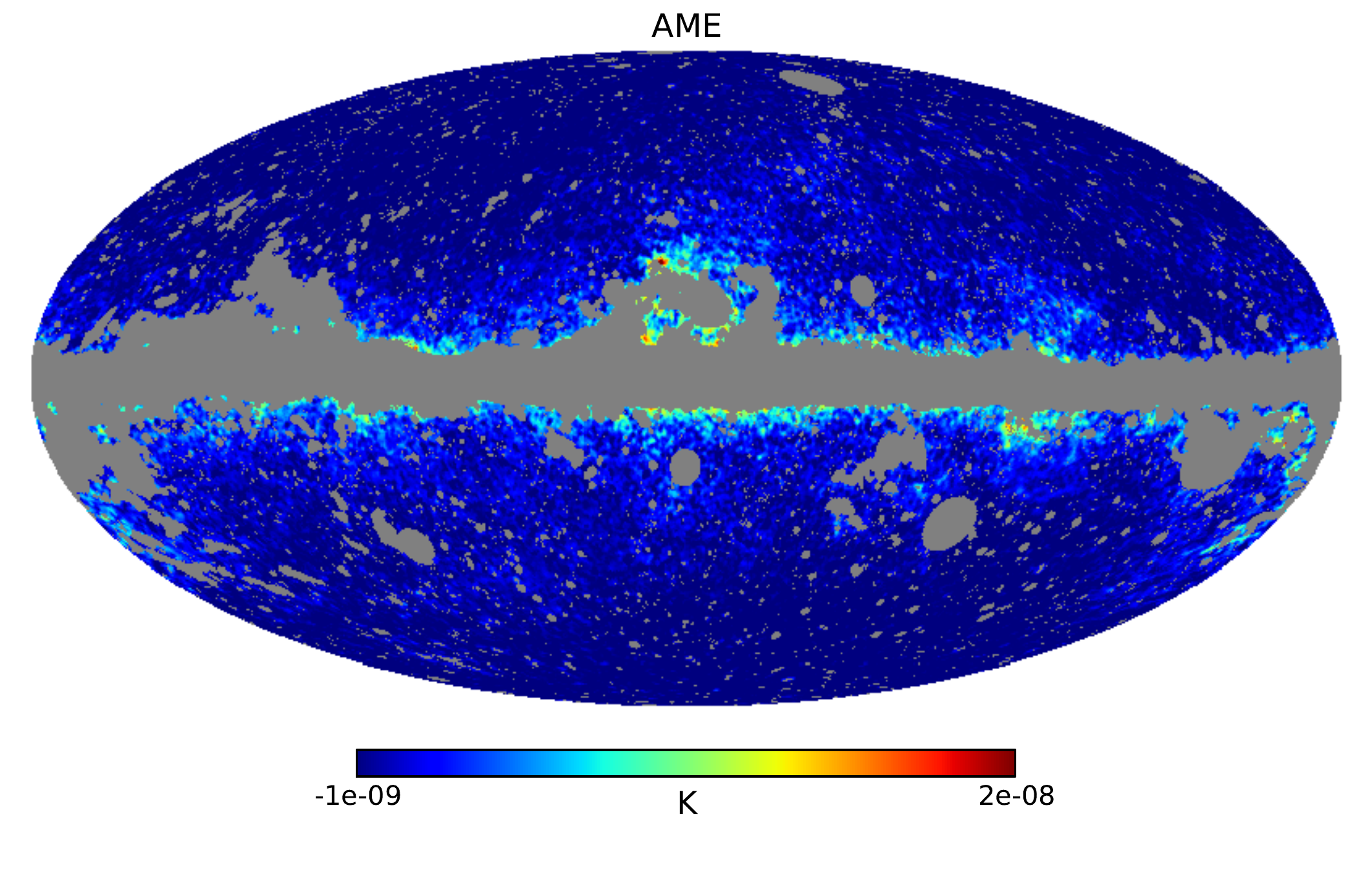}
\includegraphics[width=0.49\linewidth]{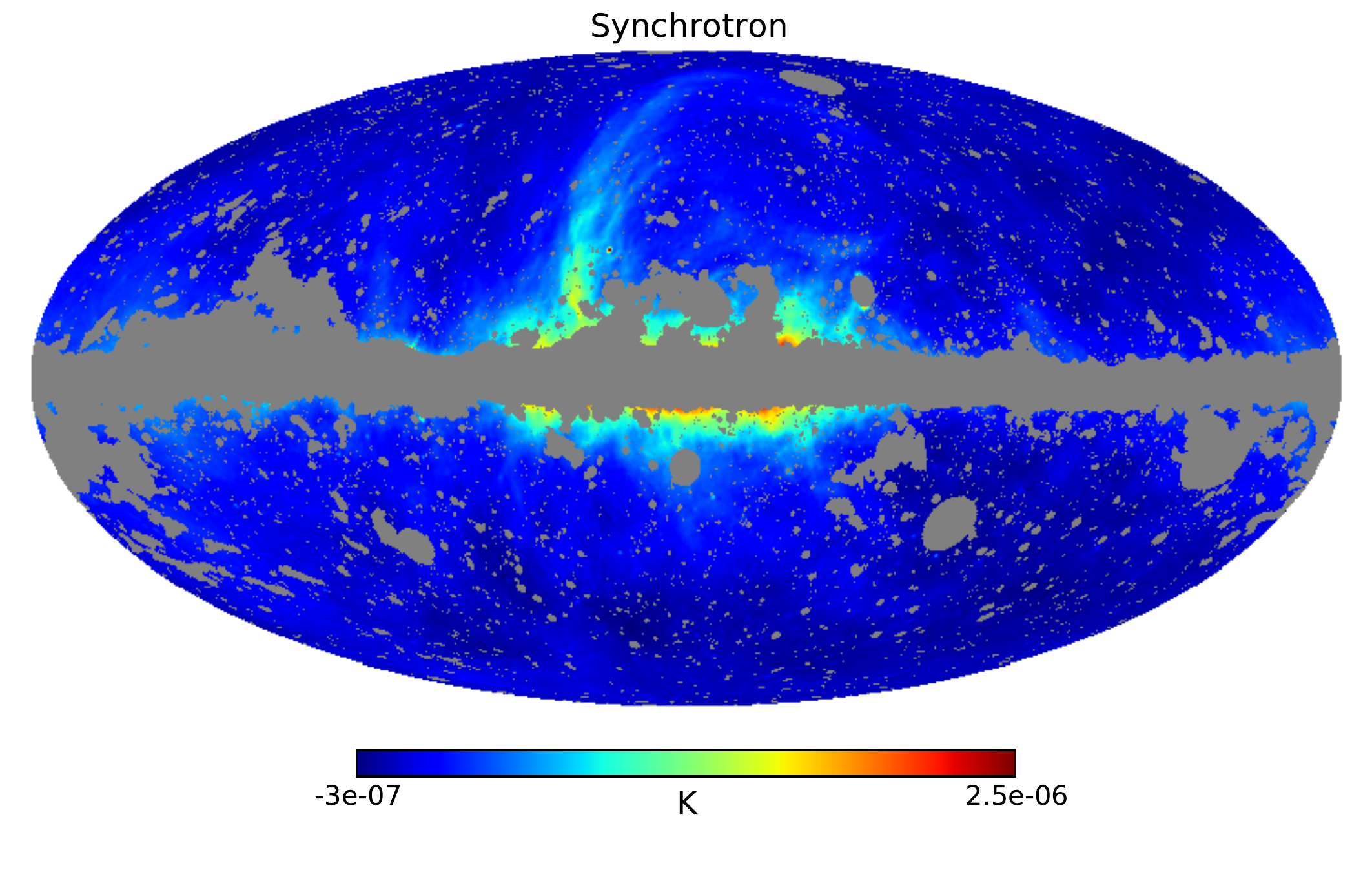}
\caption{Masked maps of the different galactic foregrounds we consider in this paper (dust, free-free, AME, synchrotron) at 143 GHz derived using the \texttt{Commander} method from Planck 2015 data. Note the different colour scales.}
  \label{fig:other-maps}
\end{figure}

\begin{figure}
  \centering
  \includegraphics[width=0.66\linewidth]{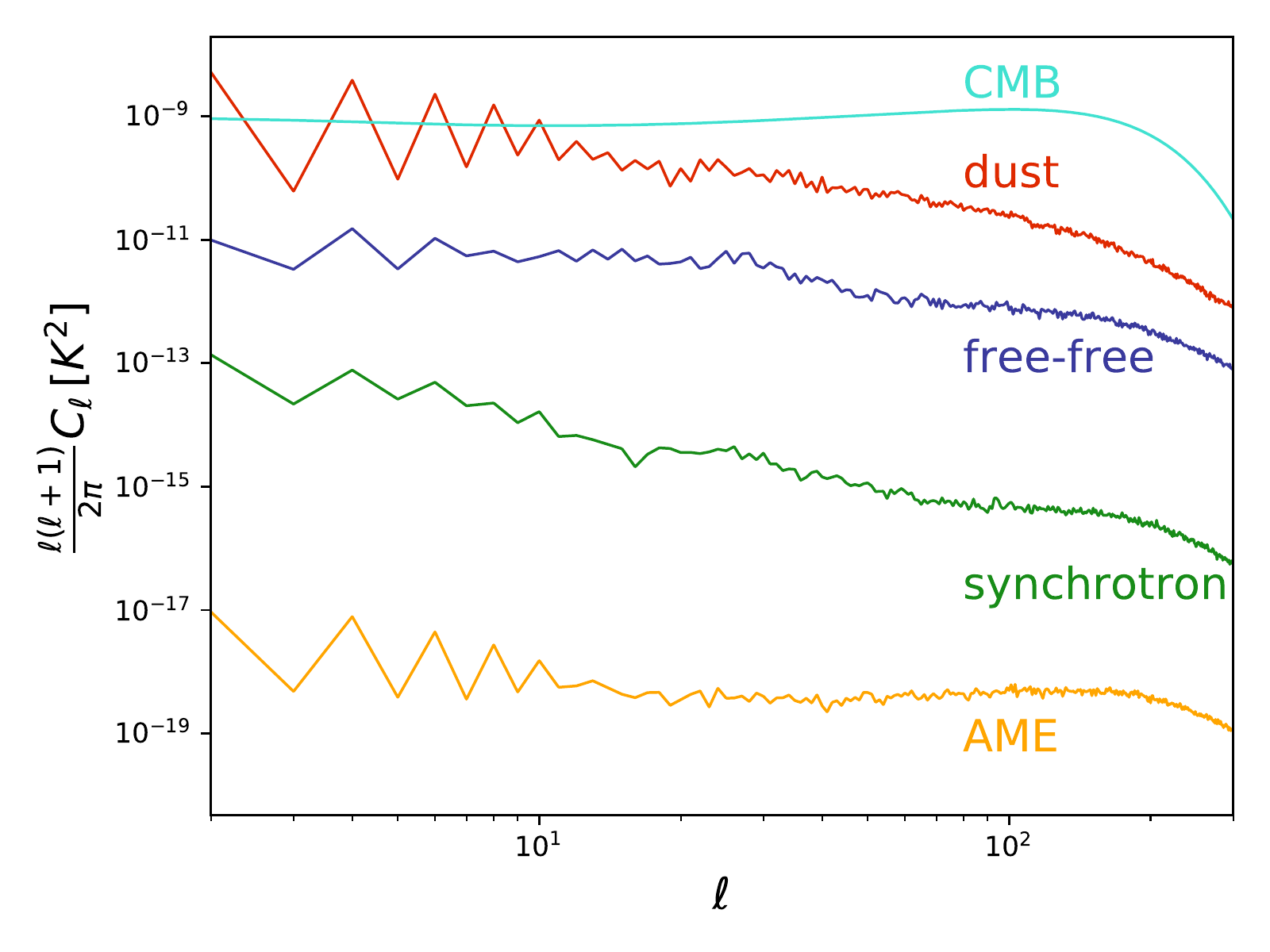}
  \caption{The different foregrounds (dust, free-free, synchrotron and AME) and the CMB power spectra at 143 GHz (including the 60 arcmin beam and the noise) as a function of the multipole $\ell$.}
  \label{fig:all-power-spectra}
\end{figure}

Figure \ref{fig:other-maps} shows the contributions of all these foregrounds at 143 GHz. Similarly to the dust in the previous section, they are all localized in the galactic plane. Moreover, we can see that the dust signal has a higher intensity and therefore is the dominant foreground contribution at 143 GHz. The same hierarchy can be seen in the power spectra, as shown in figure \ref{fig:all-power-spectra}. It is clear that at 143 GHz, the contributions of AME, synchrotron and free-free are negligible compared to the CMB (remember that the brightest parts of the sky are masked). Note that because of the low resolution of the map and the 60 arcmin beam, the range of multipoles is a lot smaller than in the previous section ($\ell_\mathrm{max}=300$ here). This also means that we were able to use smaller bins for the binned bispectrum estimator. We simply took the usual binning, with each bin split into three when possible (two otherwise). 

\begin{figure}
  \centering 
\includegraphics[width=0.32\linewidth]{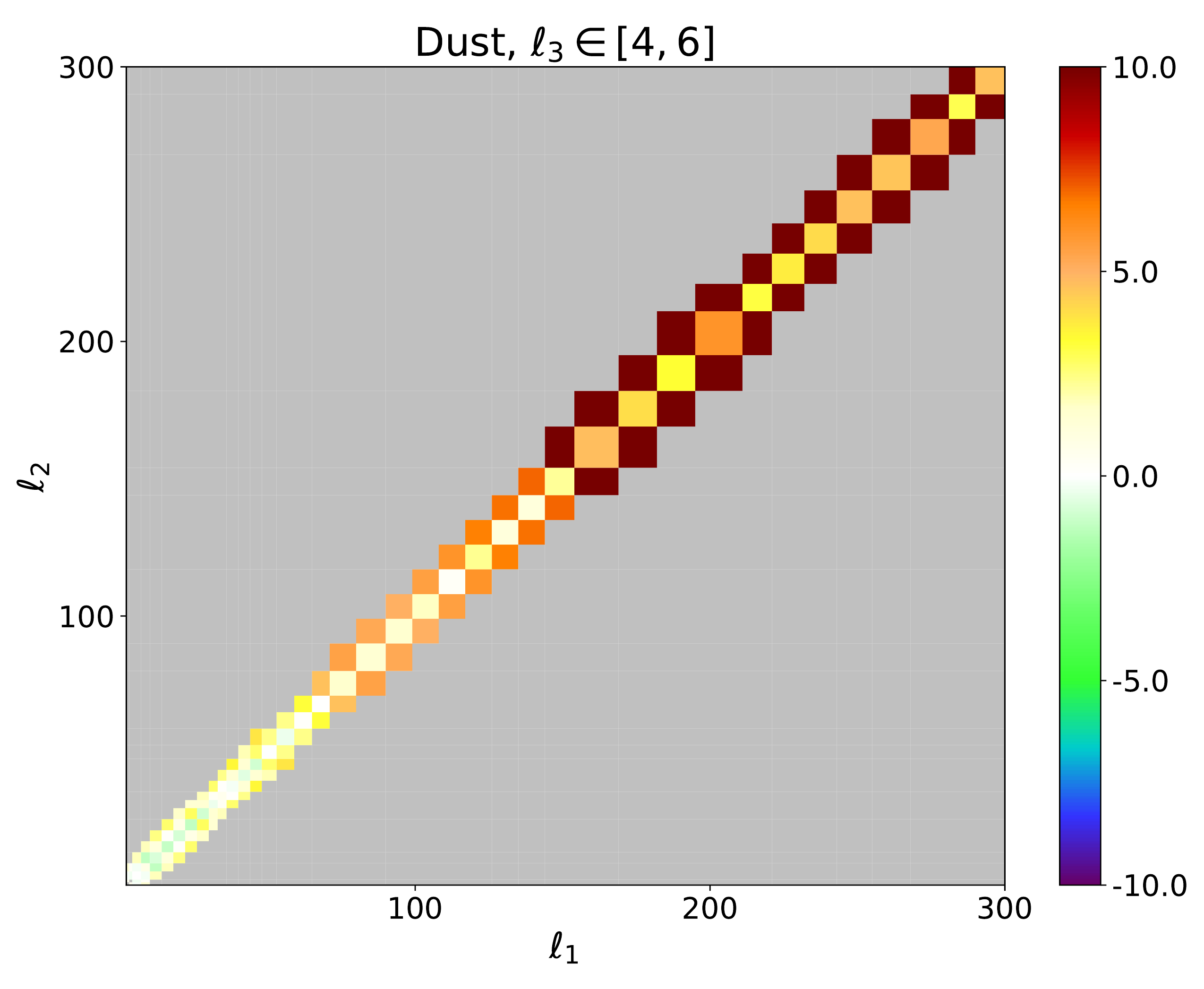}  
\includegraphics[width=0.32\linewidth]{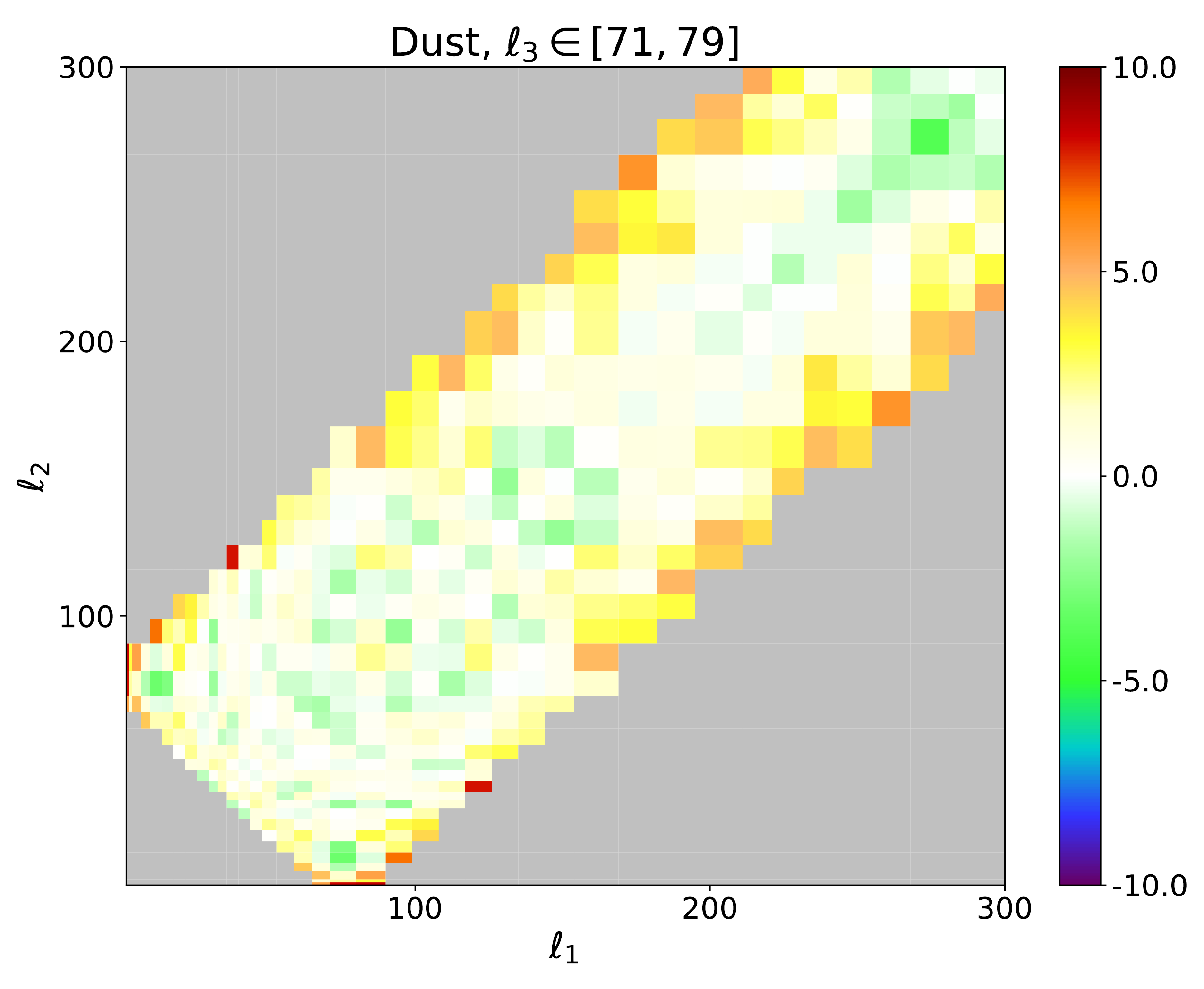} 
\includegraphics[width=0.32\linewidth]{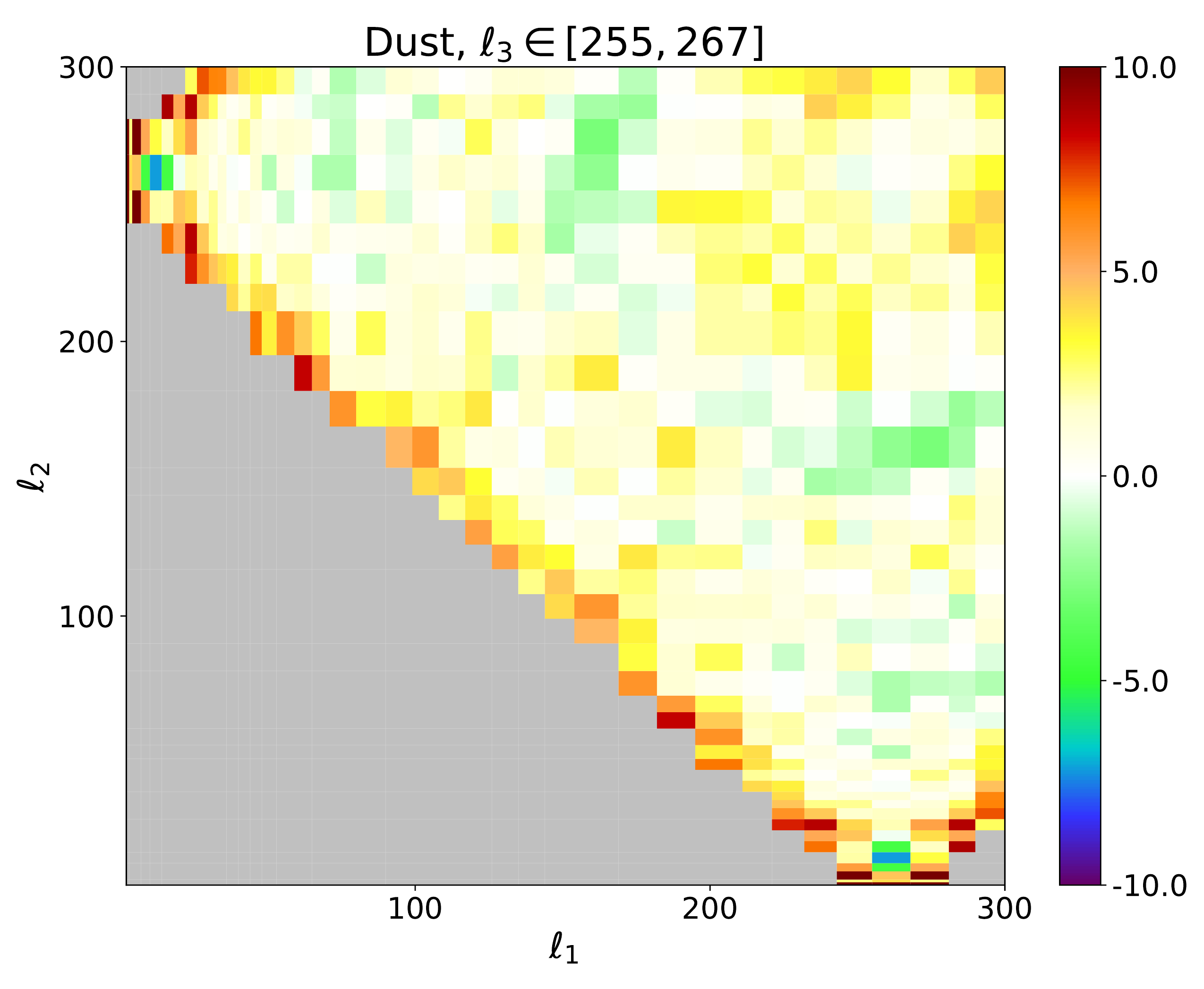}
\includegraphics[width=0.32\linewidth]{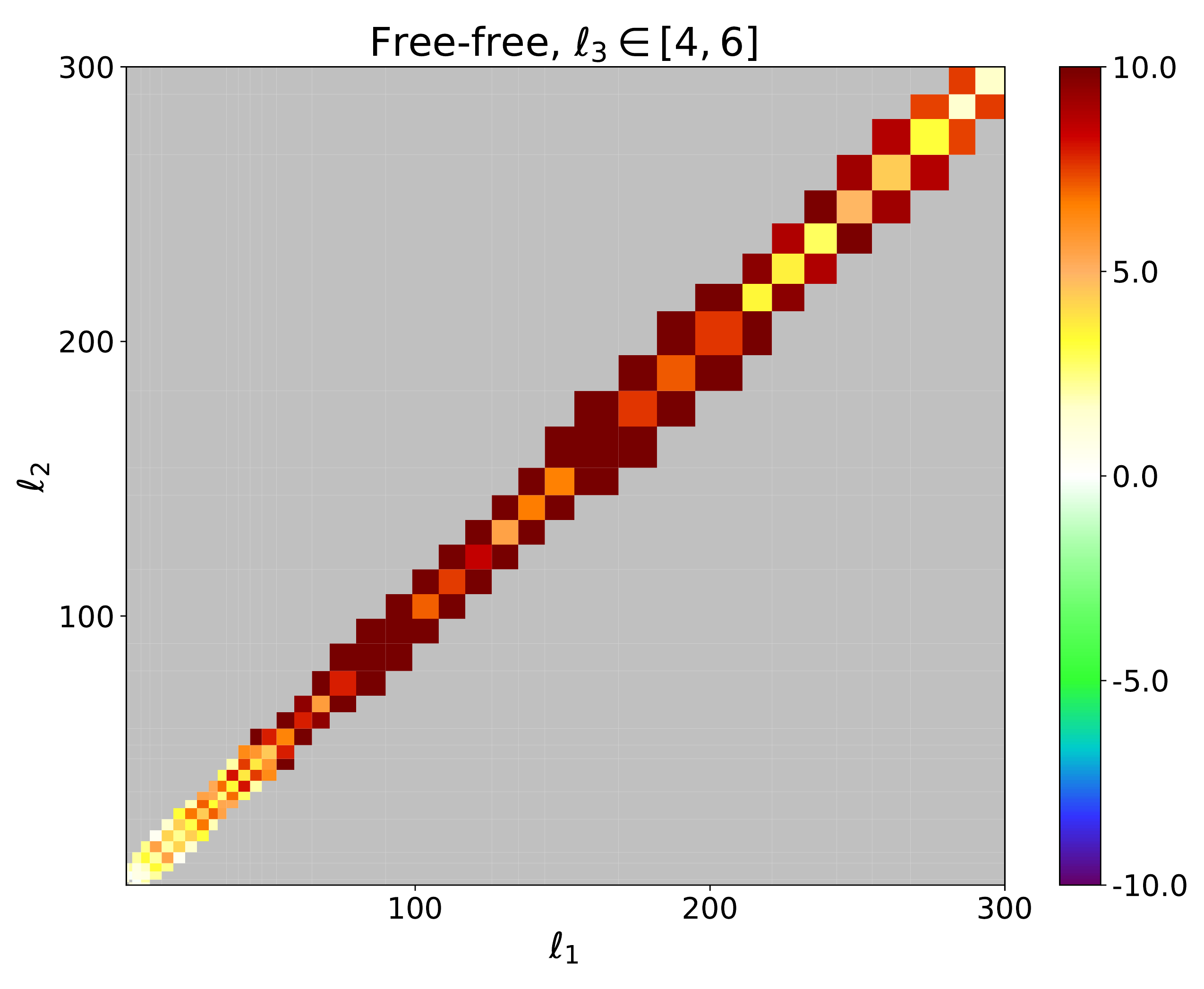}   
\includegraphics[width=0.32\linewidth]{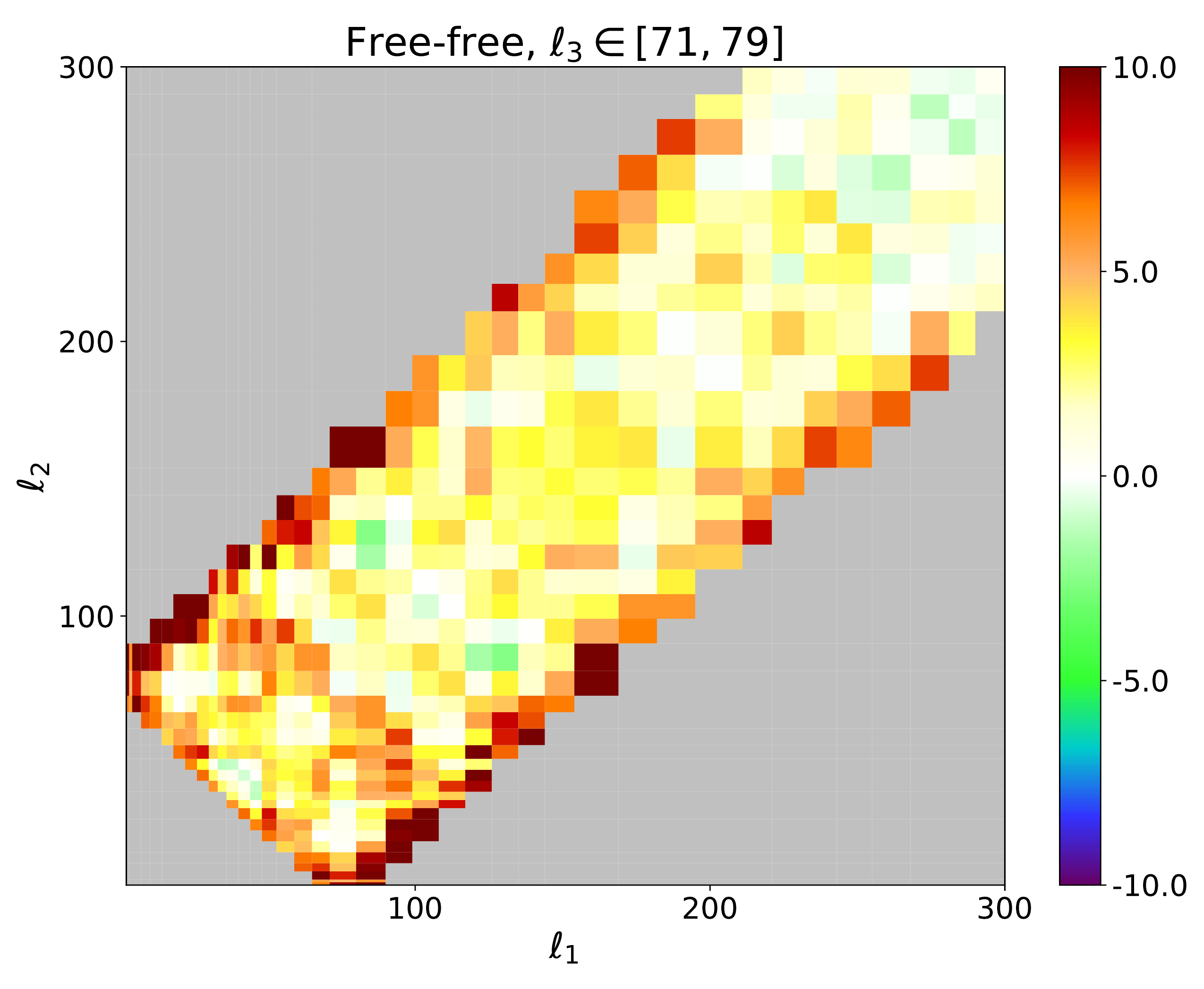}
\includegraphics[width=0.32\linewidth]{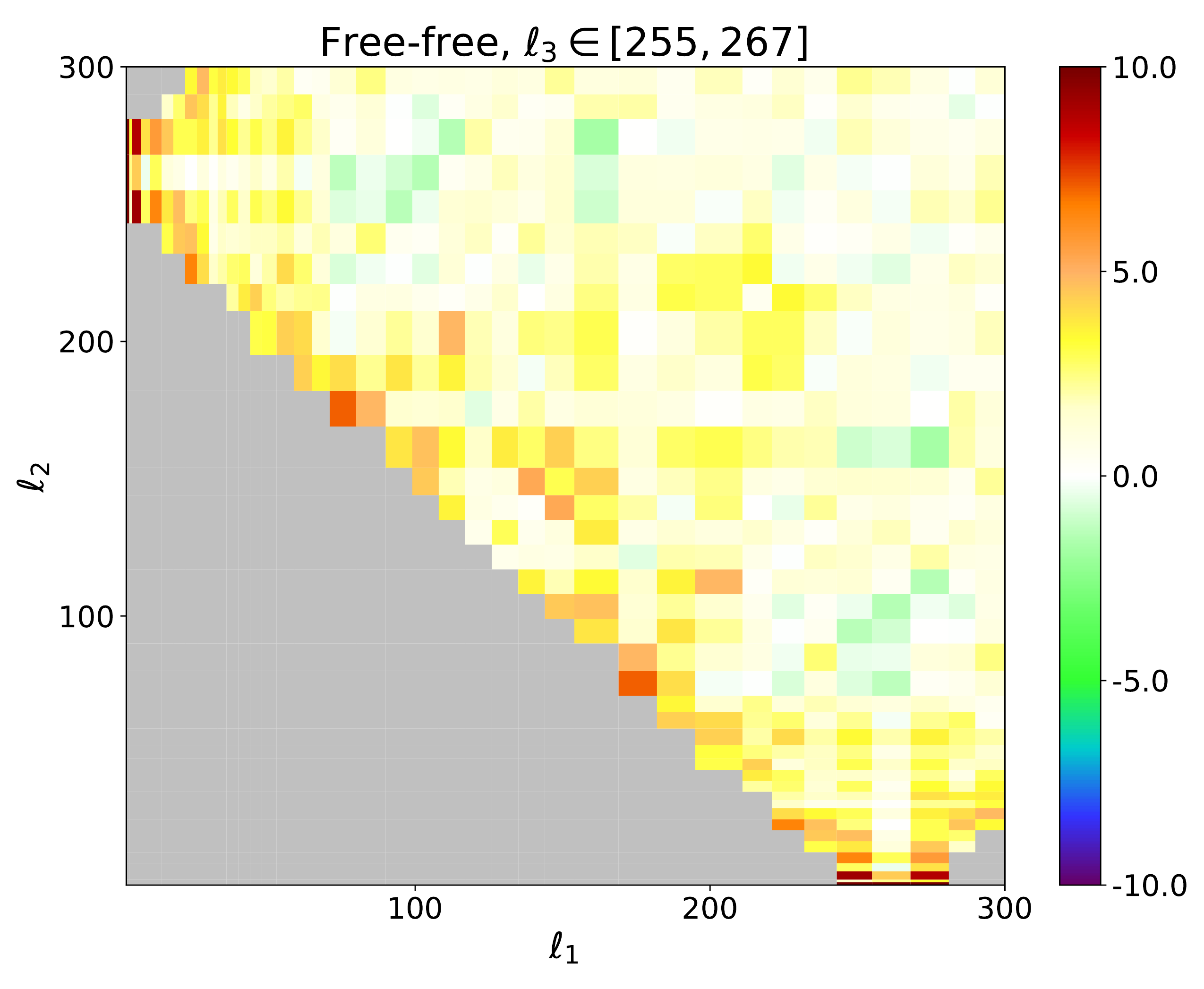}
\includegraphics[width=0.32\linewidth]{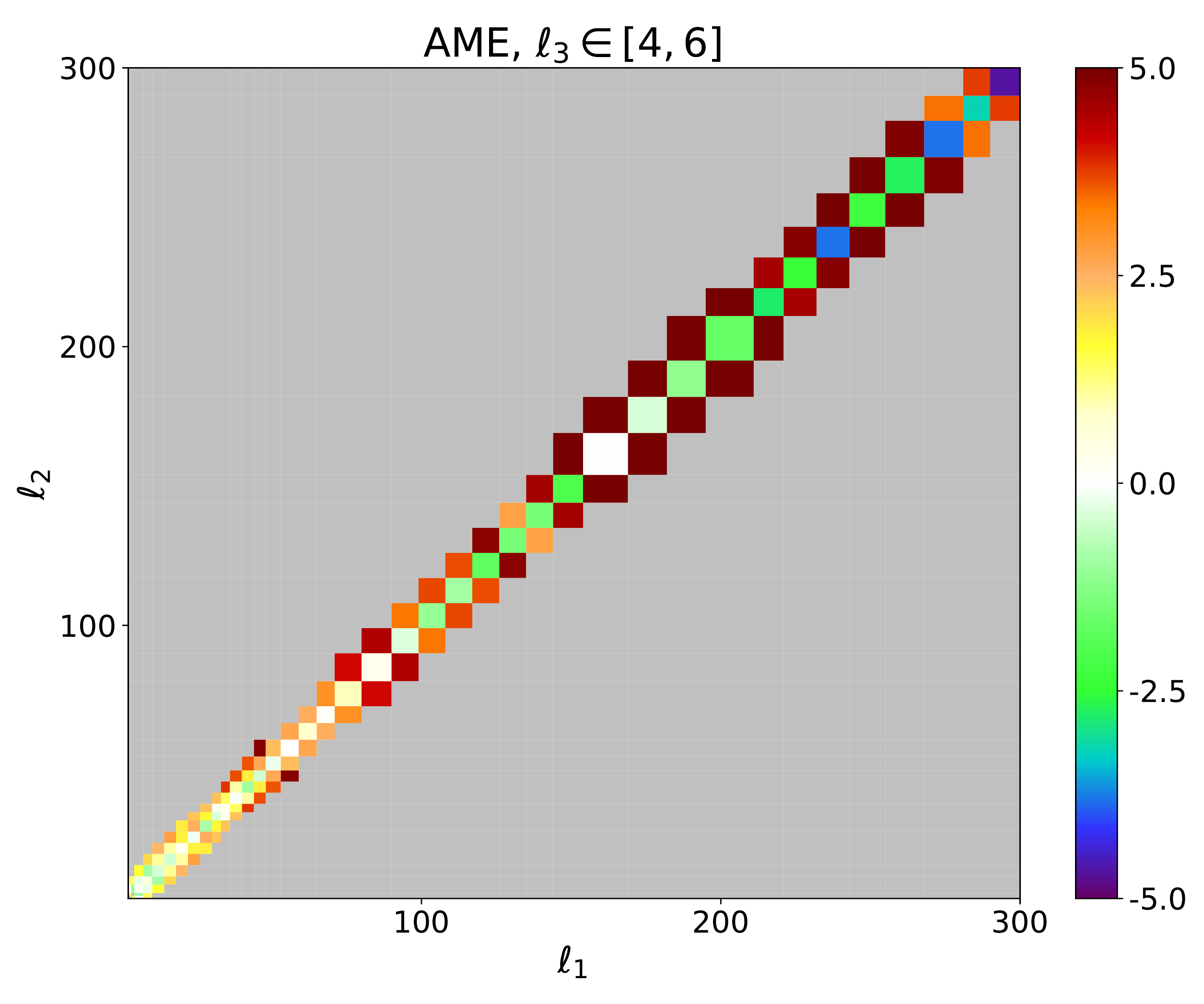} 
\includegraphics[width=0.32\linewidth]{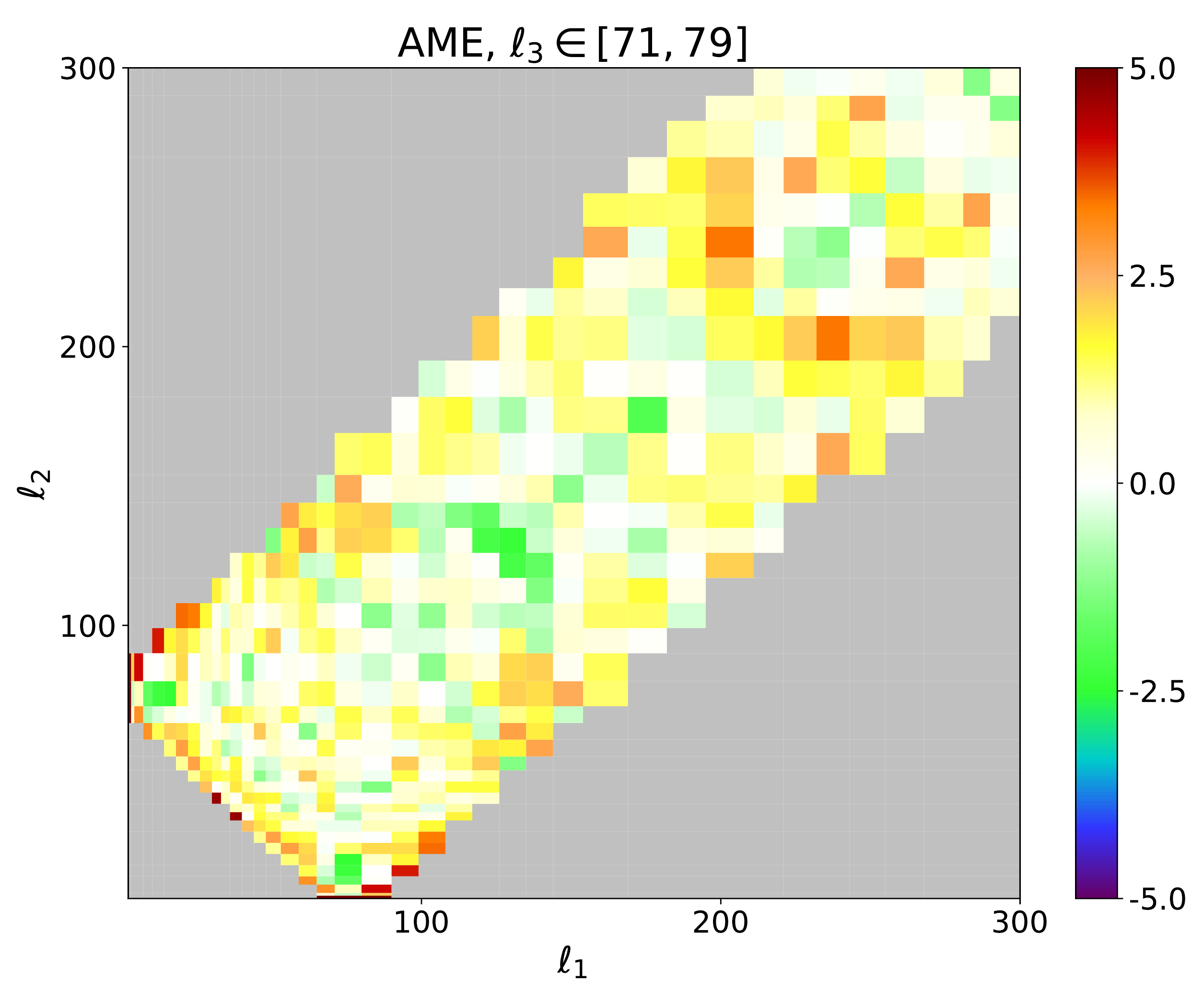}  
\includegraphics[width=0.32\linewidth]{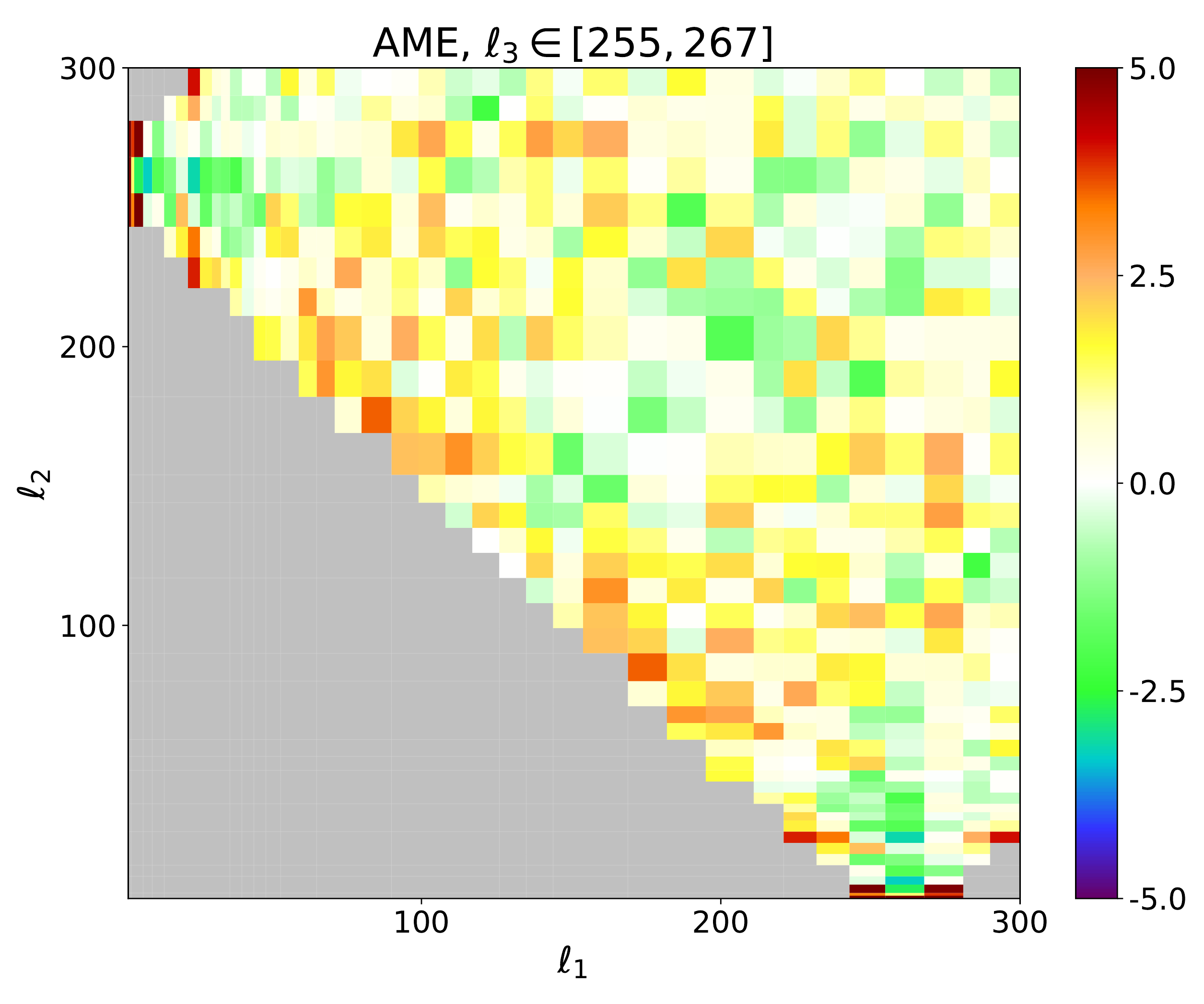}
\includegraphics[width=0.32\linewidth]{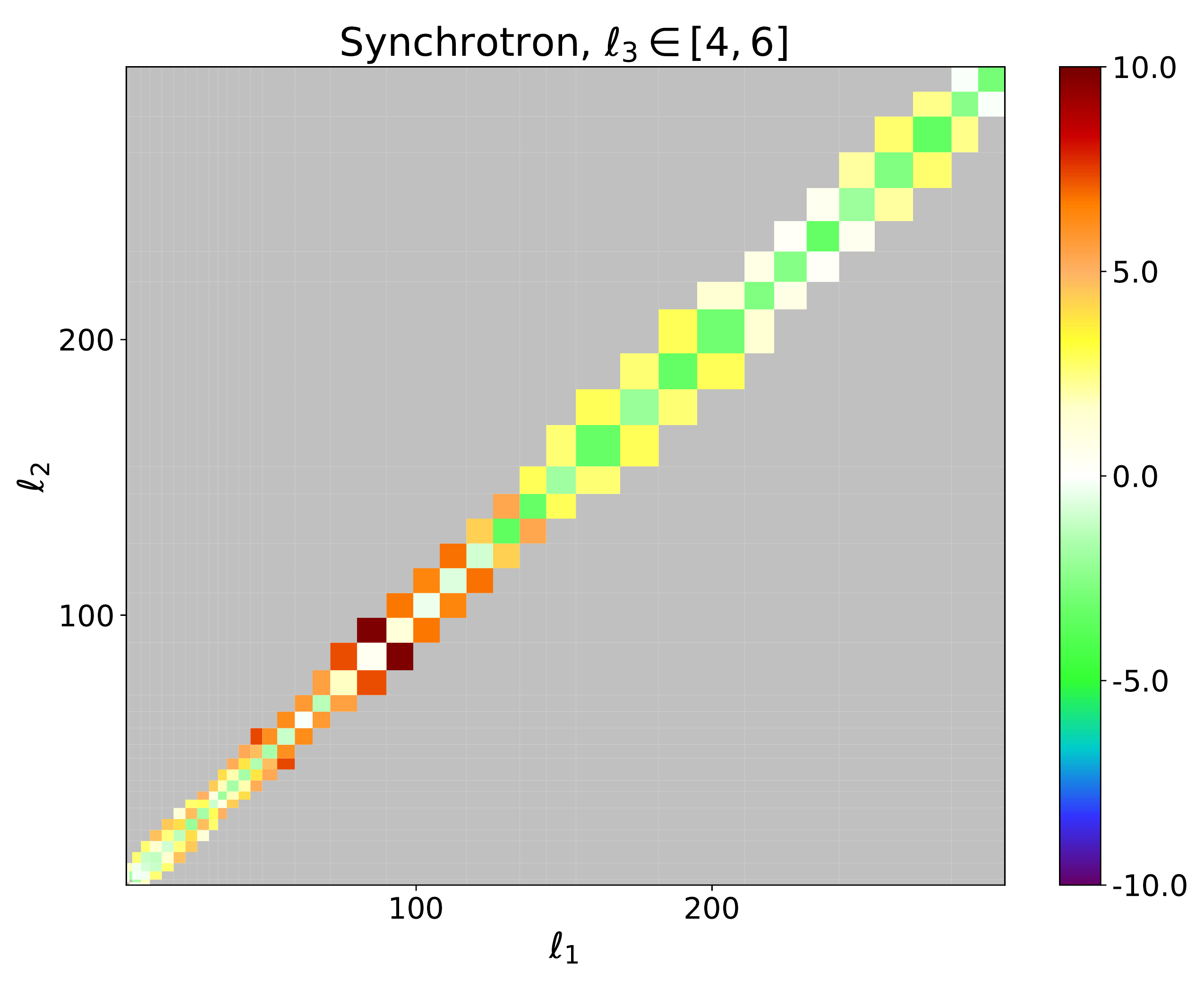}
\includegraphics[width=0.32\linewidth]{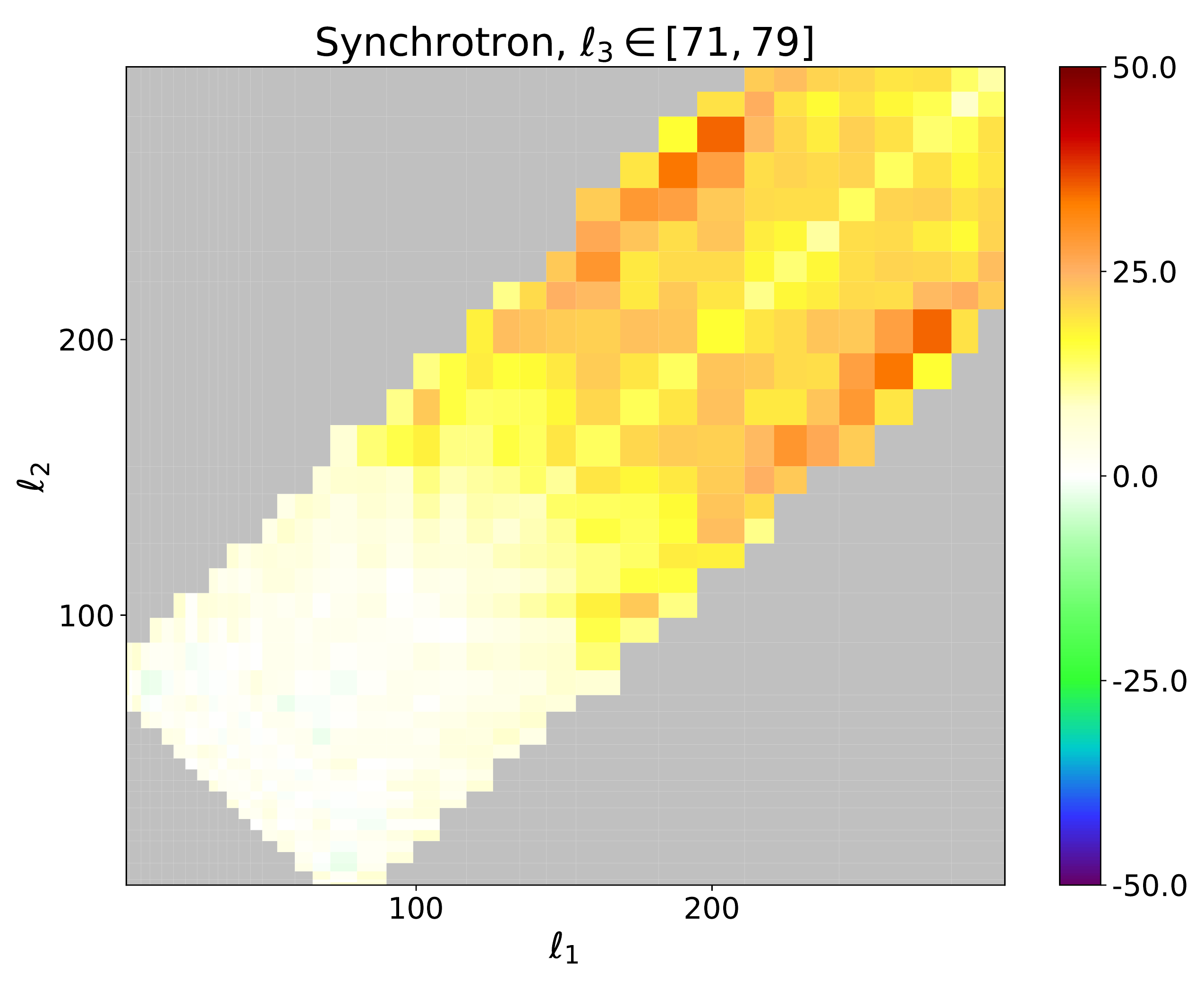}
\includegraphics[width=0.32\linewidth]{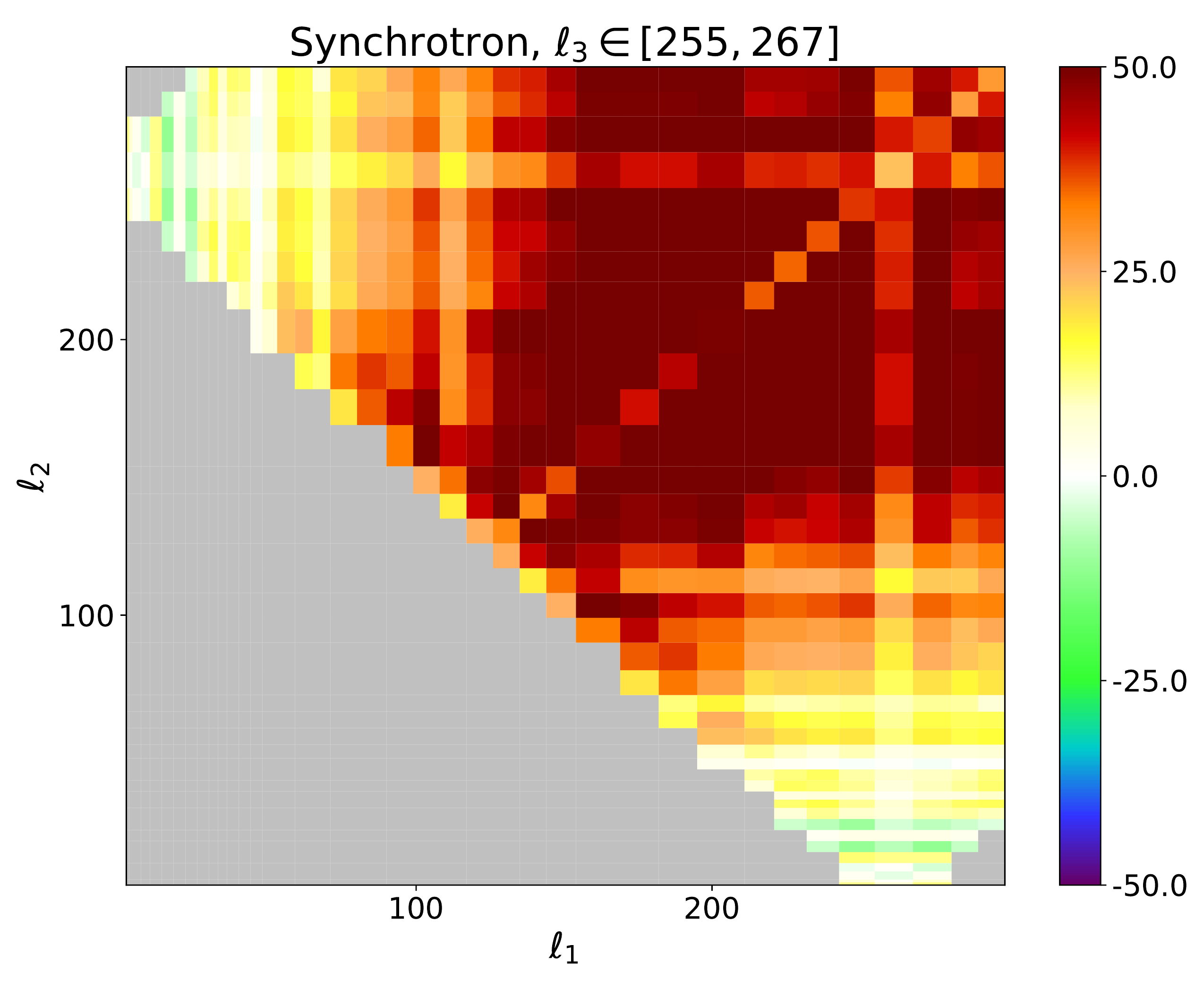}
  \caption{Bispectral signal-to-noise of the different foregrounds for $\ell_3 \in [4,6]$, $\ell_3 \in [71,79]$ and $\ell_3 \in [255,267]$. Note the different colour scales.} 
  \label{fig:other-templates}
\end{figure}

The same behaviour is of course present in the bispectra (i.e.\ the templates) where the dust dominates everything. However, as discussed in the previous section, it is more interesting to study the bispectral signal-to-noise to study the form of these bispectra. Figure \ref{fig:other-templates} shows these bispectra for three different slices of $\ell_3$. Free-free, dust and AME peak in the squeezed configuration (but for AME, the signal is so low that it could be only noise). An argument similar to the dust case described in the previous section can explain this bispectral configuration. We can also verify this in table \ref{tab:corr_coeff_others} where we have computed the correlation coefficients of these shapes with the ones previously introduced. As expected, the dust, free-free and AME bispectra are anti-correlated to the local shape (and for the other shapes see the previous section, the discussion is similar) and are correlated between themselves (they share the squeezed configuration). For a visual representation that helps to understand the correlations, see appendix \ref{ap:weights-bisp-shap}.

\begin{table}
  \begin{center}
    \small
    \begin{tabular}{l|cccccc}
      \hline
      & Local & Equilateral & Orthogonal & Lensing-ISW & Point sources & CIB \\
      \hline
      Dust (low resolution) & -0.14 & 0.0097 & 0.087 & -0.036 & 0.0083 & 0.012\\
     Free-free & -0.44 & -0.045 & 0.43 & 0.043 & 0.069 & 0.11\\
     AME & -0.23 & 0.032 & 0.052 & -0.051 & 0.033 & 0.037\\
     Synchrotron & -0.057 & 0.33 & 0.29 & 0.051 & 0.44 & 0.38\\
      \hline
    \end{tabular}
    
    \vspace{0.5cm}
    
   \begin{tabular}{l|cccc}
      \hline
      & Dust (low resolution) & Free-free & AME & Synchrotron \\
      \hline
     Dust (low resolution) & 1 & 0.24 & 0.28 & 0.56 \\
     Free-free &  & 1 & 0.37 & 0.32\\
     AME &  &  & 1 & 0.32 \\
     Synchrotron &  &  &  & 1 \\
      \hline
    \end{tabular}
  \end{center}
  \caption{Correlation coefficients between the standard theoretical templates and the observed foreground templates (low resolution) computed using the characteristics of the Planck experiment (temperature).}
  \label{tab:corr_coeff_others}
\end{table}

\begin{figure}
  \centering 
\includegraphics[width=0.32\linewidth]{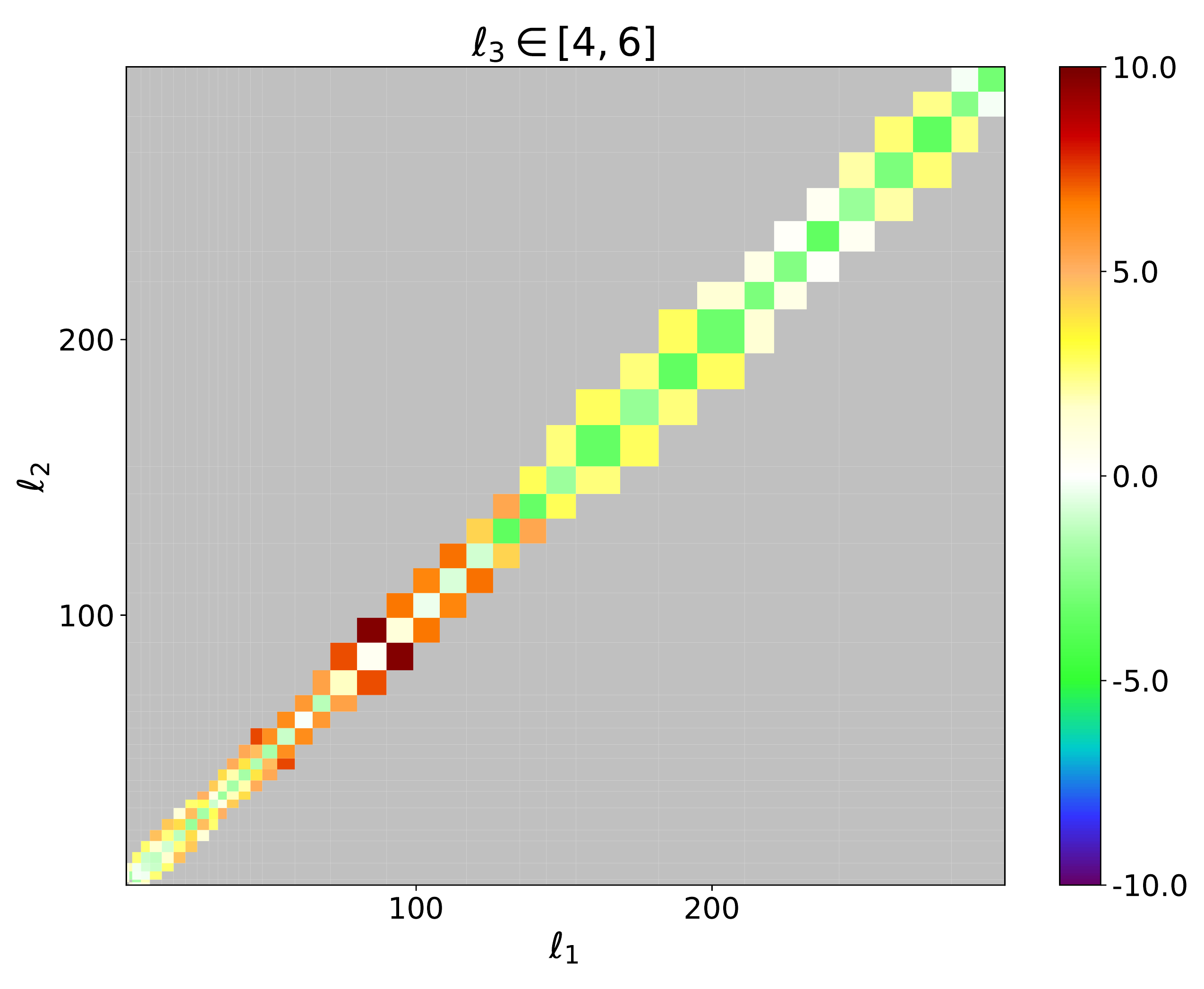}  
\includegraphics[width=0.32\linewidth]{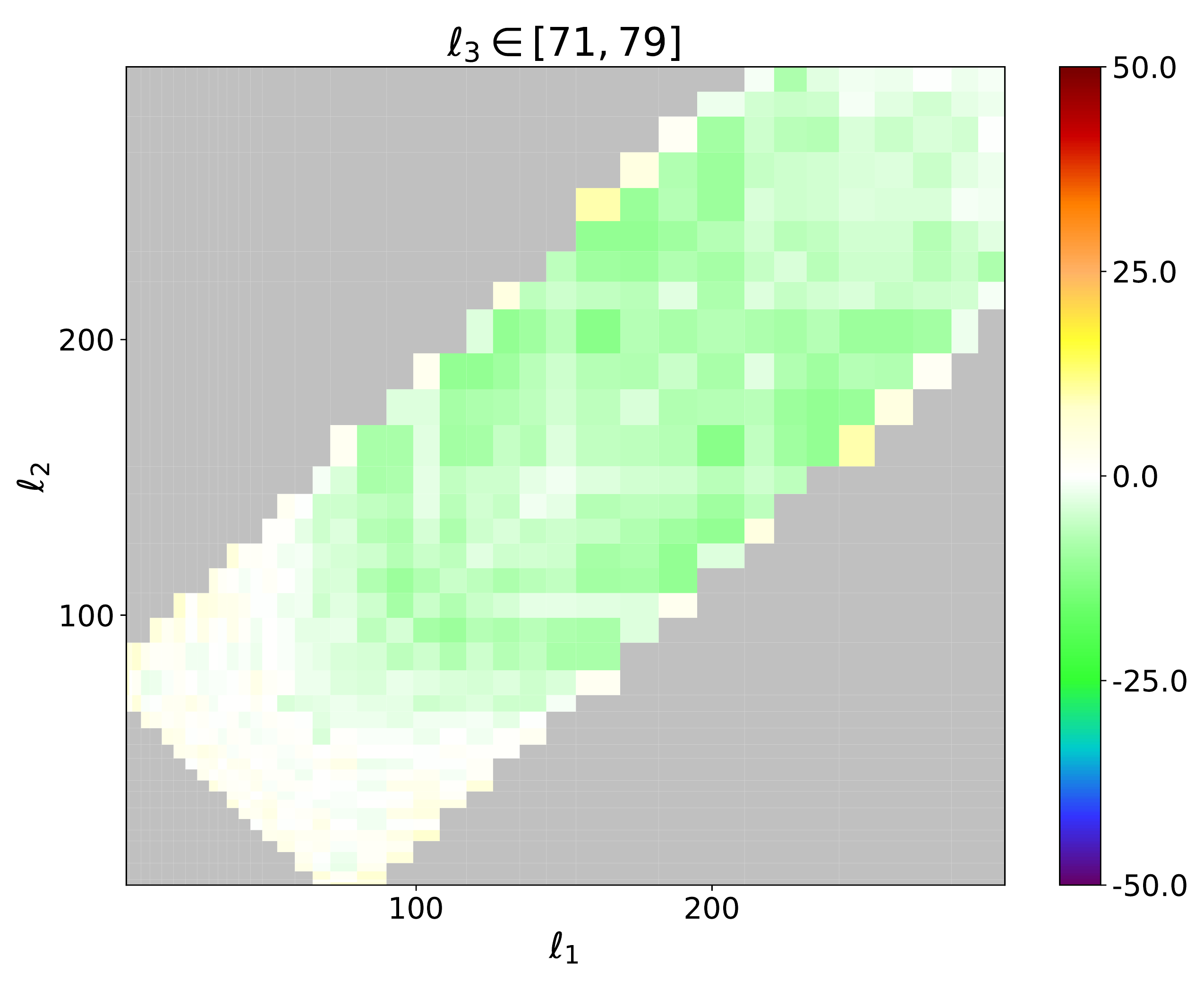} 
\includegraphics[width=0.32\linewidth]{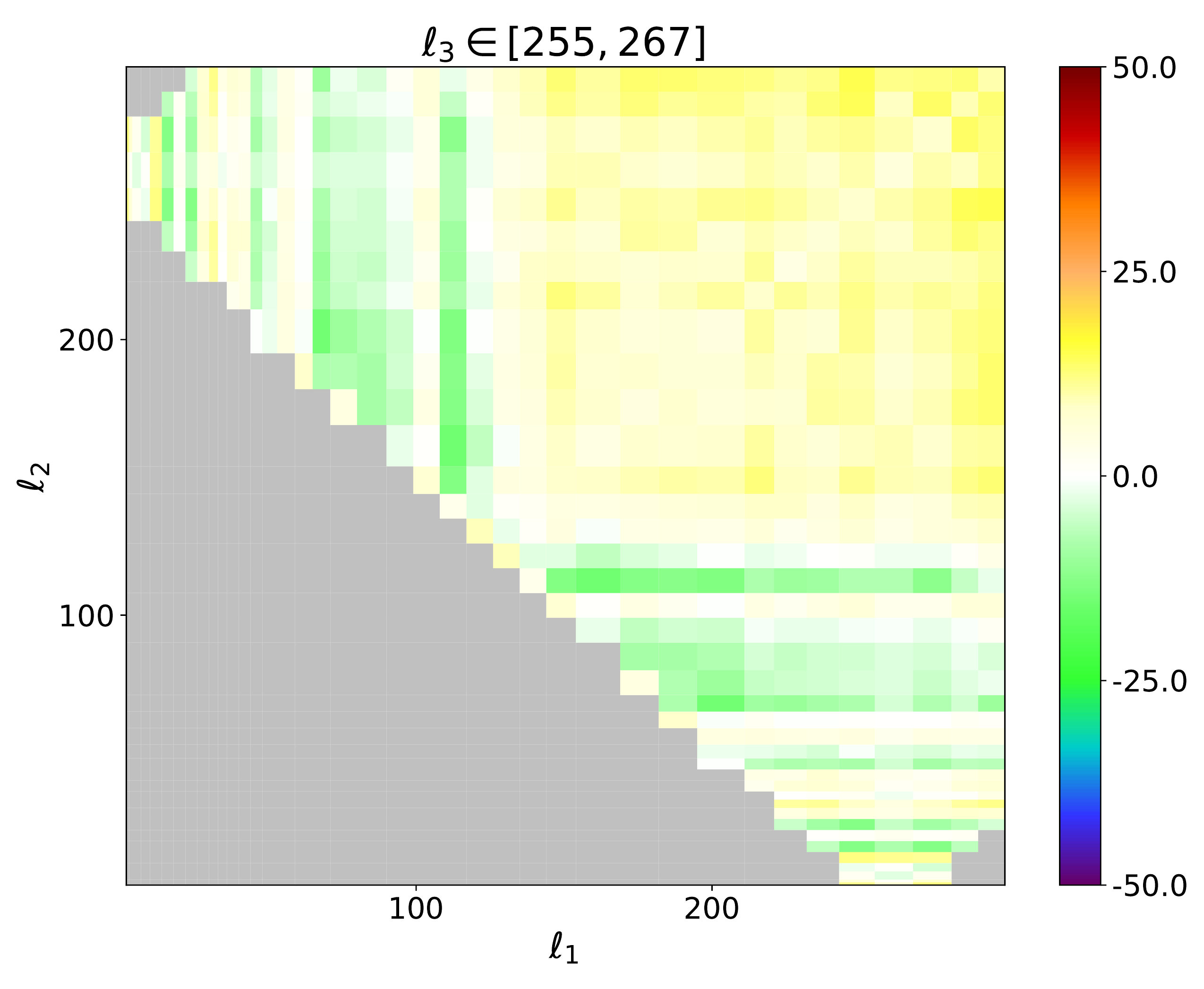}
  \caption{Bispectral signal-to-noise of the synchrotron after subtracting the unresolved point sources contamination for $\ell_3 \in [4,6]$, $\ell_3 \in [71,79]$ and $\ell_3 \in [255,267]$. Note the different colour scales.} 
  \label{fig:synch-ps-template}
\end{figure}

The case of synchrotron is different. The signal seems to be larger for three ``high'' values of $\ell$, so it is similar to the equilateral shape. This is also the typical shape produced by unresolved point sources and by the CIB. Indeed, the synchrotron is correlated (around 40~$\%$) to the point sources and CIB shapes as well as to equilateral and orthogonal (around 30~$\%$). However, it is also correlated to the other foregrounds (more than 30~$\%$), meaning that the synchrotron bispectrum also peaks in the squeezed limit, as shown in the bottom left plot of figure \ref{fig:other-templates}, even if it is not at all its dominant part. Physically that makes sense because we expect a squeezed signal for similar reasons as the other foregrounds. The simplest explanation for the equilateral shape is a contamination of the map by point sources and this possibility is mentioned in \cite{Adam:2015wua}. To verify it, we performed the simple test of subtracting the unresolved point sources bispectral template (of which the amplitude was determined using the estimator (\ref{eq:fnl-estimator})) from the bispectrum of the synchrotron map. The cleaned bispectrum is shown in figure \ref{fig:synch-ps-template} where one can see that the left plot (showing the squeezed part of the bispectrum) has not changed from the one of figure \ref{fig:other-templates}, while the other two are much less non-Gaussian (but not perfectly cleaned either). This is also illustrated in table \ref{tab:corr_coeff_synch}, where the correlation of the synchrotron bispectrum with the local shape increases (to around 15~$\%$) and becomes of the same order as for the other foreground bispectra, while the anomalous correlation with the equilateral, point sources and CIB templates vanishes. From now on, when we mention the synchrotron bispectrum, it will be the one cleaned from the unresolved point sources contamination.

\begin{table}
  \begin{center}
    \small
    \begin{tabular}{l|cccccc}
      \hline
      & Local & Equilateral & Orthogonal & Lensing-ISW & Point sources & CIB  \\
      \hline
     Cleaned synchrotron & -0.14 & 0.025 & 0.13 & -0.022 & 0.059 & 0.033\\
      \hline
    \end{tabular}
        
    \vspace{0.5cm}
    
   \begin{tabular}{l|cccc}
      \hline
      & Dust (low resolution) & Free-free & AME & Synchrotron \\
      \hline
     Cleaned synchrotron & 0.62 & 0.32 & 0.34 & 0.92 \\
      \hline
    \end{tabular}
  \end{center}
  \caption{Correlation coefficients between the synchrotron bispectrum cleaned from the point sources contamination and the templates of table \ref{tab:corr_coeff_others} computed using the characteristics of the Planck experiment (temperature).}
  \label{tab:corr_coeff_synch}
\end{table}

\subsection{Noise and masks}
\label{sec:noise-masks}

The main source of anisotropy in the foreground maps are the foregrounds themselves as they are mostly present in the galactic plane, but we still need to examine the influence of the other sources discussed in section \ref{sec:variance}.

\begin{figure}
  \centering
\includegraphics[width=0.66\linewidth]{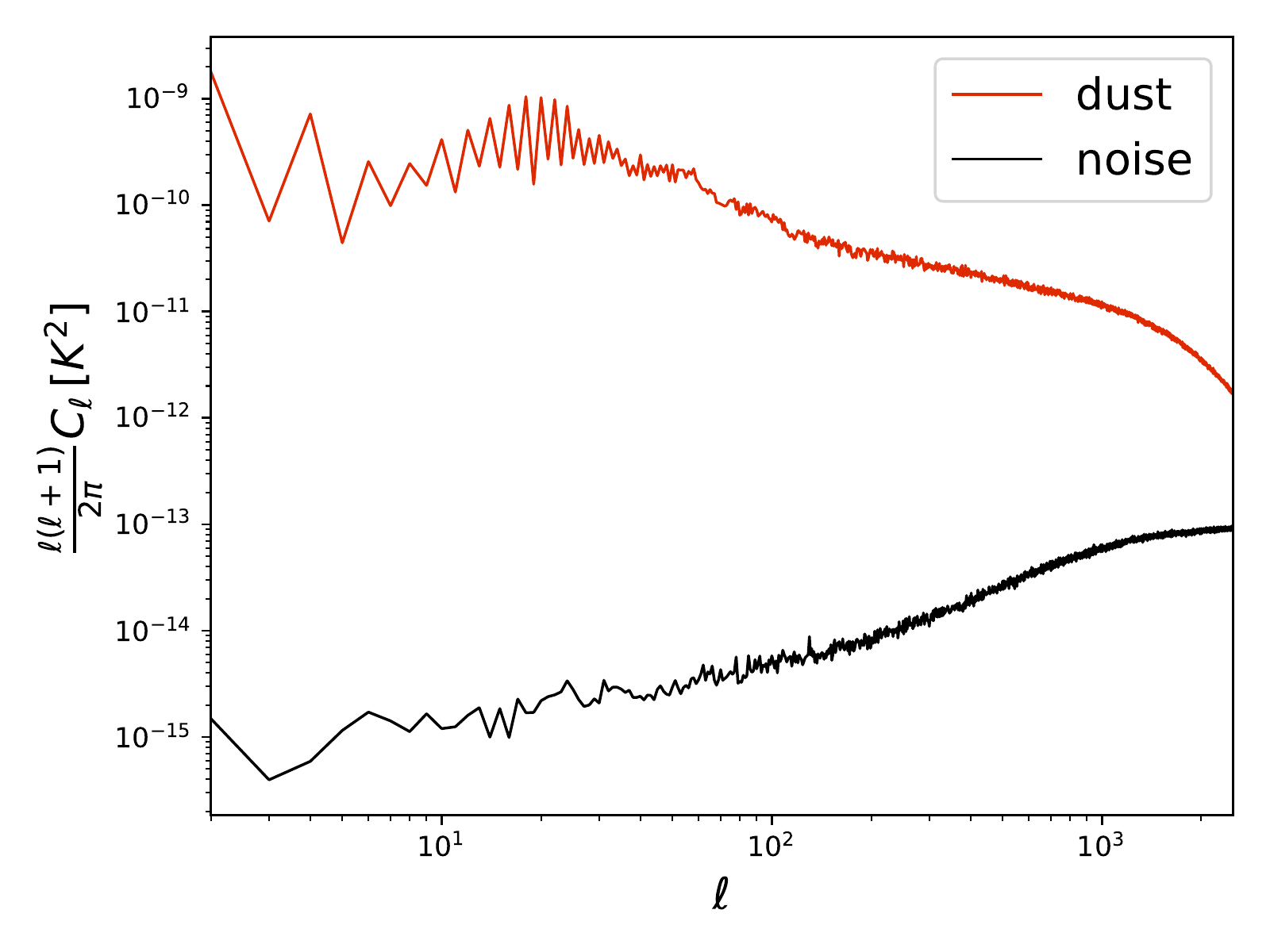}
  \caption{Dust and noise power spectra for the map studied in section \ref{sec:dust} as a function of $\ell$.}
  \label{fig:dust-noise}
\end{figure}

We start by the noise, which for the CMB has a large effect at high $\ell$. Hence, it is sufficient to look at the best resolution dust map studied in section \ref{sec:dust}. Figure \ref{fig:dust-noise} shows the noise power spectrum of the dust map evaluated using half-mission maps. Even at high $\ell$, it seems that it is small compared to the signal. Hence, we will not discuss it further in this paper.

\begin{figure}
  \centering  \includegraphics[width=0.49\linewidth]{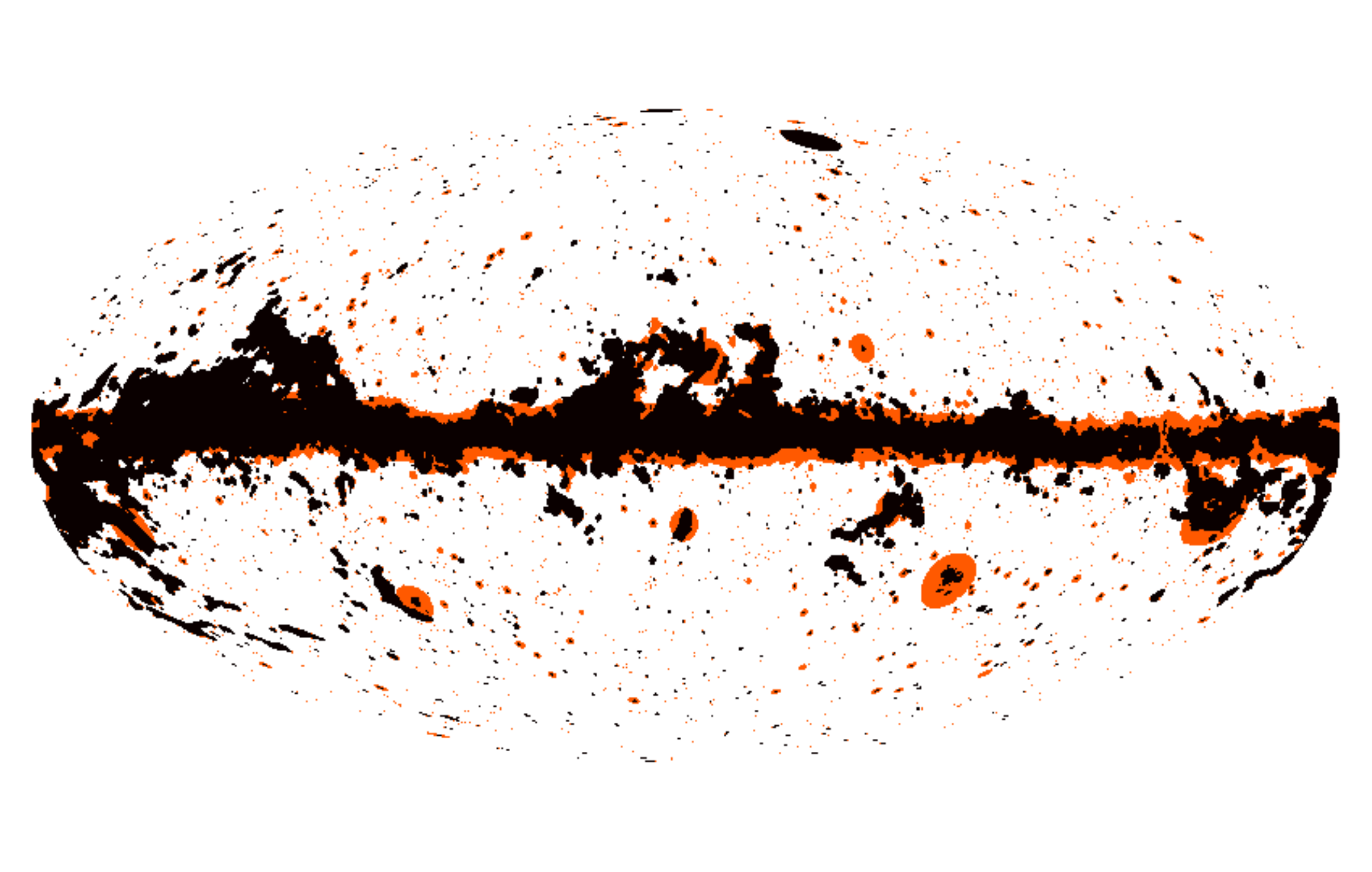} \includegraphics[width=0.49\linewidth]{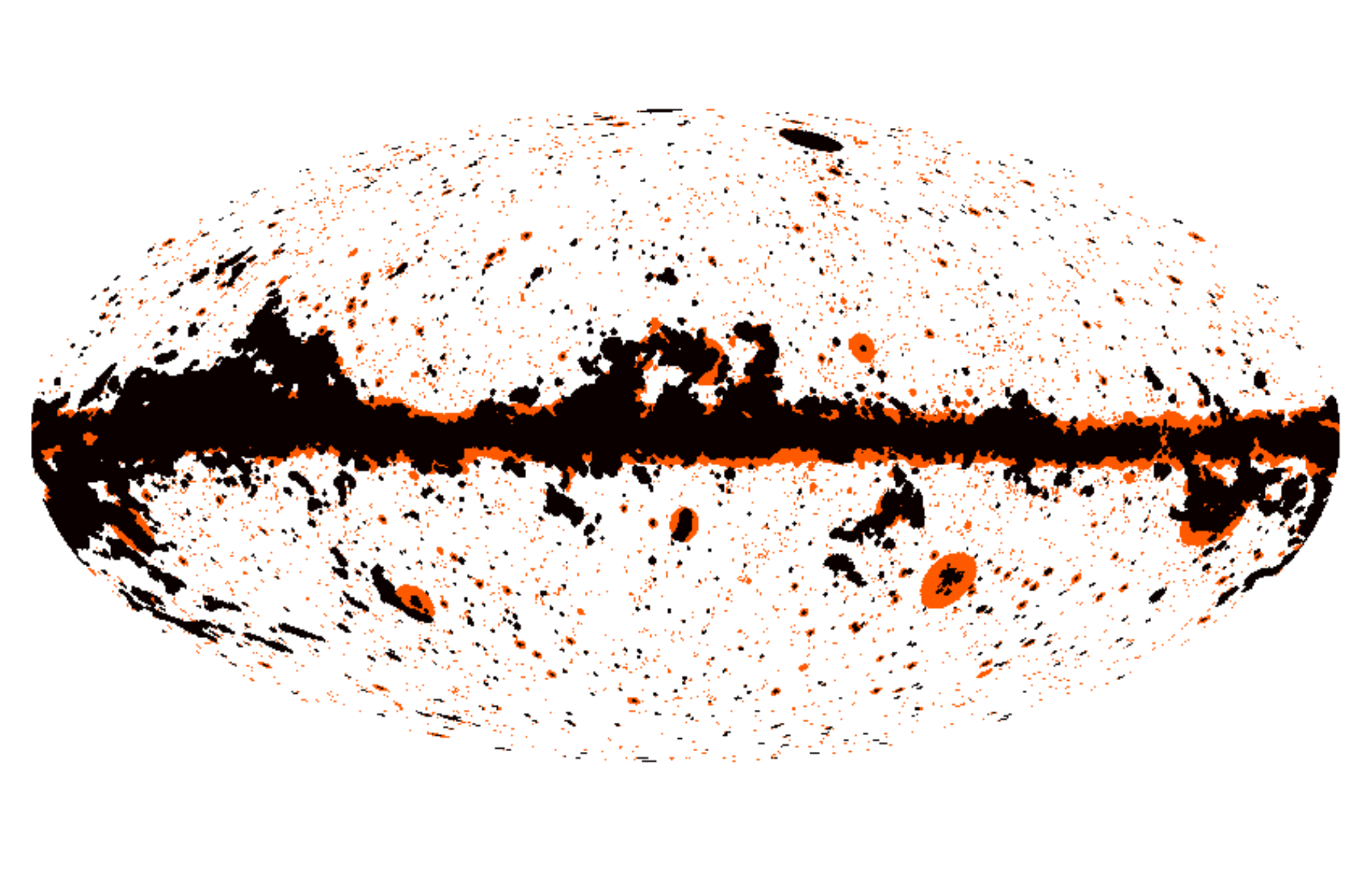}
  \caption{The different masks used in sections \ref{sec:foregrounds} and \ref{sec:analyses}. On the left, the high resolution versions ($n_\mathrm{side}=2048$) of the \texttt{Commander} mask ($f_\mathrm{sky} = 0.822$) in black only and of the common mask ($f_\mathrm{sky} = 0.776$) in black and orange. On the right, the masks have a low resolution ($n_\mathrm{side}=256$) and have been obtained by degrading the resolution of the masks on the left, in the process their size has slightly increased ($f_\mathrm{sky} = 0.804$ and $f_\mathrm{sky} = 0.745$). This effect is easily visible for the point sources.}
  \label{fig:masks}
\end{figure}

The choice of mask should also be examined more carefully. That is why here we compare our previous results obtained with the common mask ($f_\mathrm{sky} = 0.776$) to those obtained with the mask provided by the \texttt{Commander} component separation method which is slightly smaller ($f_\mathrm{sky} = 0.822$) and of course fully included in the common mask. Figure \ref{fig:masks} shows these two masks in the high and low resolution cases.

\begin{figure}
  \centering \includegraphics[width=0.49\linewidth]{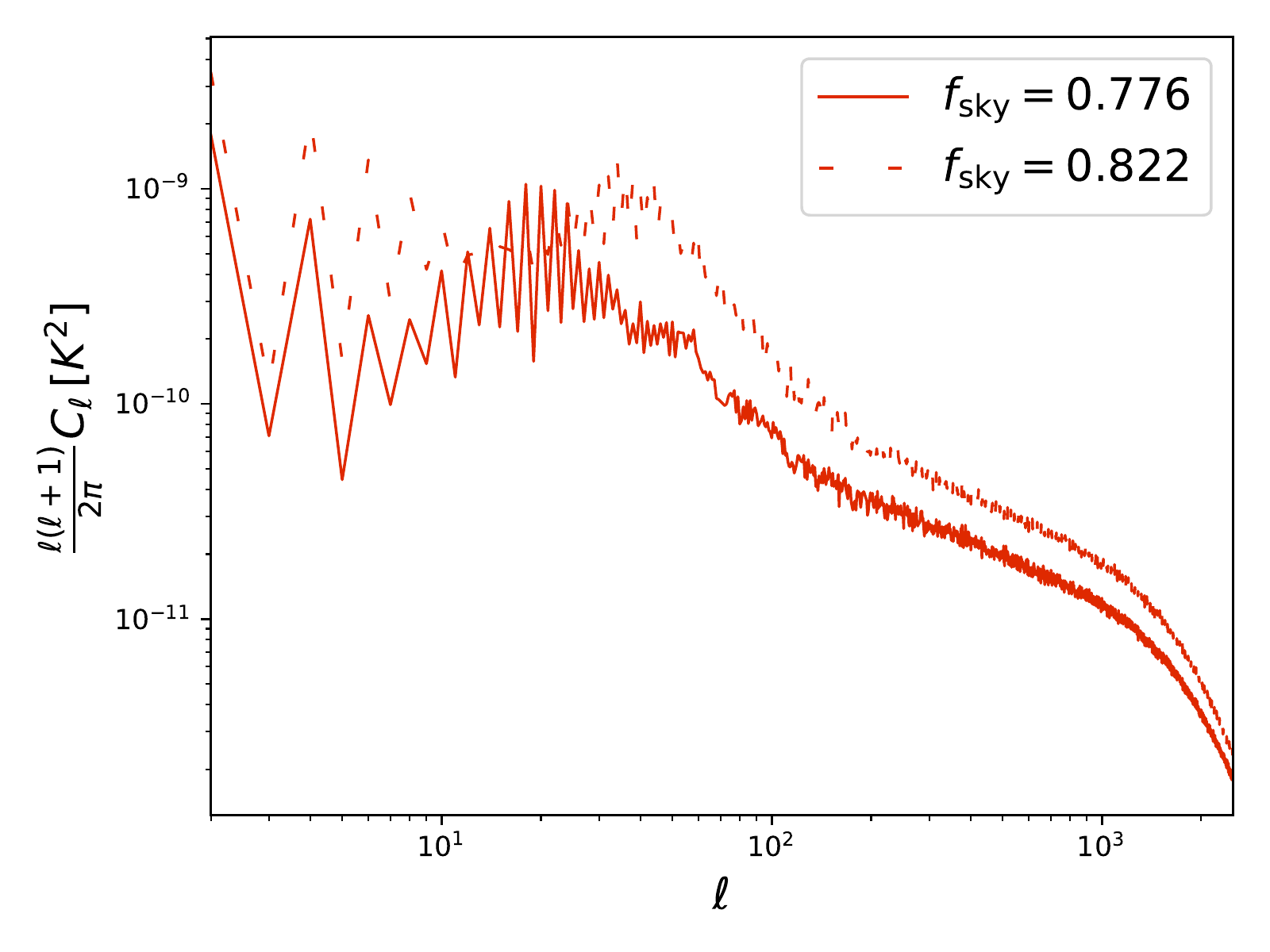}
  \includegraphics[width=0.49\linewidth]{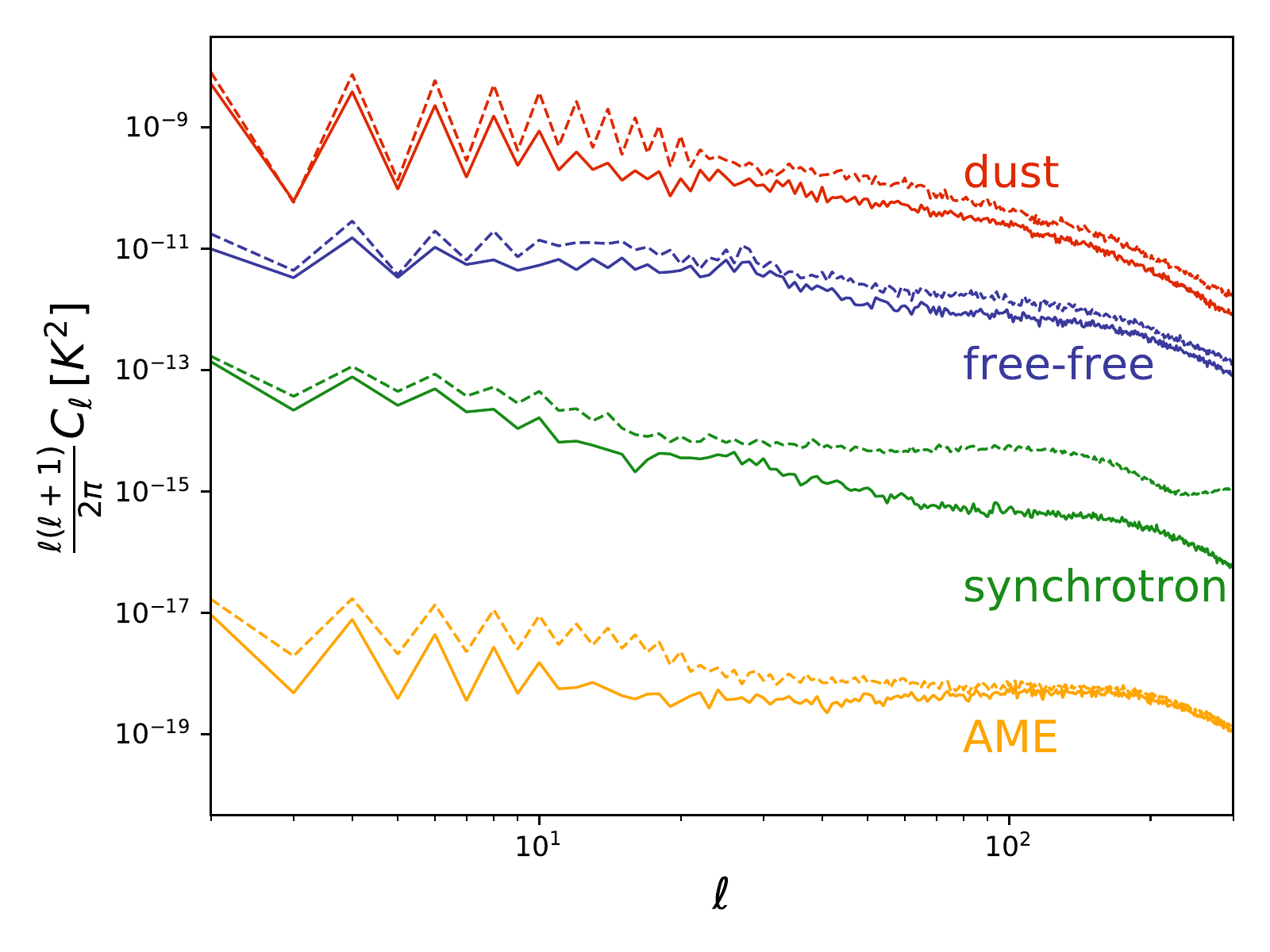}
  \caption{Power spectra of the different foregrounds using the common mask (solid line, $f_\mathrm{sky} = 0.776$) and the \texttt{Commander} mask (dashed line, $f_\mathrm{sky} = 0.822)$. On the left, the high resolution dust map ($n_{\mathrm{side}}=2048$) and on the right, the low resolution foreground maps ($n_{\mathrm{side}}=256$).}
  \label{fig:masks-power-spectra}
\end{figure}

Figure \ref{fig:masks-power-spectra} shows the power spectra of the different foregrounds with these two masks and highlights the large difference between the two cases. The reason for this difference is quite obvious because the masks have been constructed to hide most of the foregrounds, so with a smaller mask, there is a lot more of the foregrounds to detect. Moreover, as they are anisotropic, both the amplitude and the form are different depending on the mask, this is especially true for the synchrotron signal (as we will explain below).

\begin{figure}
  \centering \includegraphics[width=0.32\linewidth]{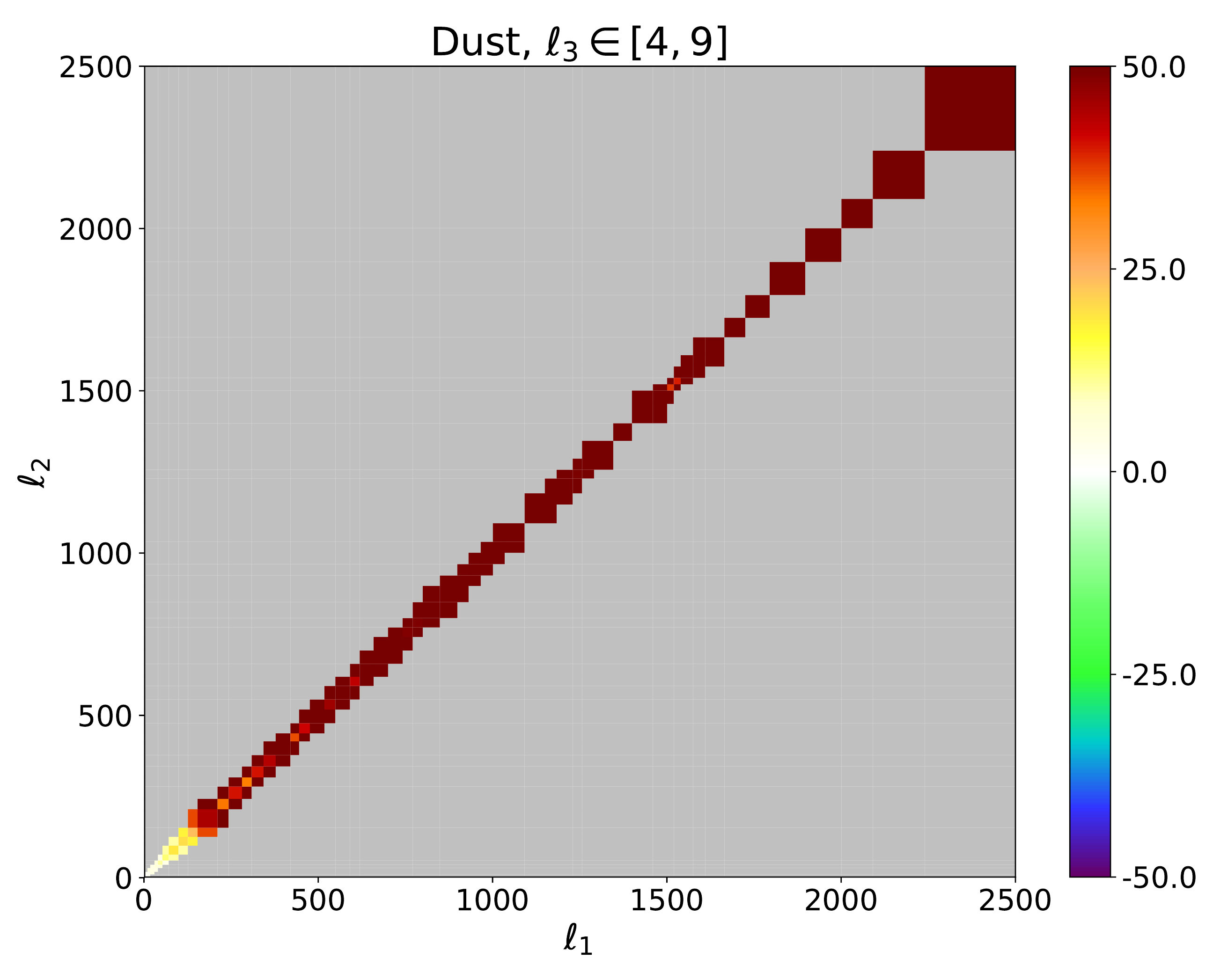}  \includegraphics[width=0.32\linewidth]{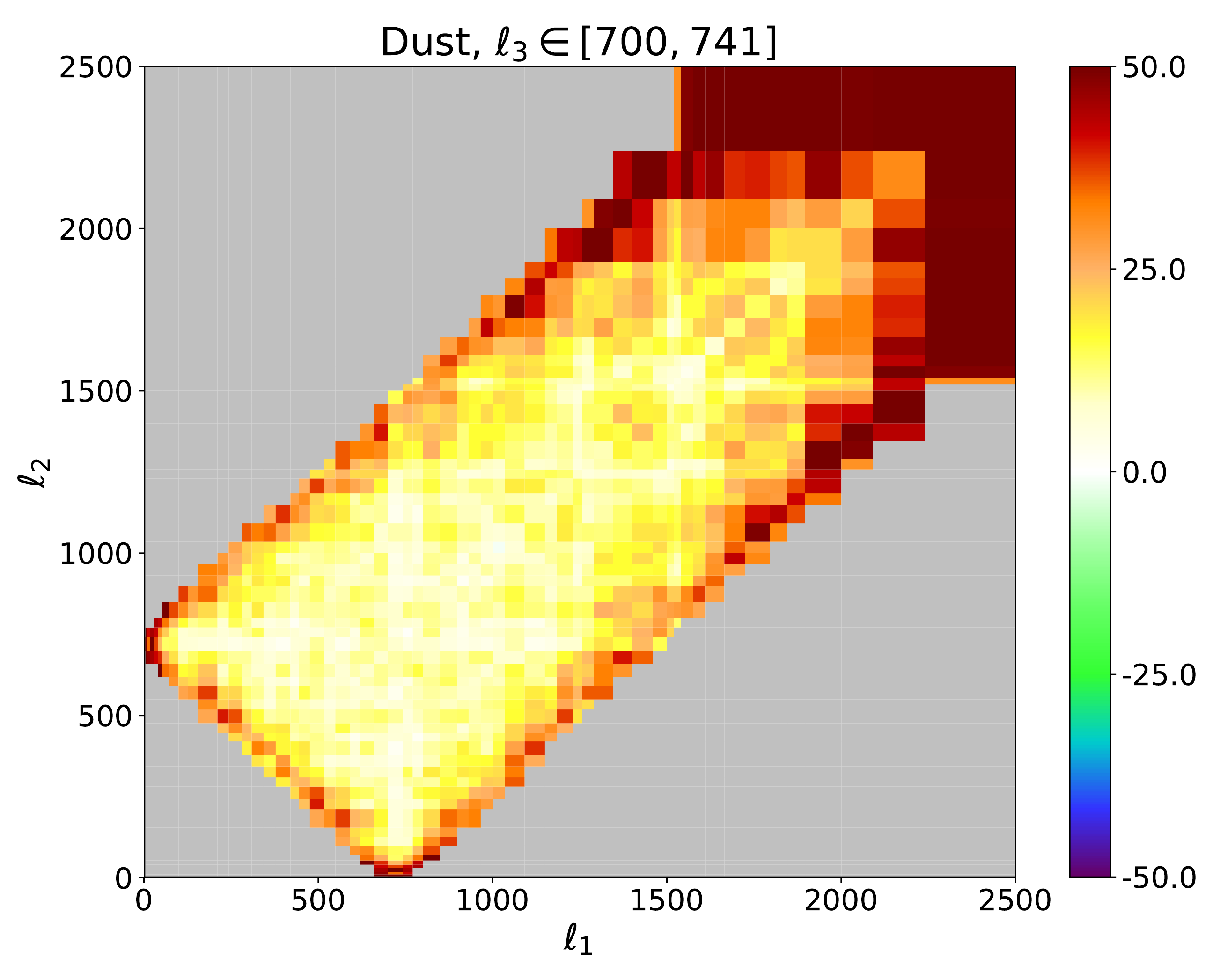} \includegraphics[width=0.32\linewidth]{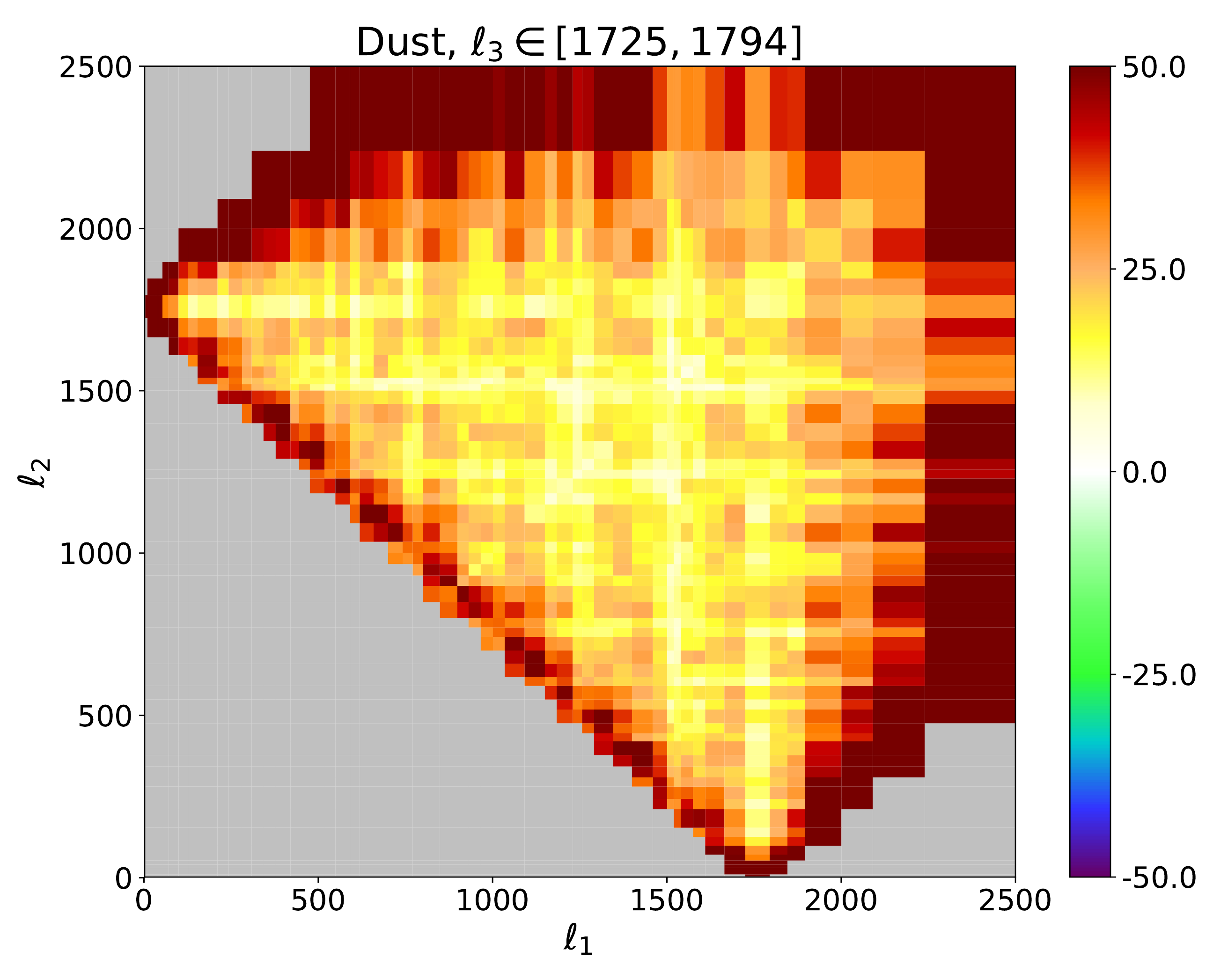}
\includegraphics[width=0.32\linewidth]{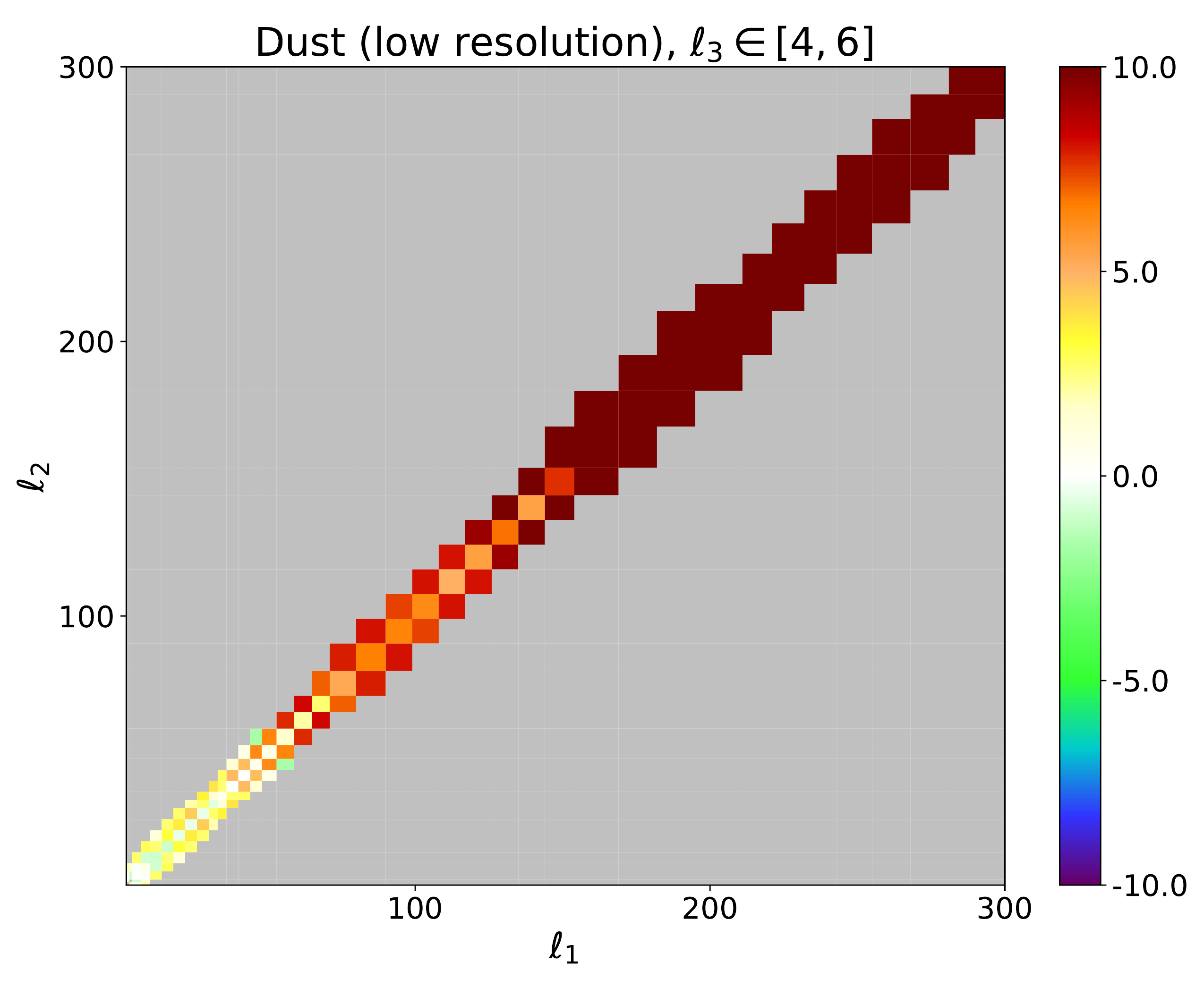}  \includegraphics[width=0.32\linewidth]{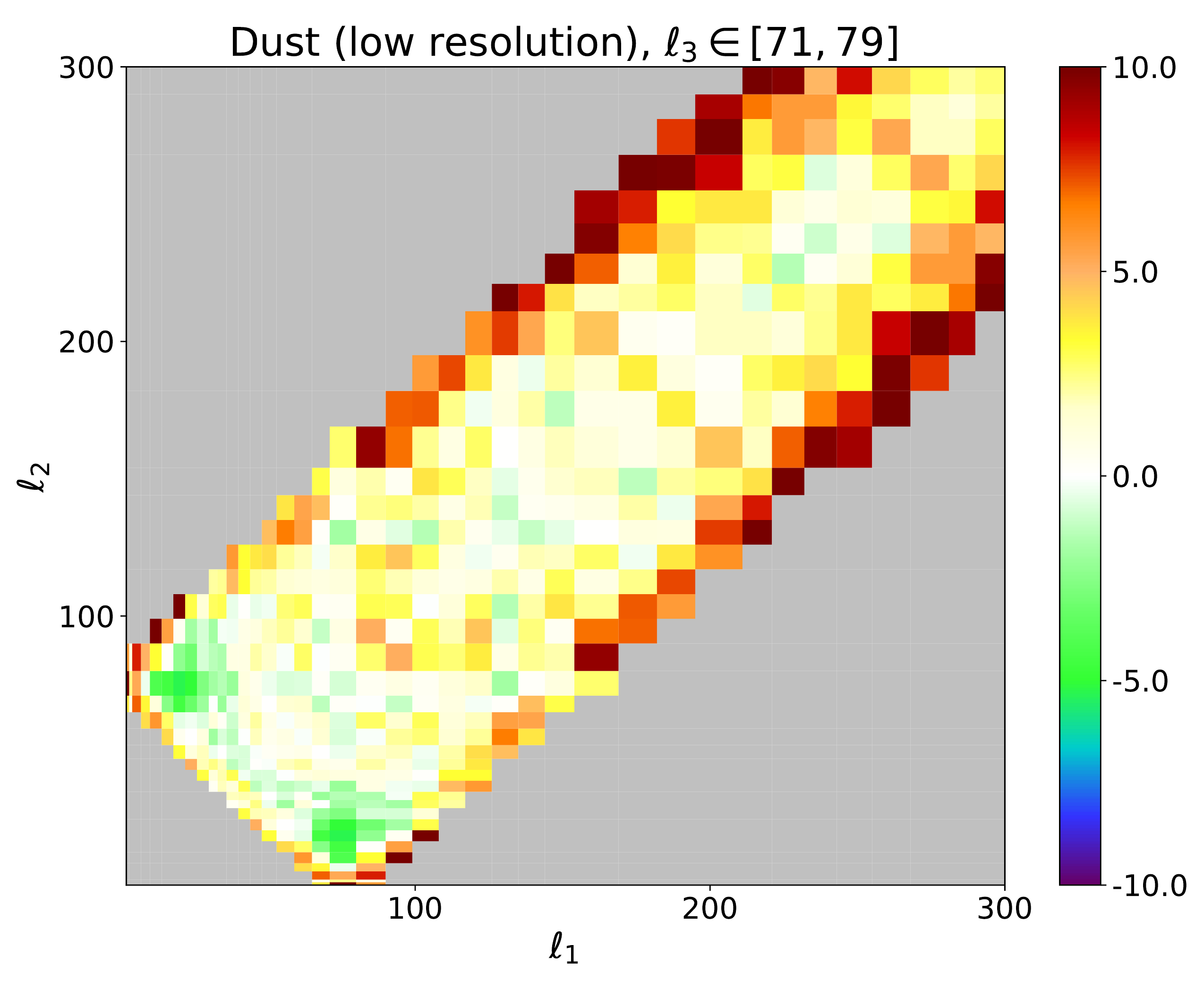} \includegraphics[width=0.32\linewidth]{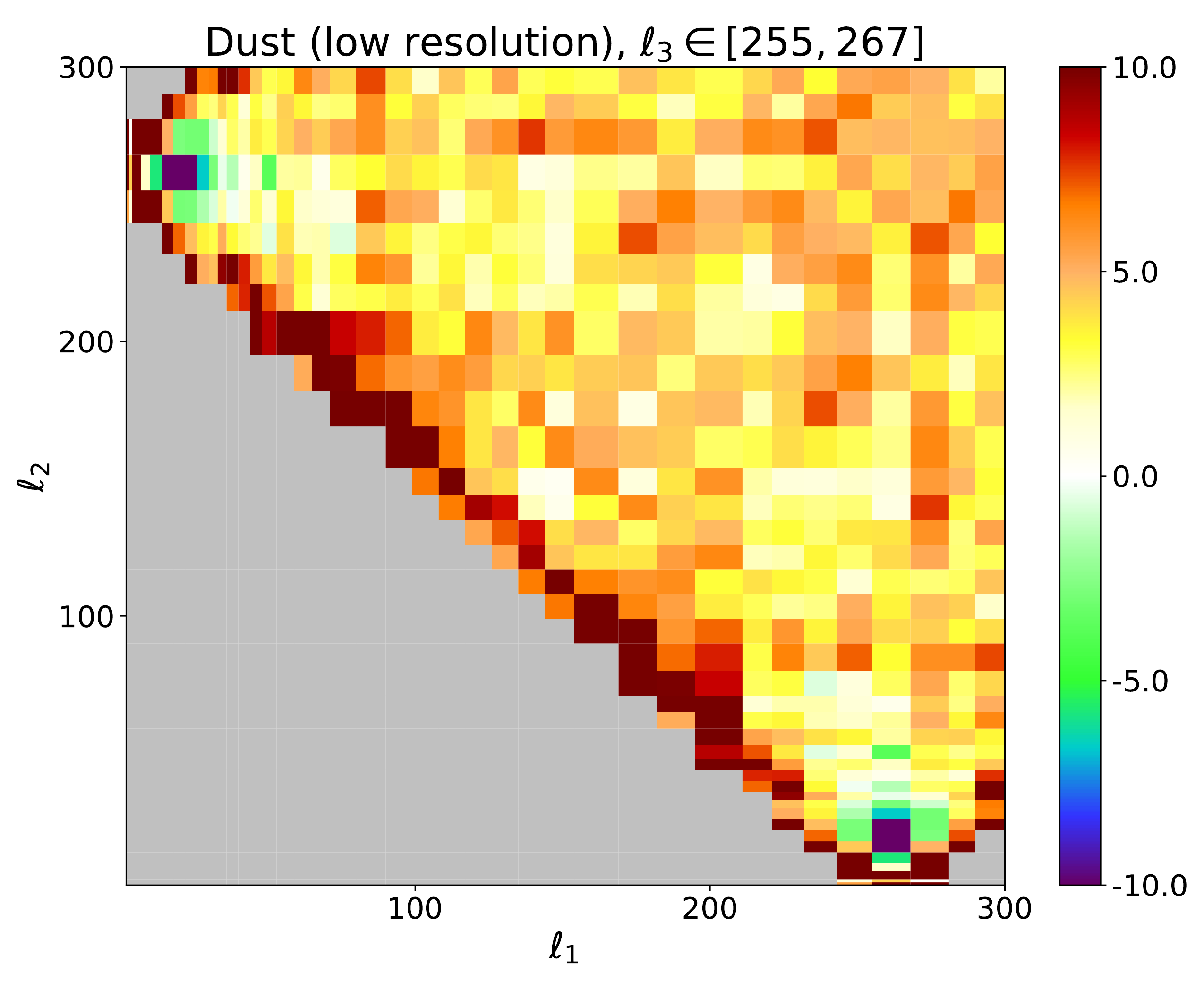}
\includegraphics[width=0.32\linewidth]{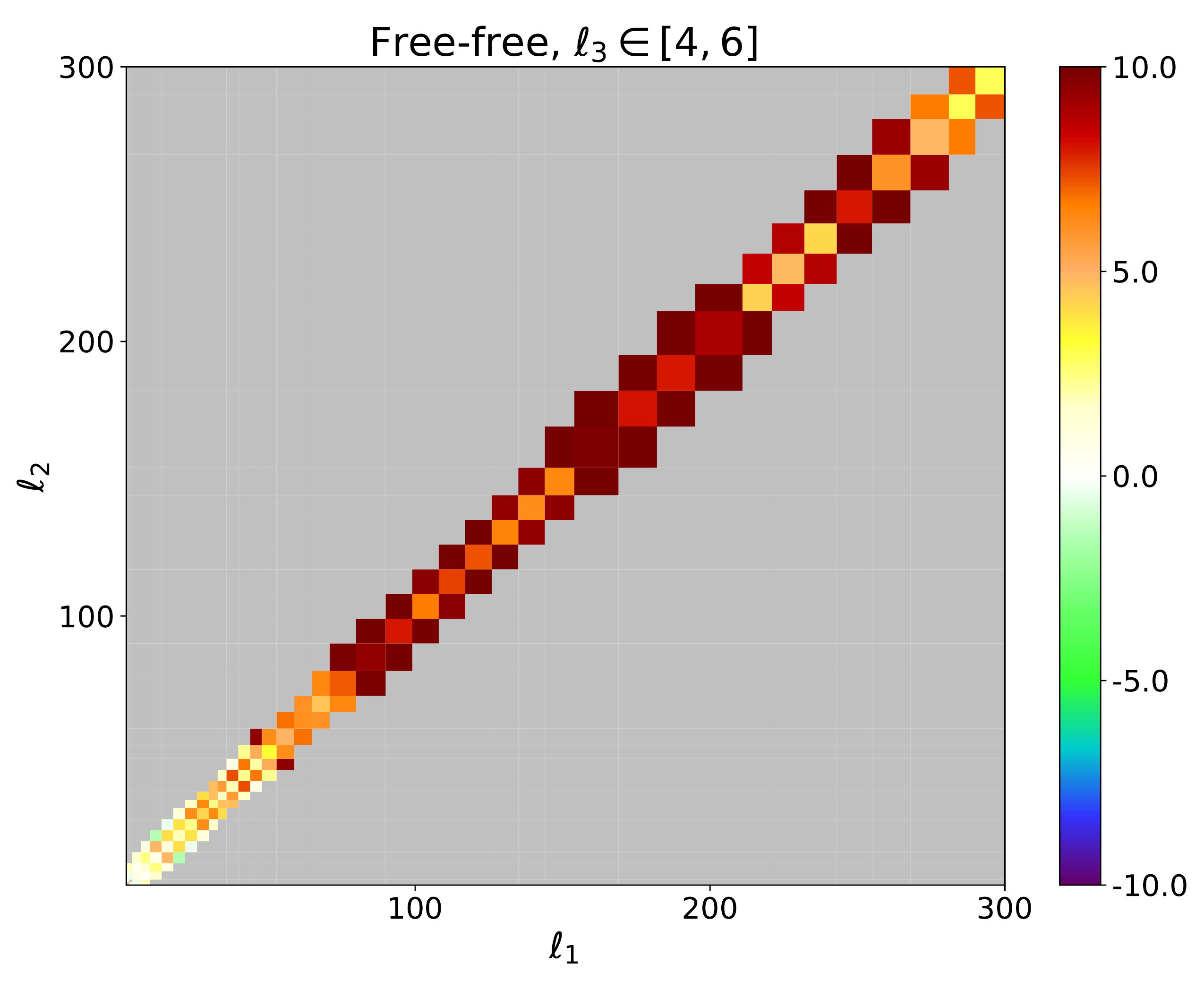}   \includegraphics[width=0.32\linewidth]{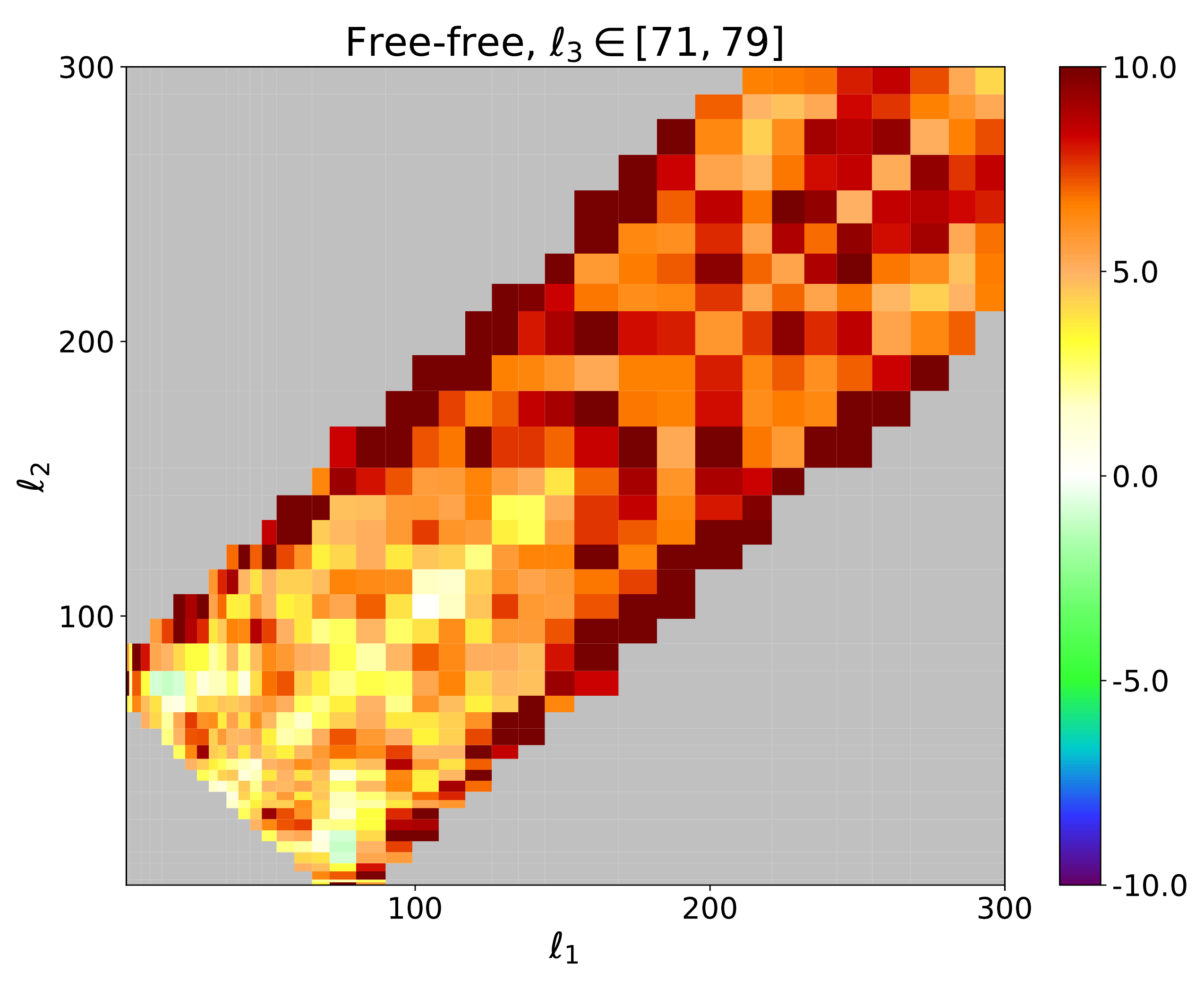} 
\includegraphics[width=0.32\linewidth]{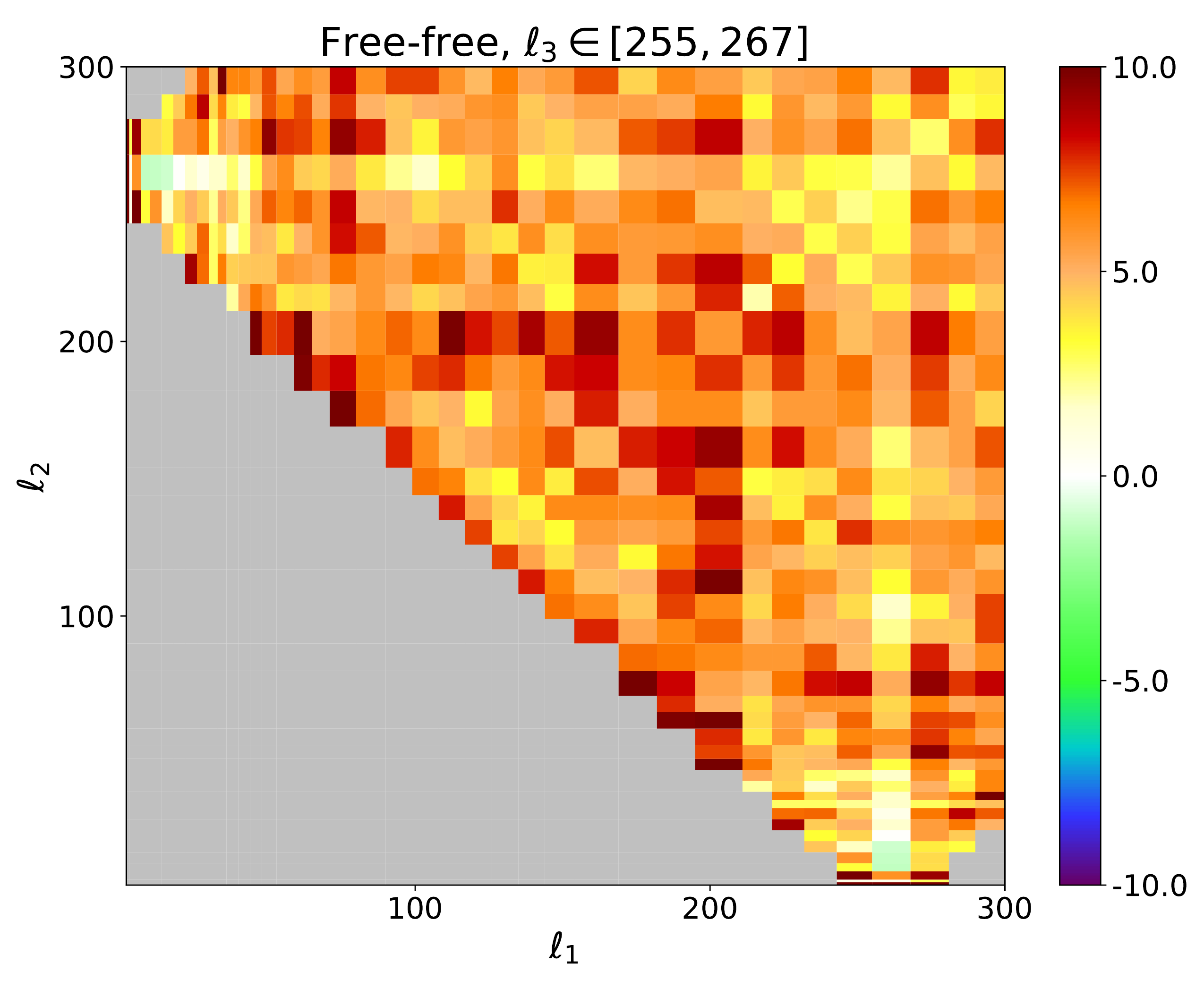}
\includegraphics[width=0.32\linewidth]{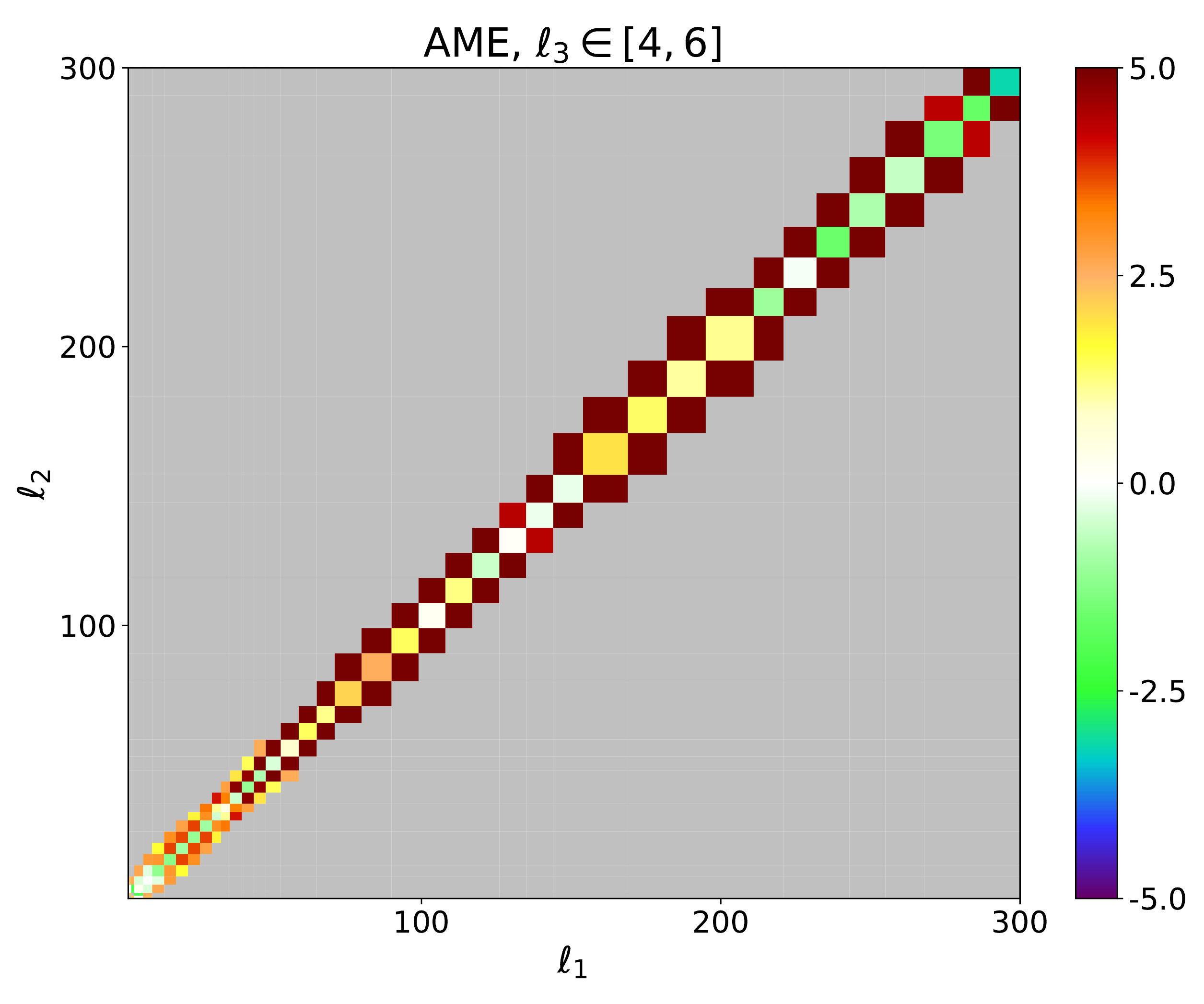} 
\includegraphics[width=0.32\linewidth]{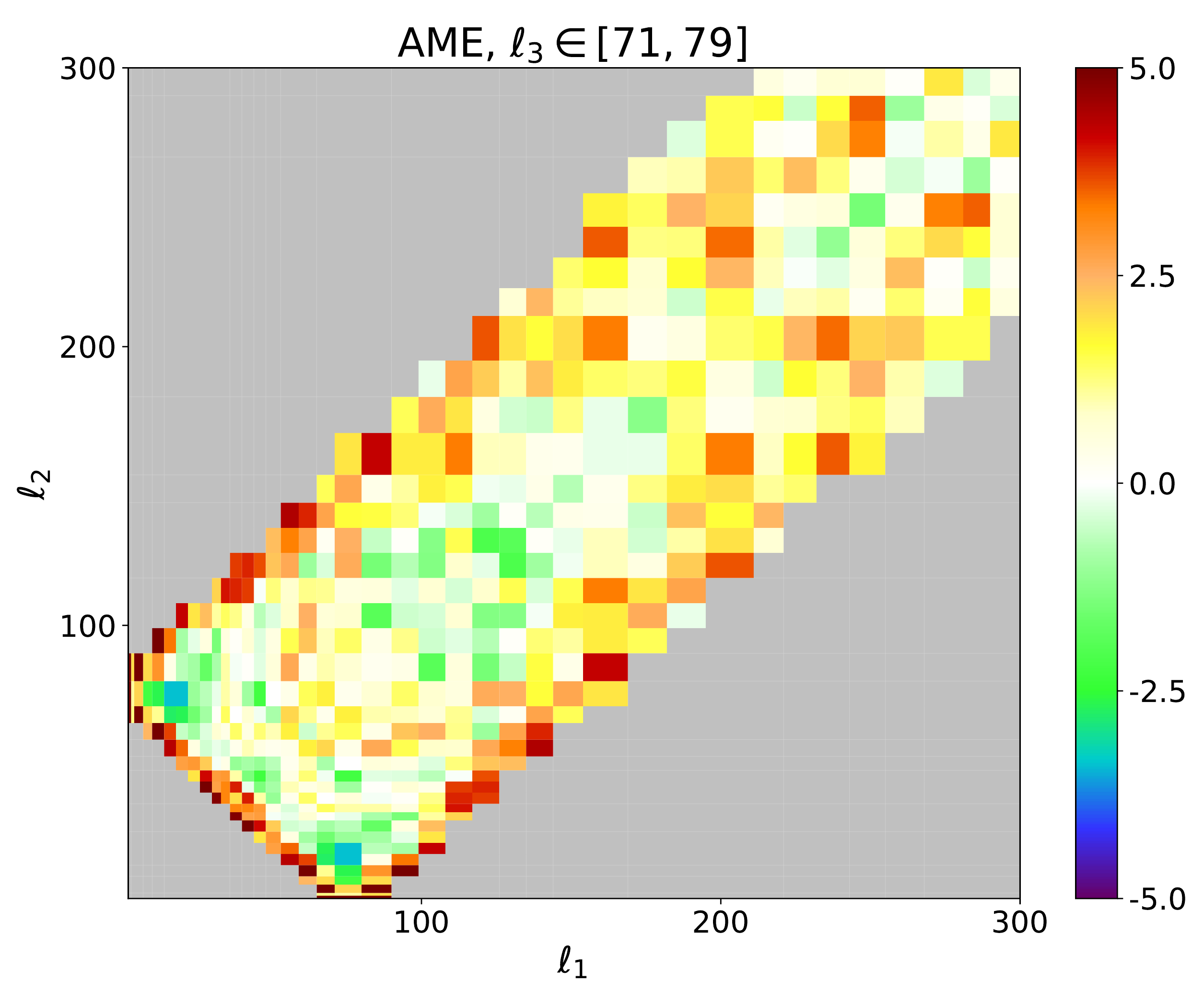}  \includegraphics[width=0.32\linewidth]{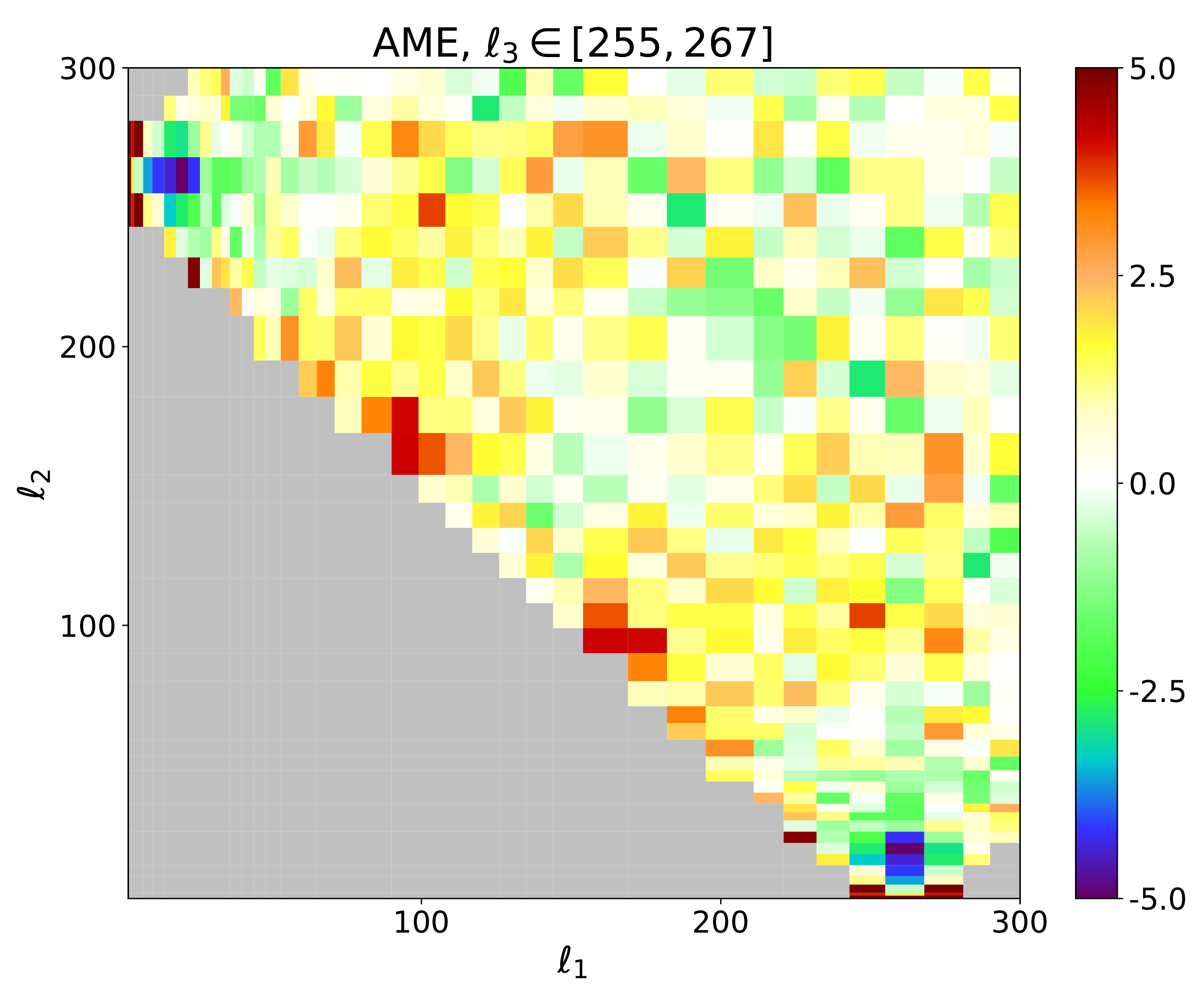}
\includegraphics[width=0.32\linewidth]{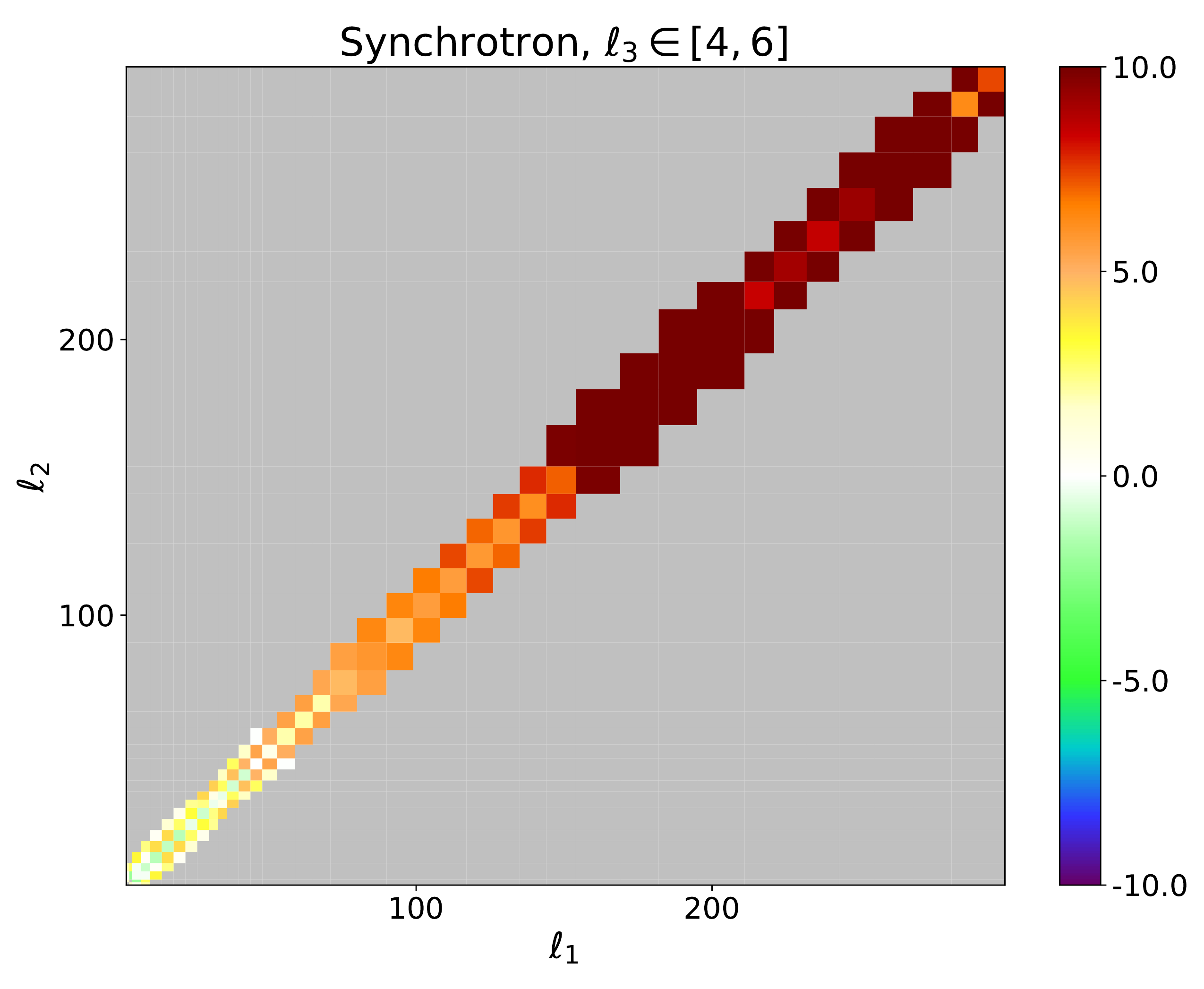} 
\includegraphics[width=0.32\linewidth]{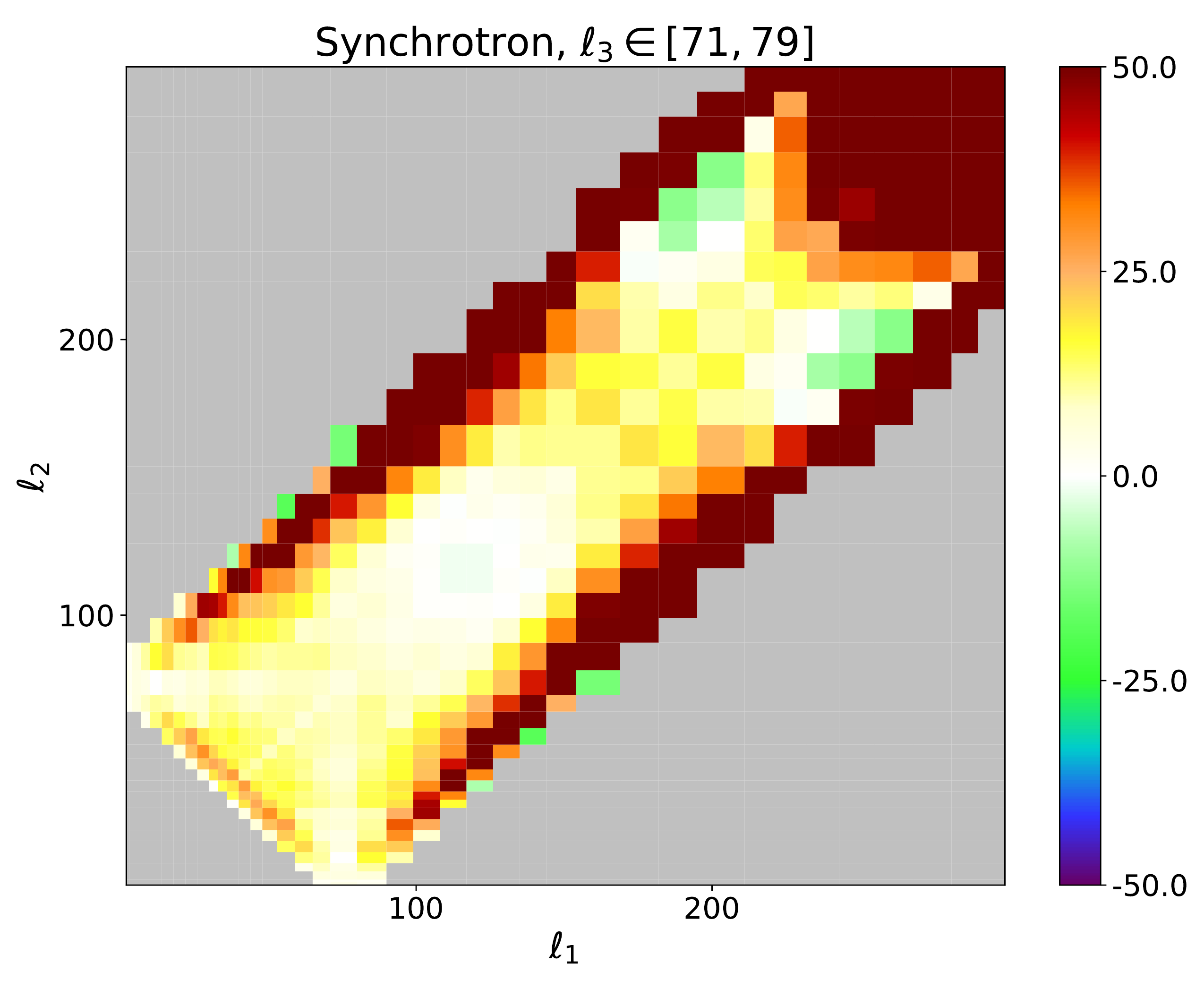}  \includegraphics[width=0.32\linewidth]{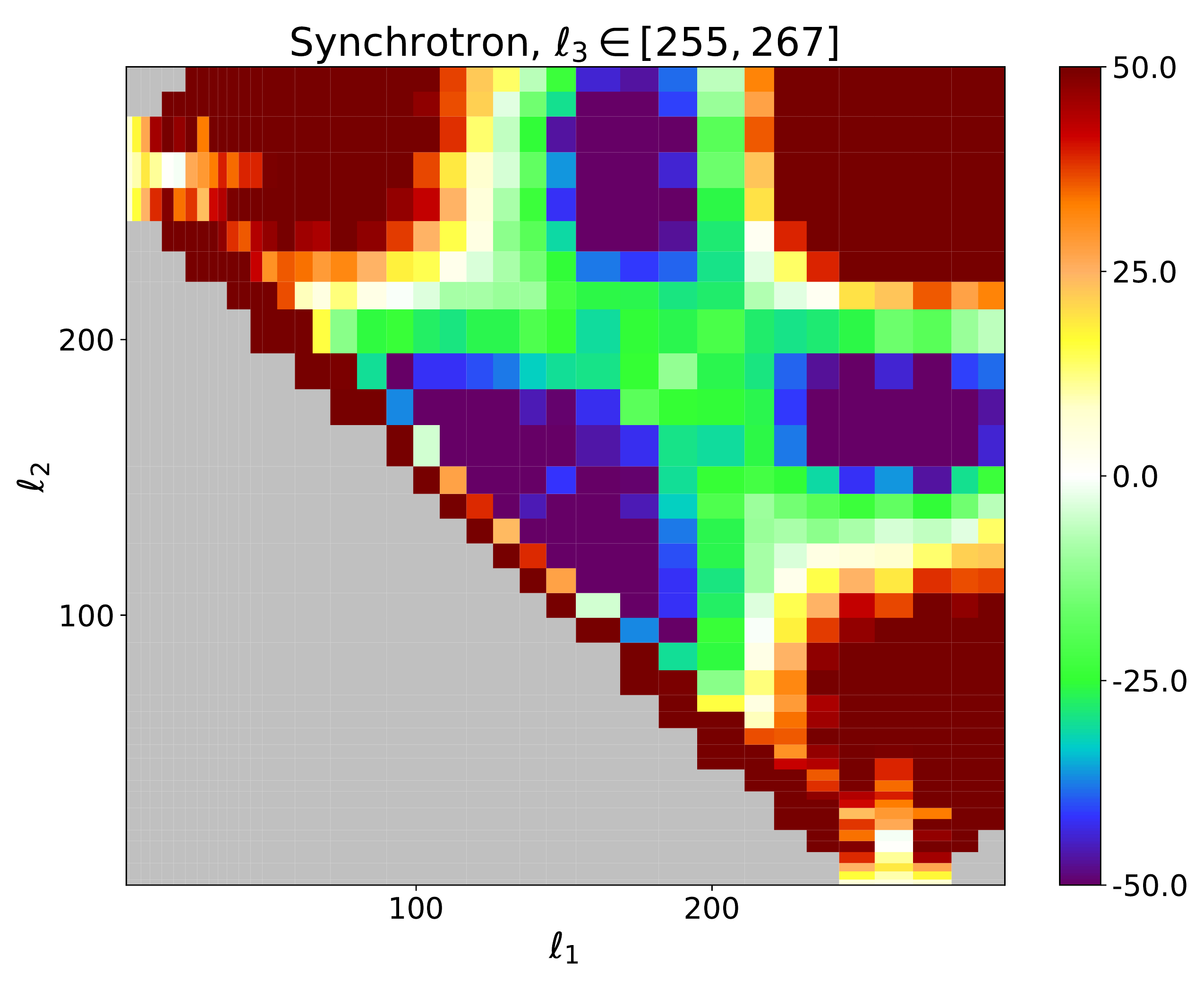}
  \caption{Bispectral signal-to-noise of the different foregrounds for the same three bins of $\ell_3$ as figures \ref{fig:dust-template} and \ref{fig:other-templates} using the \texttt{Commander} mask. Note the different colour scales.}
  \label{fig:bispectrum-masks}
\end{figure}

This is also checked for the bispectra as shown in figure \ref{fig:bispectrum-masks}. With the \texttt{Commander} mask (the smallest one), all the signals are a lot more non-Gaussian. To verify that is not only a difference of amplitude, we have at our disposal a useful tool: the correlation coefficients defined in \eqref{eq:shape-correlator}. For each foreground, we have computed the correlation between the templates determined using the two masks. The results are given in table \ref{tab:correlation-masks}. For the dust, free-free and AME emissions, the templates are correlated (above 80~$\%$) and indeed we can see that the bispectra peak in the squeezed configuration as discussed previously. However, the fact that the correlation is not 100~$\%$ shows that the difference is not only the amplitude.

\begin{table}
  \begin{center}
    \begin{tabular}{lcccc}
      \hline
      Dust & Dust (low resolution) & Free-free & AME & Synchrotron \\
      \hline
      0.90 & 0.85 & 0.88 & 0.91 & 0.11 \\
      \hline
    \end{tabular}
  \end{center}
  \caption{Correlation coefficients between the bispectral templates determined using the common mask and the \texttt{Commander} mask for each foreground.}
  \label{tab:correlation-masks}
\end{table}

\begin{figure}
  \centering
\includegraphics[width=0.49\linewidth]{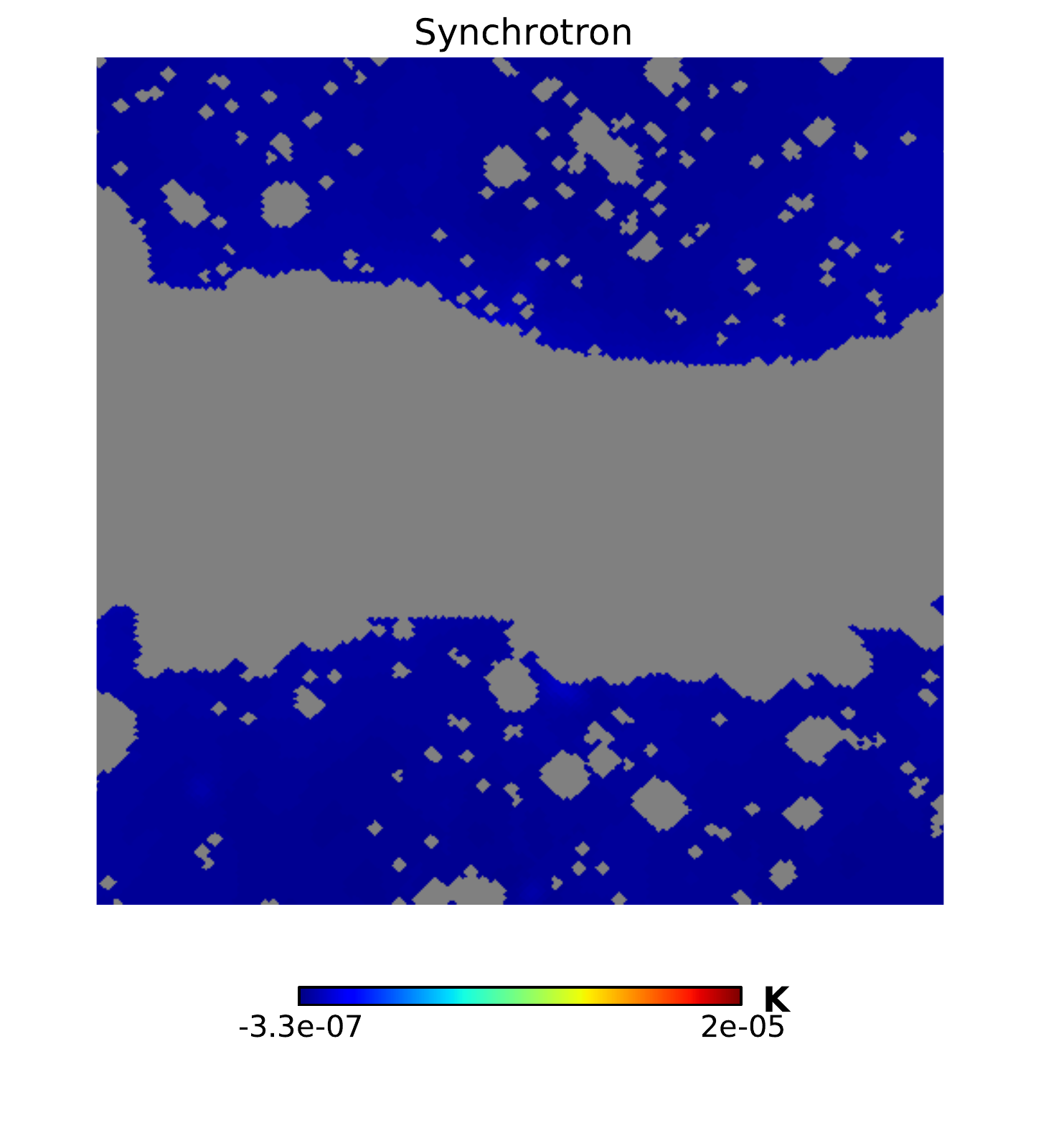}  \includegraphics[width=0.49\linewidth]{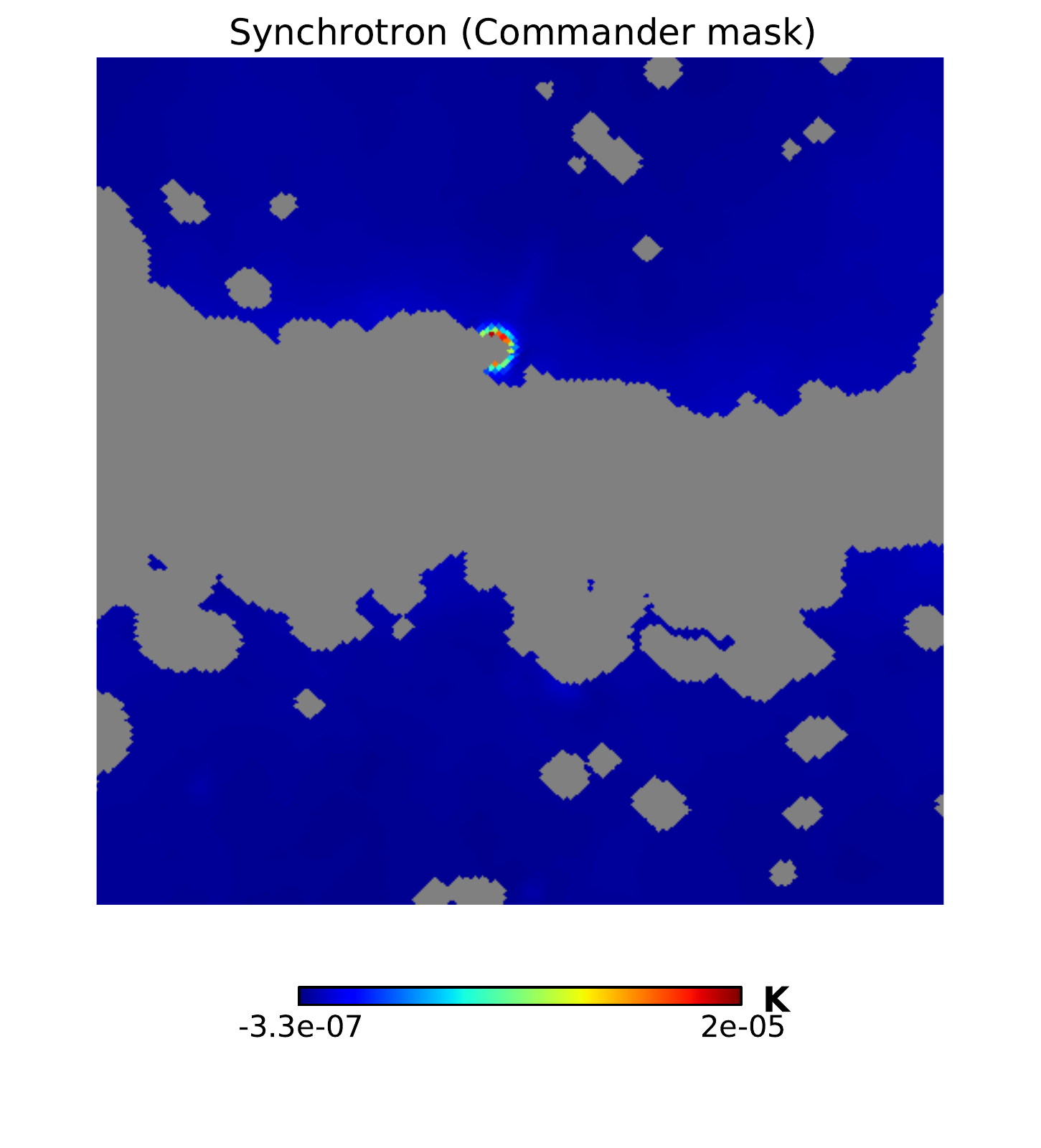}
  \caption{Zoom on the synchrotron map at 143 GHz after application of the masks (left: common mask, right: \texttt{Commander} mask).}
  \label{fig:synchrotron-commander-mask}
\end{figure}

However, for synchrotron the situation is more complicated, like in the power spectrum case. Indeed the new template is very different from the one in figure \ref{fig:synch-ps-template} and it is confirmed by the low correlation between the two synchrotron templates determined with the two different masks. To understand this result, it is interesting to examine directly the data map with the \texttt{Commander} mask in figure \ref{fig:synchrotron-commander-mask}. One can see that there are a few pixels where the intensity is ten times larger than with the common mask (where they are hidden). The influence of this very small region dominates the power spectrum and the bispectrum because the transition is so important. It could be modelized as a Heaviside step function, the Fourier transform of which is a sinc function, meaning that these two pixels have a large influence over the whole multipole space and we can see oscillations as expected in both the power spectrum (there is a minimum at $\ell \approx 240$) and the bispectrum (there are three regions of negative bispectrum with positive bispectrum around them on the plot for $\ell_3\in[255,267]$).

In conclusion, the choice of mask has a large influence on the templates we are determining because of the localization of the foregrounds in the galactic plane. This means that when we apply these templates to other maps in the next section, it is mandatory to use the same mask at every step. From now on, we will exclusively use the common mask.

\section{Data analyses}
\label{sec:analyses}

The aims of this section are twofold. First, we want to verify that the numerical templates we just determined can be used in the context of a CMB data analysis. The first basic test to check this is to apply the template to the map it comes from. The expected answer for the amplitude parameter of this specific shape is then obviously $\fnl=1$. Moreover, if we perform a correlated analysis with other shapes like the primordial ones, their own $\fnl$ has to be negligible. Indeed, that is what we observe (see table \ref{tab:dust}) and we can now discuss more interesting tests based on CMB maps.
It is important to recall that the galactic foregrounds are highly anisotropic while most other shapes have an isotropic origin (primordial, lensing-ISW or extra-galactic foregrounds). These galactic numerical templates also contain mask and noise effects, but we will show that it is not an issue. For this, we ran a series of tests with the simple idea of artificially adding dust to the maps containing CMB realizations (simulations, but also the observed data) to check that we indeed detect the right amount of dust and that it has no impact on the other shapes.

\begin{table}
  \begin{center}
    \small
  \begin{tabular}{lcccccc}
  \hline
 & Local & Equilateral & Orthogonal & P.S.$/10^{-29}$ & CIB$/10^{-27}$ & Dust\\
  \hline
  \multicolumn{1}{@{\hspace{0.5cm}}c@{\hspace{0.3cm}}}{$Indep$} & $-5.3$ & $16.5 $ & $15.4$ & $1.76$ & $1.32$ & $1.0$ \\
    \multicolumn{1}{@{\hspace{0.0cm}}c@{}}{$Joint$} & $-7 \times 10^{-11}$ & $2 \times 10^{-10}$ & $-2 \times 10^{-11}$ & $-1 \times 10^{-13}$ & $7 \times 10^{-14}$ & $1$ \\
  \hline
  \end{tabular}
  \end{center}
  \caption{Determination of $\fnl$ for the local, equilateral, orthogonal, point sources, CIB and dust shapes using the dust map studied in section \ref{sec:dust}. The only error bars at our disposal are Fisher forecasts; they are not indicated because for every case given here, they are several (at least three) orders of magnitude smaller than the determined values for $\fnl$ in the independent case, which makes them many orders of magnitude larger than the non-dust values in the joint analysis.} 
  \label{tab:dust}
  \end{table}

Then, we will focus on the second aim which is to analyze the CMB map from the 2015 Planck data. We will apply the numerical templates to the cleaned \texttt{SMICA} CMB map \cite{Adam:2015tpy}, both at low and high resolutions, for which we expect not to detect any galactic foregrounds. Finally we will perform a similar analysis on raw sky observations at 143 GHz.

\subsection{Gaussian simulations}
\label{sec:gaussian-simulations}

For the first tests, we constructed a set of 100 Gaussian simulations of the CMB obtained using the best fit of the CMB power spectrum from the 2015 Planck release \cite{Aghanim:2015xee} at the resolution $n_{\mathrm{side}}=2048$. There are several reasons to use these simulations instead of the observed CMB  map. First, it is important to check the validity of this new use of the binned bispectrum estimator with a large number of maps. Moreover, even the cleaned CMB map still contains contamination from extra-galactic foregrounds and the ISW-lensing. Here, these effects are not present. However, we need the Gaussian realizations of the CMB to have the characteristics of the \texttt{SMICA} CMB map. Hence, we smoothed the maps using a 5~arcmin FWHM Gaussian beam and we added noise based on the noise power spectrum of the \texttt{SMICA} CMB map (moreover, our choice of bins is optimal only if this noise is present in the maps, because it diminishes the weights of the bins at high $\ell$ following \eqref{eq:variance}). In this section, we will discuss two different cases for the noise. First, we will assume it has an isotropic distribution in pixel space. In the second case we will make it anisotropic by modulating it in pixel space using the hit-count map corresponding to the scanning pattern of the Planck satellite. Finally, we add some dust to these maps using the dust map at 143 GHz discussed in section \ref{sec:dust}. Every analysis presented in this section uses the common mask introduced in section \ref{sec:foregrounds}, see figure~\ref{fig:masks}.

The determination of the amplitude parameters is performed using the binned bispectrum estimator, including a linear correction term to the bispectrum as discussed in section \ref{sec:variance}. In practice, the linear correction term is computed using Gaussian simulations of the analyzed maps with the same characteristics (beam, noise, mask). We use the average power spectrum of our 100 maps (CMB + dust) to generate the maps necessary for the computation of the linear correction. Here we use 80 maps for the linear correction. We have verified that this number is sufficient to detect squeezed bispectra like the local and the dust shapes to high precision. The first analysis is performed with the same choice of 57 bins as in the 2015 Planck analysis \cite{Ade:2015ava} which was shown to be optimal to determine the primordial shapes, using multipoles from $\ell_\mathrm{min}=2$ to $\ell_\mathrm{max}=2500$ (remember that our analysis is temperature only). We add the dust map to the simulations of the CMB, thus the expected value of the $\fnl$ for the dust template is $1$. We also determine the amplitude parameters $\fnl$ for the primordial shapes, the point sources and CIB bispectra in both the independent and the joint case.

\begin{table}
\begin{center}
  \small
\begin{tabular}{lcccccc}
\hline
& Local & Equilateral & Orthogonal & P.S.$/10^{-29}$ & CIB$/10^{-27}$ & Dust\\
\hline
\multicolumn{4}{l}{Dust 100$\%$, 57 bins (expected $\fnl^{\mathrm{dust}}=1$)} &&& \\
\multicolumn{1}{@{\hspace{0.7cm}}c@{\hspace{0.5cm}}}{$Indep$} & $-86 \pm 14$ & $27 \pm 67$ & $103 \pm 38$ & $1.4 \pm 0.9$ & $1.1\pm 0.5$ & $1.03 \pm 0.20 $ \\
  \multicolumn{1}{@{\hspace{0.0cm}}c@{}}{$Joint$} & $-6 \pm 14$ & $16 \pm 77$ & $-10 \pm 45$ & $0.1 \pm 2.6$ & $0.0 \pm 1.5$ & $1.00 \pm 0.24$ \\
\multicolumn{4}{l}{Dust 100$\%$, 70 bins (expected $\fnl^{\mathrm{dust}}=1$)} &&& \\
\multicolumn{1}{@{\hspace{0.7cm}}c@{\hspace{0.5cm}}}{$Indep$} &  -67 $\pm$ 11 & 20 $\pm$ 68 & 92 $\pm$ 34 & 1.4 $\pm$ 1.0 & 1.0 $\pm$ 0.5 & 1.00 $\pm$ 0.20 \\
\multicolumn{1}{@{\hspace{0.0cm}}c@{}}{$Joint$} & 0 $\pm$ 14 & -5 $\pm$ 75 & -1 $\pm$ 39 & 0.0 $\pm$ 2.6 & 0.0 $\pm$ 1.4 & 1.01 $\pm$ 0.24 \\
  \multicolumn{4}{l}{Dust 75$\%$, 70 bins (expected $\fnl^{\mathrm{dust}}=0.42$)} &&& \\
\multicolumn{1}{@{\hspace{0.7cm}}c@{\hspace{0.5cm}}}{$Indep$} & -30 $\pm$ 8 & 11 $\pm$ 66 & 41 $\pm$ 36 & 0.6 $\pm$ 0.9 & 0.4 $\pm$ 0.5 & 0.42 $\pm$ 0.12 \\
\multicolumn{1}{@{\hspace{0.0cm}}c@{}}{$Joint$} & 0 $\pm$ 9 & 1 $\pm$ 70 & -2 $\pm$ 42 & 0.0 $\pm$ 2.6 & 0.0 $\pm$ 1.4 & 0.42 $\pm$ 0.13 \\
  \multicolumn{4}{l}{Dust 0$\%$, 70 bins (expected $\fnl^{\mathrm{dust}}=0$)} &&& \\
\multicolumn{1}{@{\hspace{0.7cm}}c@{\hspace{0.5cm}}}{$Indep$} & -0.1 $\pm$ 0.5 & -1.7 $\pm$ 6.1 & -3.1 $\pm$ 3.4 & -0.03 $\pm$ 0.09 & -0.01 $\pm$ 0.05 & 0.001 $\pm$ 0.003 \\
\multicolumn{1}{@{\hspace{0.0cm}}c@{}}{$Joint$} & -0.3 $\pm$ 0.7 & -1.6 $\pm$ 6.4 & -4.2 $\pm$ 4.1 & -0.15 $\pm$ 0.26 & 0.06 $\pm$ 0.13 & 0.001 $\pm$ 0.003 \\
\hline
\end{tabular}
\end{center}
\caption{Determination of $\fnl$ for the local, equilateral, orthogonal, point sources, CIB and dust shapes using a set of 100 Gaussian simulations of the CMB with isotropic noise to which we added a known amount of dust (the dust map of section \ref{sec:dust} multiplied by a factor 1 or 0.75, or no dust at all). The analysis is performed using 57 bins or 70 bins and the error bars are given at 1$\sigma$. For the reason behind the much smaller error bars in the 0 $\%$ dust case, see the main text.} 
\label{tab:gaussian-isotropic}
\end{table}

\begin{figure}
  \centering
 \includegraphics[width=0.49\linewidth]{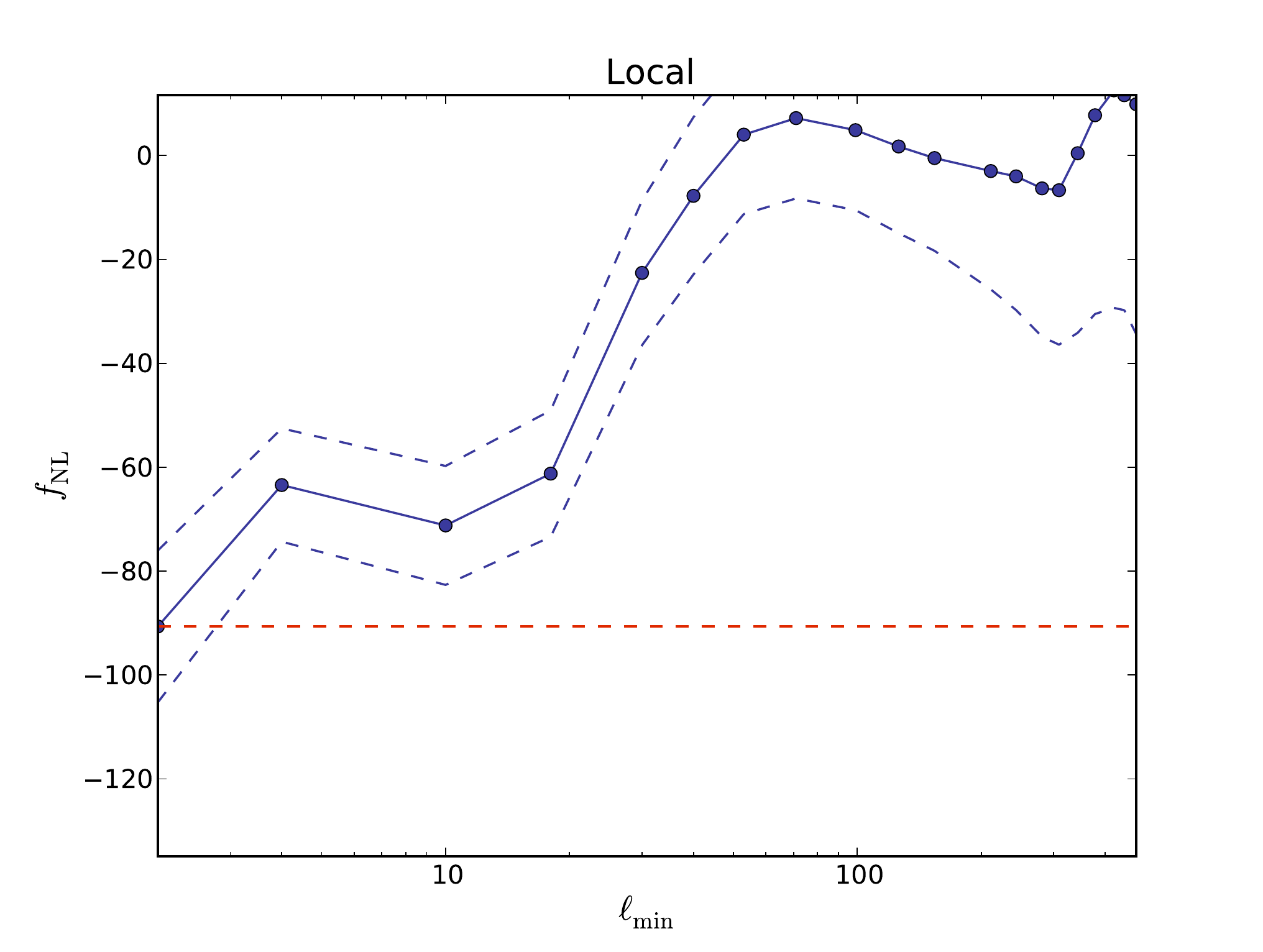}
  \includegraphics[width=0.49\linewidth]{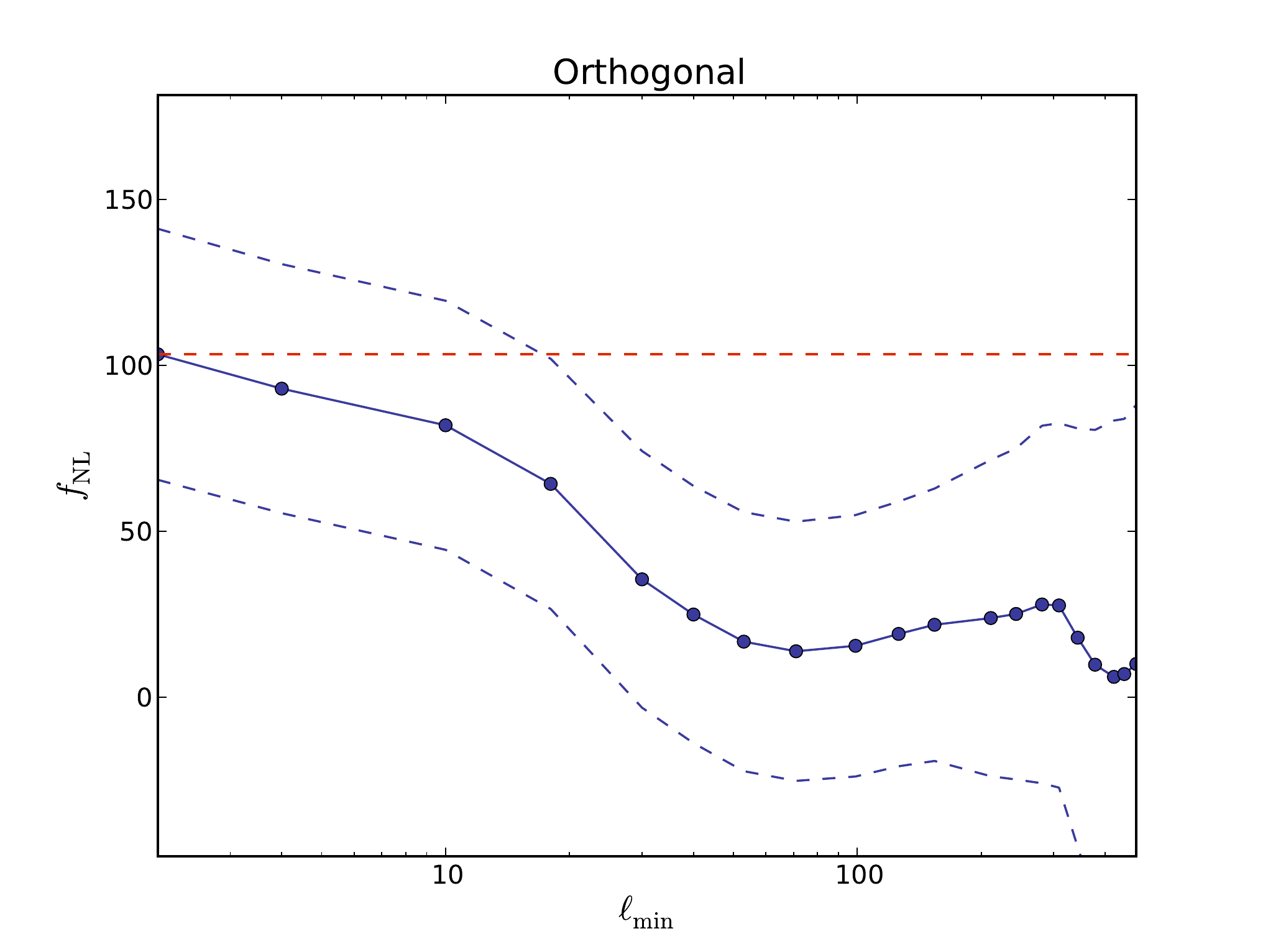}
  \includegraphics[width=0.49\linewidth]{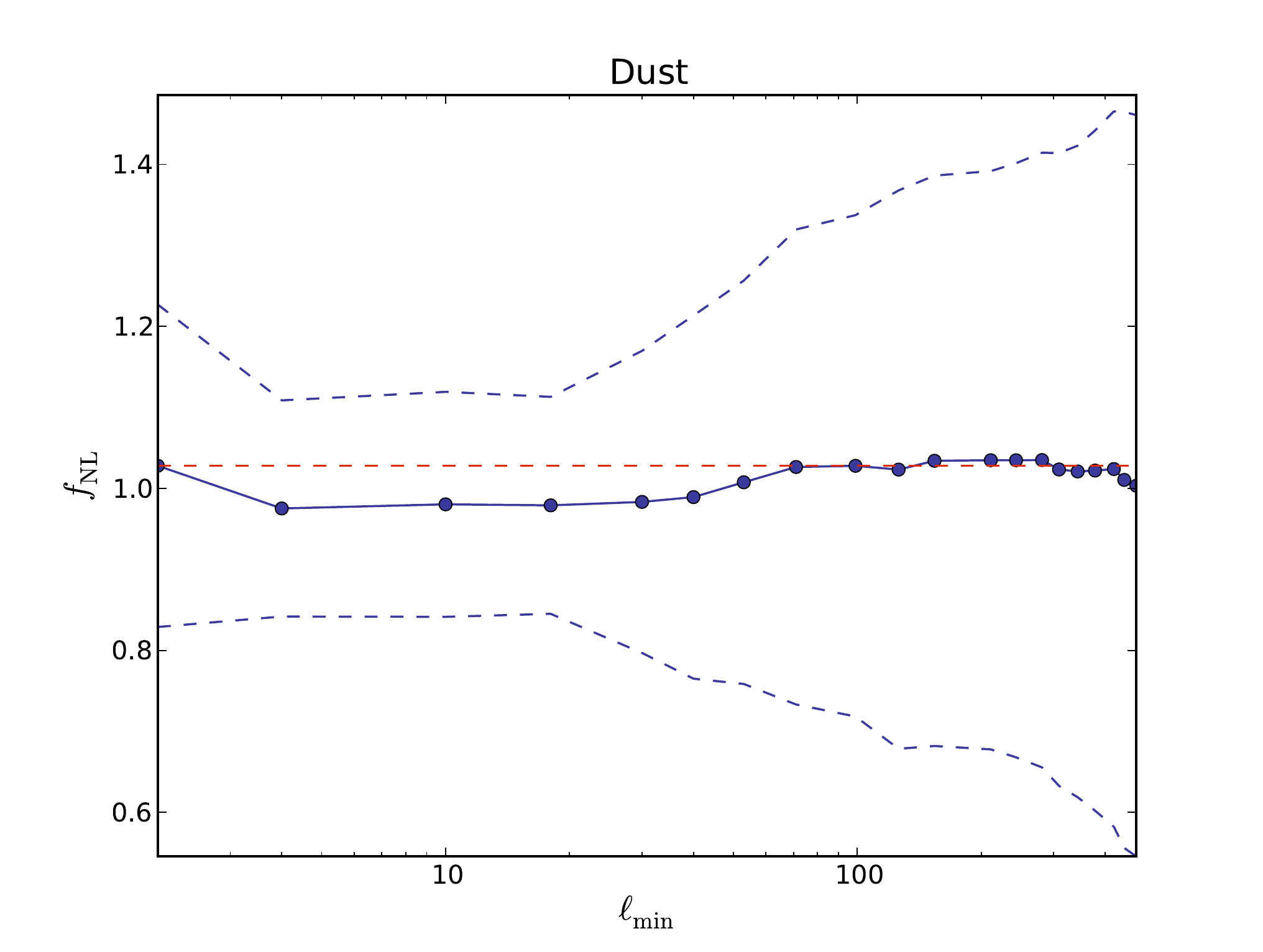}
  \includegraphics[width=0.49\linewidth]{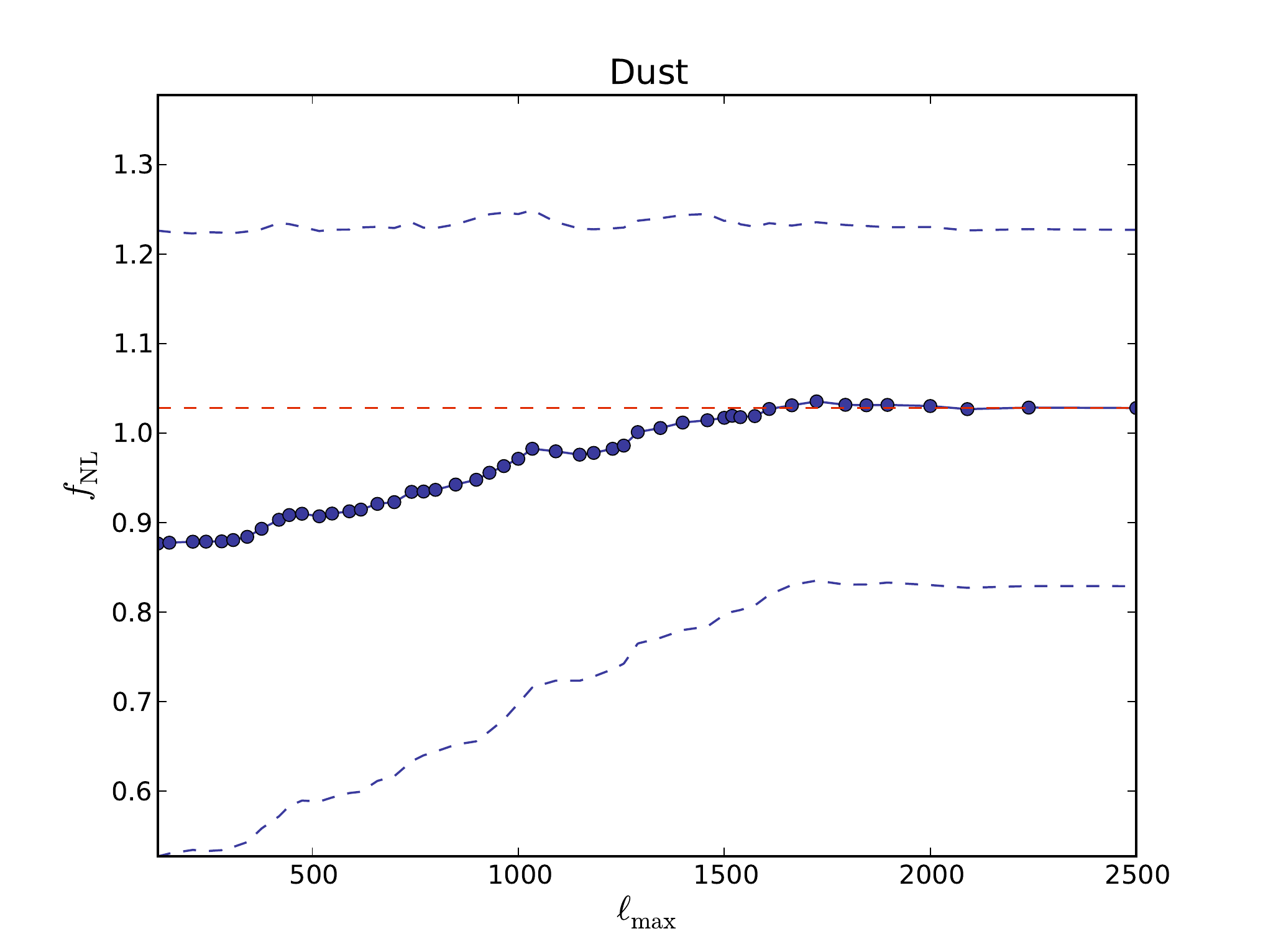}
  \caption{Convergence of $\fnl$ in the independent analysis of the 100 Gaussian CMB simulations + dust map as a function of  $\ell_\mathrm{min}$ for the local (top left), orthogonal (top right) and dust (bottom left) shapes, and as a function of $\ell_\mathrm{max}$ for the dust shape (bottom right). The blue dots correspond to the values determined by the binned bispectrum estimator when $\ell_\mathrm{min}$ (or $\ell_\mathrm{max}$ for the fourth plot) is inside the corresponding bin. The $68\%$ confidence interval is given by the blue dashed lines. The horizontal red dashed line corresponds to the determined value of $\fnl$ using the whole multipole interval (from 2 to 2500 with 57 bins).}
  \label{fig:convergence-fnl}
\end{figure}

Results are given in table \ref{tab:gaussian-isotropic}. First, we see that we detect the expected amount of dust with a good accuracy. We also observe that the shapes correlated to the dust template (see table~\ref{tab:corr_coeff_dust}), because they also peak in the squeezed configuration, are strongly detected in the independent case. However, in the joint analysis all the non-Gaussianity of the maps is attributed to the dust, with only a small impact on the error bars of the primordial shapes, meaning that this test is successful. However, this choice of bins is only optimized to detect the primordial bispectra and not the dust. Then, it is important to verify if the results can be improved by adding a few bins at very low $\ell$ (below $30$) to better measure the dust contribution. This can be seen in appendix~\ref{ap:weights-bisp-shap} where we observe that only the very low $\ell_1$ are important for the template (it is more squeezed than the local shape). Figure~\ref{fig:convergence-fnl} can also be used to highlight this effect. It shows the convergence of $\fnl$ when using a smaller multipole interval to determine $\fnl$. In the two top plots, we can see that if we exclude the very low $\ell$ (below 30) both the local and orthogonal $\fnl$ are consistent with 0. If we exclude the region of multipole space where the dust template is the strongest, there is no detection of the primordial shapes, even in an independent analysis. The two bottom plots are interesting as they show that the determination of $\fnl$ for the dust template is very stable when increasing $\ell_{\mathrm{min}}$ or decreasing $\ell_{\mathrm{max}}$. Note however that the error bars on $\fnl^{\mathrm{dust}}$ increase a lot if we use $\ell_{\mathrm{min}}>30$. This is visible with the dashed blue lines which correspond to the 68$\%$ confidence intervals.

There is another important effect in the dust template when a very large scale (small $\ell$) is concerned: the sawtooth pattern in the dust power spectrum (see figure \ref{fig:dust-power-spectrum}) is also expected in the dust bispectrum for the same reason (the only large harmonic coefficients describing the dust at low $\ell$ are the $a_{\ell 0}$ with $\ell$ even). In principle, it could be used to differentiate between the dust and the local shapes, but this effect is hidden if the bins are large because it is averaged over several $\ell$'s, thus providing another motivation to add some bins at low $\ell$. One issue when adding bins is that the memory constraints on the computer system we use limit us to a number of bins between 50 and 60 at most at the Planck resolution when including the polarization too. Here we can use 70 bins, because we only look at the temperature data and because we only add bins for the largest scales, where it is possible to downgrade the resolution of the filtered maps. With this new binning, the correlation coefficient between the local and dust shapes becomes $-0.48$ (instead of $-0.60$ for 57 bins, see table~\ref{tab:corr_coeff_dust}). So indeed adding a few bins at low $\ell$ helps to differentiate these squeezed shapes. The results of the same test with 70 bins are also given in table~\ref{tab:gaussian-isotropic}. In the independent analysis, the amount of local non-Gaussianity and its error bar decreases which is consistent with the fact that the dust and the local templates are easier to differentiate with the new binning. However, in the joint analysis there is no clear difference, except that the different central values are now very close to the expected values.

We also have to note that the approximation of weak non-Gaussianity, which is needed for the validity of the linear correction of the bispectrum to take into account the effects of the mask here, starts to break down when we observe a local shape at more than $6\sigma$ (independent case). This is why it is important to verify how a similar analysis works with a smaller amount of dust in the map. Hence, with the same choice of 70 bins, we perform two other tests with the 100 CMB simulations. For one we multiply the dust map by a factor 0.75 before adding it to the CMB realizations and the expected value of $\fnl$ is then $0.75^3 \approx 0.42$. For the other test, we use the the Gaussian CMB maps without adding dust, to verify that we do not detect any bispectral shape. These results are also given in table \ref{tab:gaussian-isotropic} and are exactly as expected.

Note that the error bars for the case of the CMB only are roughly one order of magnitude smaller than for the rest. The reason is that we made a distinction between the standard deviation (square root of the variance) and the standard error (standard deviation divided by the square root of the number of maps, so divided by 10 here). The standard error gives the expected error on the determination of the mean value of $\fnl$ with our sample of 100 Gaussian maps. The standard deviation gives the $1\sigma$ interval in which we would detect $\fnl$ if we study one map. It is clear that the standard error has to be used in the CMB-only analyses because we determine the mean value of each $\fnl$ from a sample of 100 maps. However, when we add dust to these maps, the situation is different because we only have one realization of the dust so the standard error cannot be used. We are however very conservative by using the standard deviation, the real error bars on the mean values of the different $\fnl$ are probably between the standard error and the standard deviation (the more dust in the map, the closer to the standard deviation it will be). However, the fact that for the two amounts of dust with 70 bins the central values in the joint analysis are so close to the expected values is an indication that the error bars are likely overestimated for these two cases (the results with dust would still be correct if we divided the standard deviation by 10 to obtain the standard error, which is not true with 57 bins).

\begin{figure}
  \centering \includegraphics[width=0.49\linewidth]{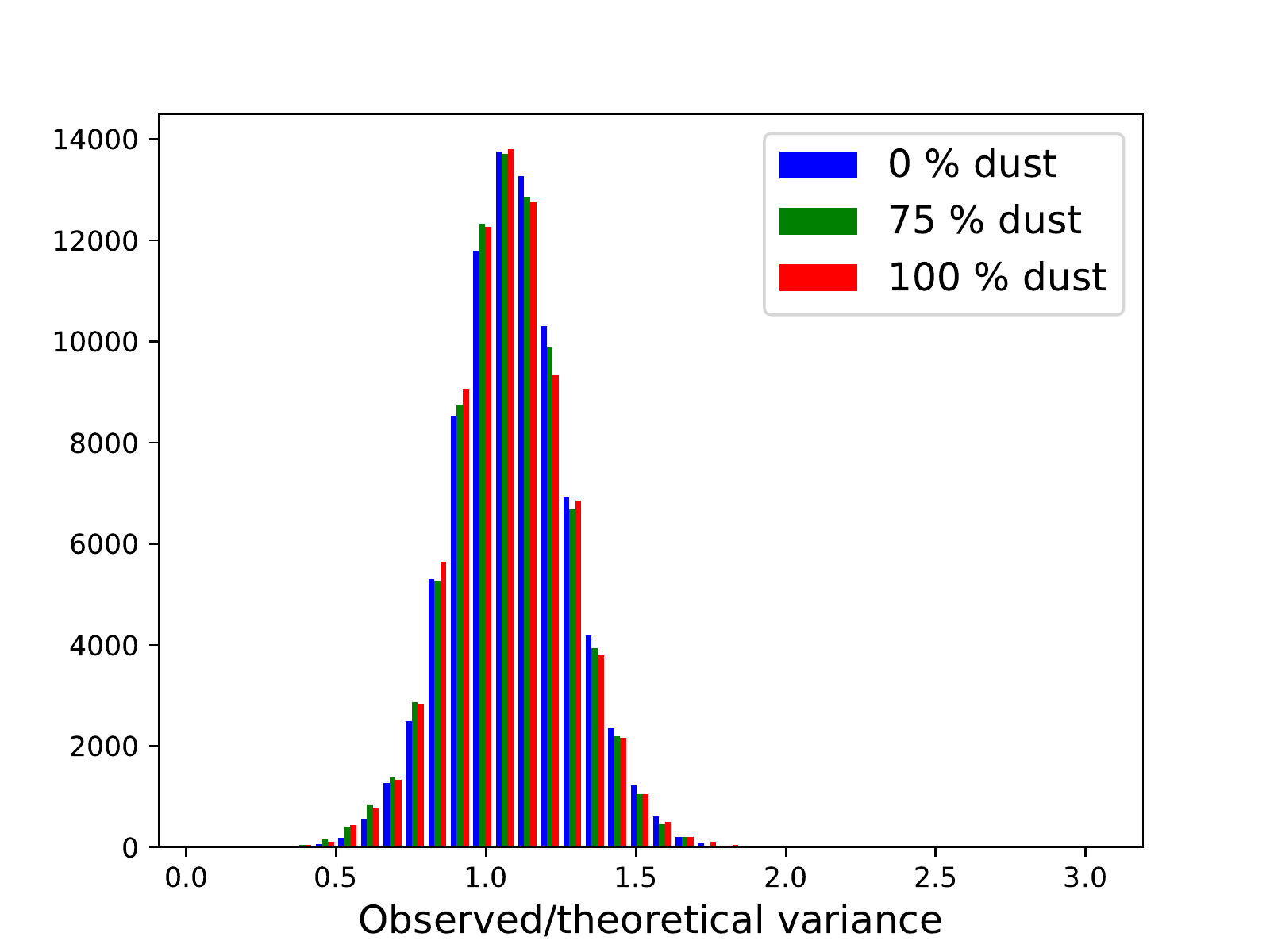}
  \includegraphics[width=0.49\linewidth]{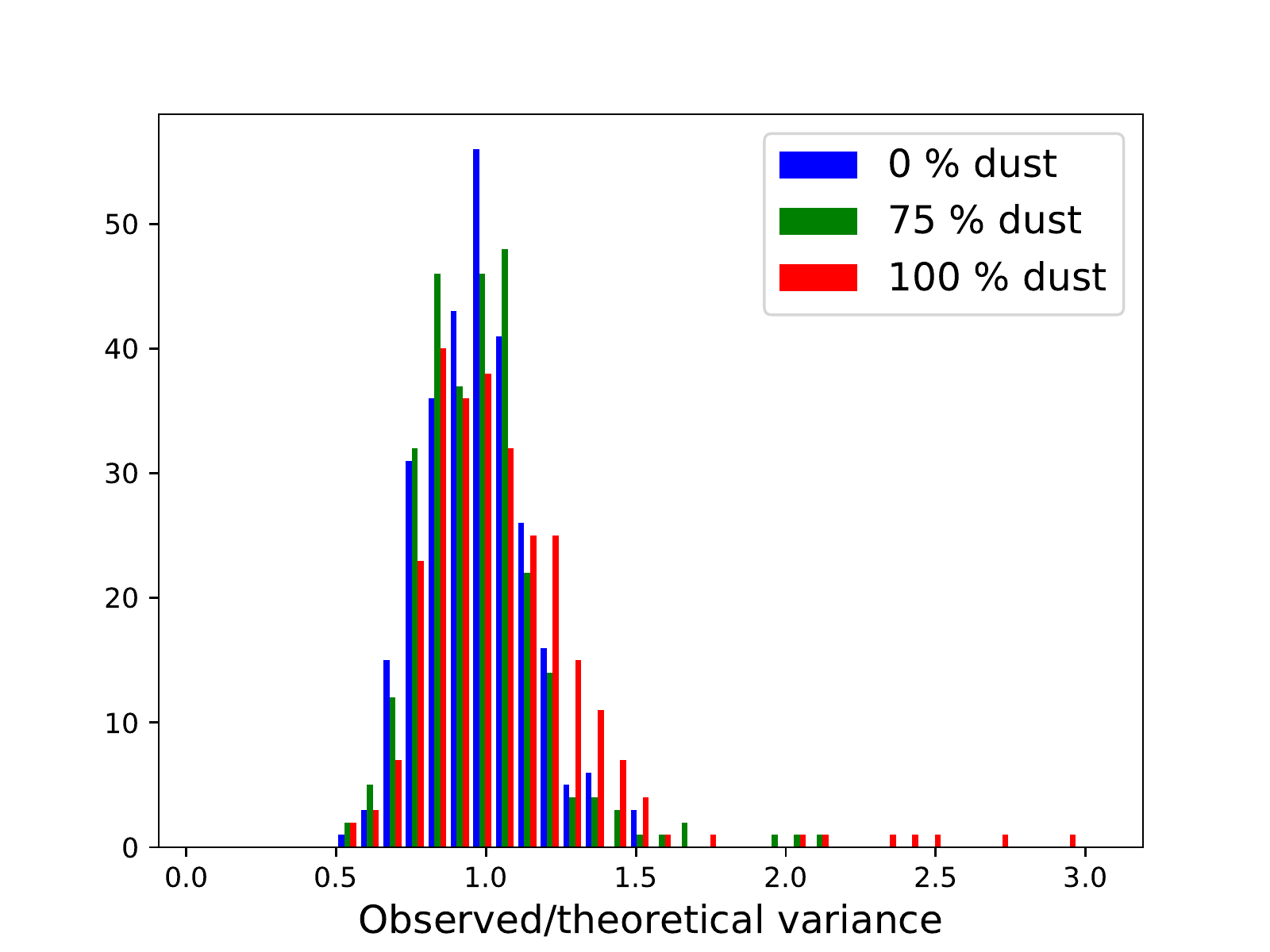}
  \caption{Distribution of the ratio of the observed variance over the theoretical prediction for the three different amounts of dust (0 $\%$ in blue, 75 $\%$ in green and 100 $\%$ in red). On the left, all the valid $\ell$-triplets are included while on the right only very squeezed $\ell$-triplets are shown (one small multipole $\ell_1 \leq 13$ and two large ones $\ell_{2,3} \geq 742$).}
  \label{fig:ratio-variance}
\end{figure}

We can illustrate the breakdown of the weak non-Gaussianity approximation using the variance of the bispectrum. Indeed, we have at our disposal a theoretical prediction for the variance, given in \eqref{eq:variance}, that scales as the power spectrum cubed and for which the derivation relies on the weak non-Gaussianity approximation. However, we can also directly compute the variance of the bispectrum from our 100 maps, which we call here observed variance. Figure \ref{fig:ratio-variance} shows the distribution of the ratio of the observed variance over the theoretical variance for the three different amounts of dust in two different configurations. First, we examine this ratio over the whole triplet space (on the left) where there is no difference between the three cases and the values are distributed around 1 as expected. That is logical because the non-Gaussianity of the dust is very localized in multipole space; the bispectrum is large only in the very squeezed configuration. This is why on the right we consider only the triplets where one $\ell$ is very small (in the first five bins i.e.\ $\ell \leq 13$) and the two others large (in the last 30 bins, i.e.\ $\ell \geq 742$). Adding or removing a few bin triplets here does not change the results. Here we can see that if there is more dust (in red),  there are several values which strongly deviate from one. This effect is even more obvious when we examine the mean and the standard deviation of these distributions, which are given in table \ref{tab:ratio-variance}. When considering the full space of multipole triplets, there is no significant difference between the three cases. However, when we examine only the squeezed part of the bispectrum, the standard deviation increases slightly with a small amount of dust (75 $\%$), and is three times larger for 100 $\%$ dust compared to the CMB-only case. Hence, the weak non-Gaussianity approximation stops being valid, but not enough to invalidate the results (only a few bin-triplets deviate strongly). However, if we were to add even more dust, we would have to take this effect into account.

\begin{table}
  \begin{center}
    \small
    \begin{tabular}{l|cc|cc|cc}
      \hline
      & \multicolumn{2}{c|}{0 $\%$ dust} & \multicolumn{2}{c|}{75 $\%$ dust} & \multicolumn{2}{c}{100 $\%$ dust}\\
  & \multicolumn{1}{c}{Full} & \multicolumn{1}{c|}{Squeezed} & \multicolumn{1}{c}{Full} & \multicolumn{1}{c|}{Squeezed} & \multicolumn{1}{c}{Full} & \multicolumn{1}{c}{Squeezed}   \\ 
      \hline
     Mean & 1.09 & 0.97 & 1.08 & 0.97 & 1.08 & 1.11\\
     Standard deviation & 0.19 & 0.16 & 0.19 & 0.21 & 0.20 & 0.51\\
     \hline
    \end{tabular}
    \caption{Means and standard deviations of the distributions of the ratio of the observed variance over the theoretical prediction, shown in figure \ref{fig:ratio-variance}, for the three amounts of dust, including the full bispectrum or only a very squeezed part of it.}
    \label{tab:ratio-variance}
  \end{center}
\end{table}

In addition to looking at the variance of the bispectrum itself, we can also investigate the variance of the $\fnl$ parameters with regard to the validity of the weak non-Gaussianity approximation.
Every error bar given in table \ref{tab:gaussian-isotropic} was computed from the observed variance of the set of 100 maps. However, we can also compute Fisher error bars from the theoretical prediction of the variance and they are given in table~\ref{tab:error-bars-gaussian} for the three cases studied in this section for the local and dust shapes. We can see that for both, the more non-Gaussian the map is, the more important is the difference between Fisher and observed error bars. This is related to the breakdown of the weak non-Gaussianity approximation. For a local $|\fnl|$ of around 70 (corresponding to 100 $\%$ dust), the difference is a factor 2 between the two kinds of error bars. The difference is larger for the dust template, where for this case the observed error bars are four times larger than the Fisher forecasts. For both templates, when there is no dust (so purely Gaussian maps), the observed error bars agree with the Fisher forecasts up to the expected precision (the relative error in the standard deviation is $1/\sqrt{2(N-1)}$, which is $7~\%$ for 100 maps).

\begin{table}
\begin{center}
  \small
  \begin{tabular}{l|ccc}
    \hline
        & 100 $\%$ dust & 75 $\%$ dust & 0 $\%$ dust \\
        \hline
    Local &&& \\
    \;\;{\em Fisher} & 6.6 & 6.4 & 5.6 \\
    \;\;{\em Observed} & 14 & 7.7 & 5.2 \\
    Dust &&& \\
    \;\;{\em Fisher} & 0.05 & 0.04 & 0.031 \\
    \;\;{\em Observed} & 0.20 & 0.12 & 0.030 \\
    \hline
  \end{tabular}
\end{center}
\caption{Fisher and observed standard deviations on $\fnl^{\mathrm{local}}$ and $\fnl^{\mathrm{dust}}$ (independent analysis) determined from 100 Gaussian simulations of the CMB with isotropic noise to which we added a known amount of dust (100 $\%$, 75 $\%$ or 0 $\%$ of the dust map of figure  \ref{fig:dust-map}) using 70 bins.} 
\label{tab:error-bars-gaussian}
\end{table}

\begin{figure}
  \centering
   \includegraphics[width=0.49\linewidth]{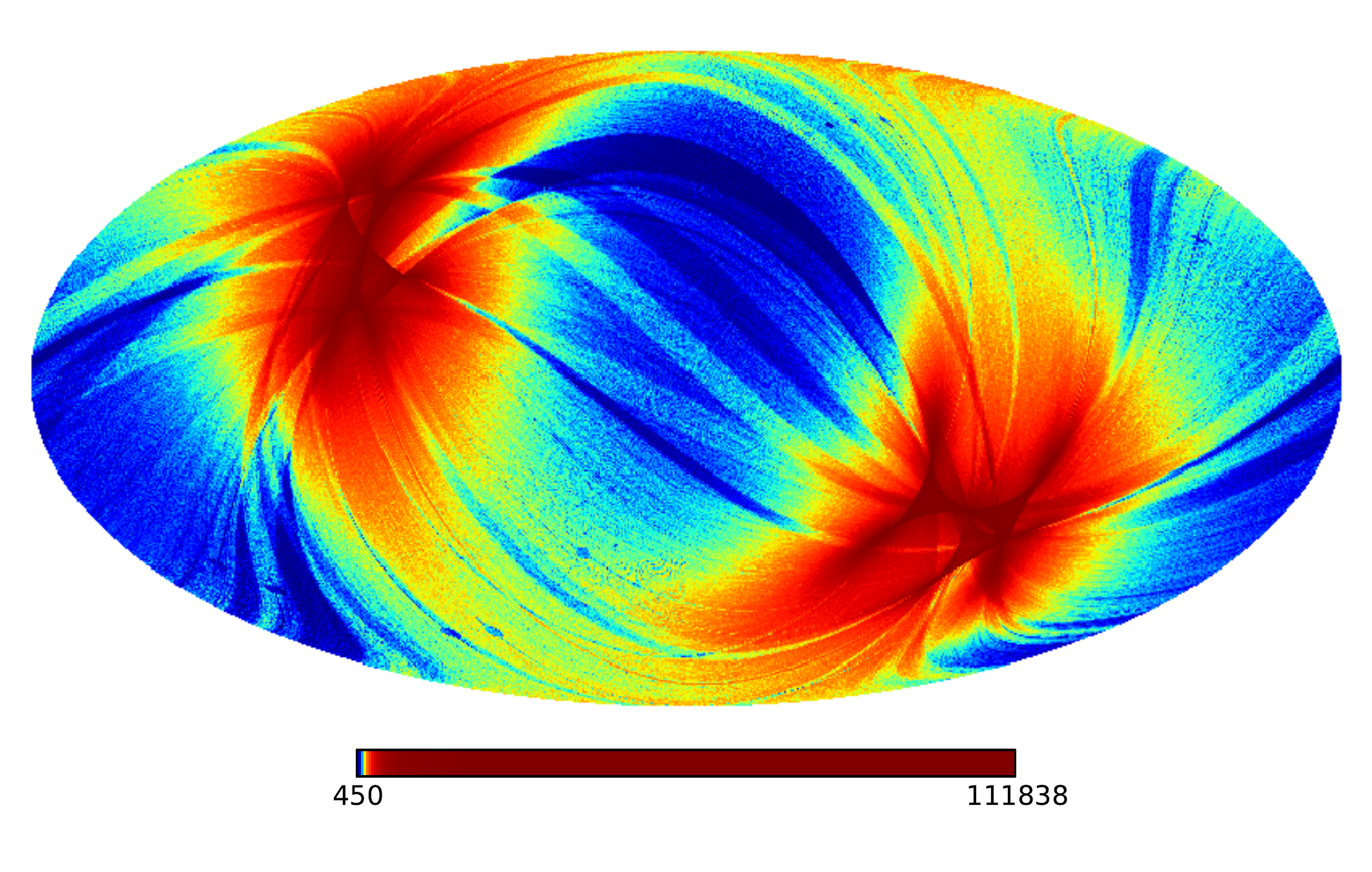}
 \includegraphics[width=0.49\linewidth]{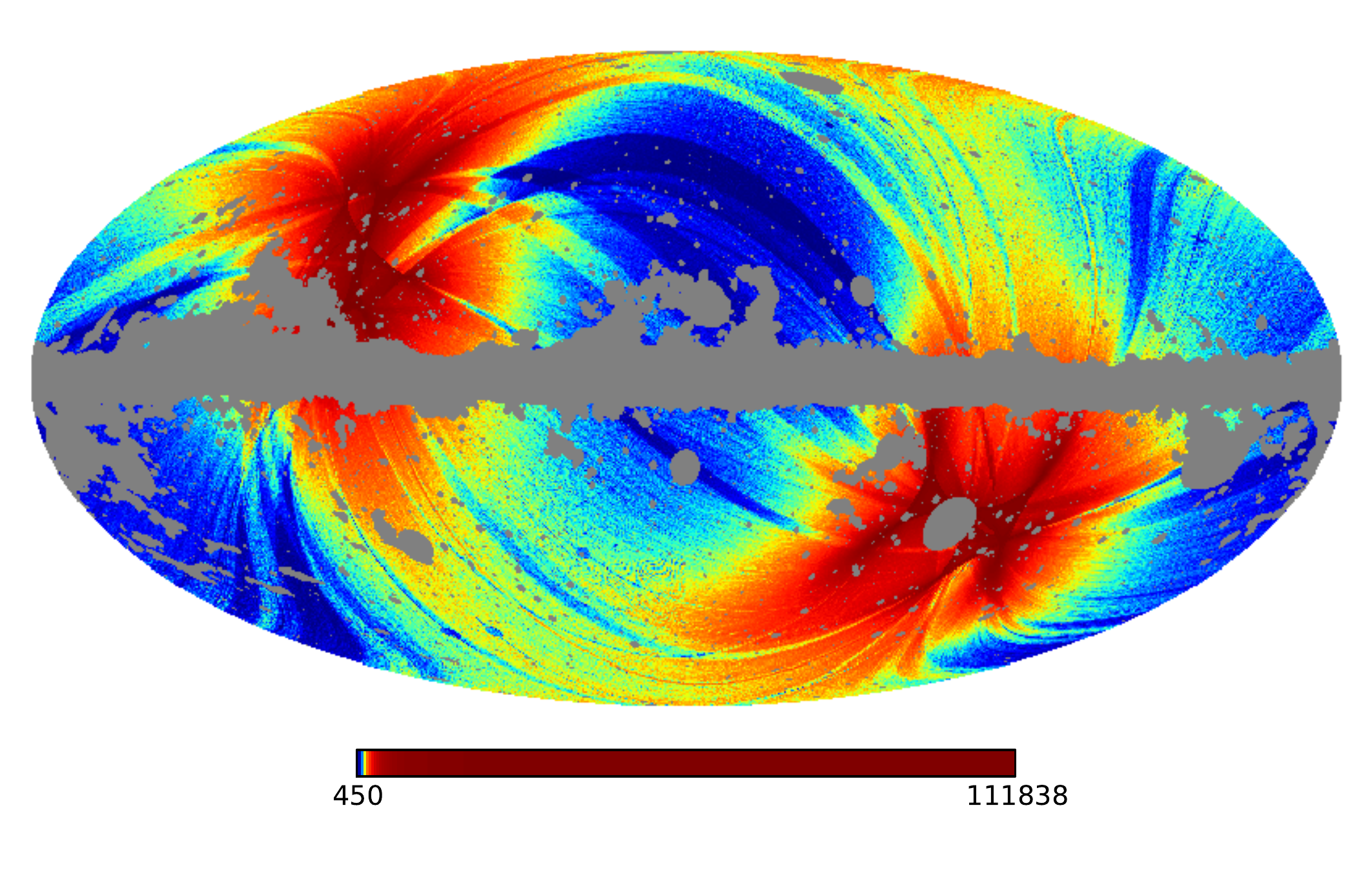}
  \caption{Unmasked (left) and masked (right) hit-count map of Planck (number of observation samples per pixel).}
  \label{fig:hitcount-map}  
\end{figure}

As explained before, adding noise realizations with the correct power spectrum to the CMB simulations is necessary for the optimization of the binning and to make the simulations more realistic. However, the real instrument noise does not have an isotropic distribution in pixel space because some parts of the sky are observed more often than others, as shown in figure \ref{fig:hitcount-map}. Without a linear correction, the anisotropic noise also gives a large squeezed contribution to the bispectrum for the usual reason: small-scale fluctuations are larger (more noise) in the large-scale regions which are less observed and vice versa.
This is why we also verify the previous results with an anisotropic distribution of the noise following the scanning pattern of the Planck satellite. The results are given in table~\ref{tab:gaussian-anisotropic}. Here we use only the best choice of bins (70 bins) and the results are given for the same three amounts of dust as in table~\ref{tab:gaussian-isotropic}. Results are very similar with isotropic and with anisotropic noise for the three cases; each time we detect successfully the amount of dust we added to the maps. 

\begin{table}
\begin{center}
  \small
\begin{tabular}{lcccccc}
\hline
& Local & Equilateral & Orthogonal & P.S.$/10^{-29}$ & CIB$/10^{-27}$ & Dust\\
\hline
\multicolumn{4}{l}{Dust 100$\%$ (expected $\fnl^{\mathrm{dust}}=1$)} &&& \\
\multicolumn{1}{@{\hspace{0.7cm}}c@{\hspace{0.5cm}}}{$Indep$} & -67 $\pm$ 11  & 24 $\pm$ 65 & 93 $\pm$ 34 & 1.5 $\pm$ 1.1 & 1.1 $\pm$ 0.5 & 1.00 $\pm$ 0.20\\
  \multicolumn{1}{@{\hspace{0.0cm}}c@{}}{$Joint$} & -1 $\pm$ 14 & 4 $\pm$ 73 & -4 $\pm$ 40 & 0.2 $\pm$ 2.8 & 0.0 $\pm$ 1.4 & 1.00 $\pm$ 0.24\\
\multicolumn{4}{l}{Dust 75$\%$ (expected $\fnl^{\mathrm{dust}}=0.42$)} &&& \\
\multicolumn{1}{@{\hspace{0.7cm}}c@{\hspace{0.5cm}}}{$Indep$} & -30 $\pm$ 7 & 4 $\pm$ 62 & 44 $\pm$ 33 & 0.5 $\pm$ 1.1 & 0.4 $\pm$ 0.5 & 0.42 $\pm$ 0.12 \\
\multicolumn{1}{@{\hspace{0.0cm}}c@{}}{$Joint$} & 0 $\pm$ 10 & -5 $\pm$ 69 & 1 $\pm$ 38 & -0.1 $\pm$ 2.8 & 0.0 $\pm$ 1.4 & 0.42 $\pm$ 0.14\\
  \multicolumn{4}{l}{Dust 0$\%$ (expected $\fnl^{\mathrm{dust}}=0$)} &&& \\
\multicolumn{1}{@{\hspace{0.7cm}}c@{\hspace{0.5cm}}}{$Indep$} & -0.15 $\pm$ 0.50 & 0.3 $\pm$ 6.6 & -1.4 $\pm$ 3.8 & -0.08 $\pm$ 0.10 & -0.05 $\pm$ 0.05 & 0.000 $\pm$ 0.003\\
\multicolumn{1}{@{\hspace{0.0cm}}c@{}}{$Joint$} & -0.30 $\pm$ 0.64 & 1.2 $\pm$ 6.8 & -2.2 $\pm$ 4.3 & 0.08 $\pm$ 0.23 & -0.08 $\pm$ 0.11 & 0.000 $\pm$ 0.004 \\
\hline
\end{tabular}
\end{center}
\caption{Determination of $\fnl$ for the local, equilateral, orthogonal, point sources, CIB and dust shapes using a set of 100 Gaussian simulations of the CMB with anisotropic noise to which we added a known amount of dust (the dust map of section \ref{sec:dust} multiplied by a factor 1 or 0.75, or no dust at all). The analysis is performed using 70 bins and the error bars are given at 1$\sigma$.} 
\label{tab:gaussian-anisotropic}
\end{table}

With these different tests, we have proven that the binned bispectrum estimator can be used to detect a galactic foreground shape that we determined numerically. It works well with the amount of dust that is expected at 143 GHz, the dominant frequency channel in the cleaned CMB map. The next logical step is then to use the template on real data.

\subsection{CMB analyses}
\label{sec:cmb-analyses}

The previous tests have shown that detecting the dust is possible when there is a large amount of it. We can now apply the dust template to a real CMB analysis. Here, we follow the analysis of the Planck 2015 paper~\cite{Ade:2015ava} (note that we only study the temperature bispectrum, while for Planck the polarization was also taken into account). We use a set of 160 simulation maps for the computation of the error bars and the linear correction. The power spectrum is the best fit cosmological model from the 2015 Planck analysis. This time, we also include the ISW-lensing shape in the analysis because it is present in the data. The amplitude of this template is known, so it can be used to subtract the bias (see equation \eqref{eq:bias}) from the bispectral non-Gaussianity of the map. Results are given in table \ref{tab:cmb-high}.\footnote{The difference between the values in this table and those in the Planck paper~\cite{Ade:2015ava}, in particular for equilateral, is mainly due to our use here of a slightly different mask (the preferred temperature mask from~\cite{Adam:2015tpy} instead of the slightly extended mask used in~\cite{Ade:2015ava}).} We include the results with and without taking into account the ISW-lensing bias, and we perform two different joint analyses for comparison, with and without the dust. As expected, there is no detection of the primordial shapes or the dust. However, it is important to note that the error bars of the local and dust shapes in the joint analysis increase because these shapes are correlated. Similarly to the previous section, one way to improve the situation would be to find a binning that is optimal for both shapes. 

\begin{table}
\begin{center}
  \footnotesize
\begin{tabular}{lccccccc}
\hline
& Local & Equilateral & Orthogonal & P.S.$/10^{-29}$ & CIB$/10^{-27}$ & Dust/$10^{-2}$ & Lensing-ISW\\
\hline
\multicolumn{4}{l}{No ISW-lensing bias subtraction} &&&& \\
\multicolumn{1}{@{\hspace{0.3cm}}c@{\hspace{0.2cm}}}{$Indep$} & 8.7 $\pm$ 5.5 & 8 $\pm$ 67 & -34 $\pm$ 33 & 9.6 $\pm$ 1.0 &  4.6 $\pm$ 0.5 & -0.8 $\pm$ 3.8 & 0.59 $\pm$ 0.29 \\
  \multicolumn{1}{@{\hspace{0.0cm}}c@{}}{$Joint$} & 6 $\pm$ 8 & -21 $\pm$ 69 & -3 $\pm$ 38 & 7.3 $\pm$ 2.7 & 1.2 $\pm$ 1.4 & -2.2 $\pm$ 5.2 & 0.57 $\pm$ 0.31 \\
  \multicolumn{1}{@{\hspace{0.0cm}}c@{}}{$Joint\setminus dust$} & 4.2 $\pm$ 6.7 & -15 $\pm$ 68 & -6.6 $\pm$ 37 & 7.2 $\pm$ 2.7 & 1.3 $\pm$ 1.4 & & 0.55 $\pm$ 0.31\\
\multicolumn{4}{l}{ISW-lensing bias subtracted} &&&& \\
\multicolumn{1}{@{\hspace{0.3cm}}c@{\hspace{0.2cm}}}{$Indep$} &  1.2 $\pm$ 5.5 & 6 $\pm$ 67 & -8 $\pm$ 33 &  9.6 $\pm$ 1.0 & 4.6 $\pm$ 0.5 &  -4.0 $\pm$ 3.8 &  \\
  \multicolumn{1}{@{\hspace{0.0cm}}c@{}}{$Joint$} &  -5 $\pm$ 8 & -16 $\pm$ 69 & 1 $\pm$ 38 & 7.1 $\pm$ 2.7 & 1.3 $\pm$ 1.4 & -3.5 $\pm$ 5.1 & \\
   \multicolumn{1}{@{\hspace{0.0cm}}c@{}}{$Joint\setminus dust$} & 1.0 $\pm$ 6.3 & -7 $\pm$ 68 & -5 $\pm$ 37 & 7.0 $\pm$ 2.7 & 1.4 $\pm$ 1.4 & & \\
\hline
\end{tabular}
\end{center}
\caption{Determination of $\fnl$ for the local, equilateral, orthogonal, point sources, CIB, dust and ISW-lensing shapes in the cleaned \texttt{SMICA} CMB map from the 2015 Planck release. In the three first lines, the ISW-lensing shape is considered as the others. In the last three, the ISW-lensing bias is subtracted. The joint analysis is performed with and without the dust template. The binning consists of 57 bins.} 
\label{tab:cmb-high}
\end{table}

We also performed a similar analysis on a low resolution cleaned CMB map ($n_\mathrm{side}=256$) with a 60 arcmin FWHM Gaussian beam to look for the other foreground templates with the usual choice of bins. Results are given in table \ref{tab:cmb-low}. Because of the resolution and the beam, we only analyze multipoles in the interval $[2, 300]$, which is the reason for the very large error bars. As that would leave only few bins from the original binning, we split all the bins below $\ell=300$ into three (where possible, two otherwise), which gives 39 bins in total. Moreover, we did not subtract the ISW-lensing bias as its contribution is small compared to the error bars. The point sources and the CIB are not given in the table because they were not observed here. The results are consistent with zero non-Gaussianity in the map. But the new foreground shapes (AME, free-free and synchrotron) have very large error bars and even if they were present in the map, it would not be possible to detect them.

\begin{table}
\begin{center}
  \small
\begin{tabular}{lccccccc}
\hline
& Local & Equilateral & Orthogonal & Dust & Free-free & Synch$/10^5$ & AME$/10^{10}$\\
\hline
\multicolumn{1}{@{\hspace{0.3cm}}c@{\hspace{0.2cm}}}{$Indep$} & 13 $\pm$ 30 & 49 $\pm$ 155 & 66 $\pm$ 130 & -0.01 $\pm$ 0.07 & -1 $\pm$ 32 & -0.1 $\pm$ 4.4 & -10 $\pm$ 7 \\
  \multicolumn{1}{@{\hspace{0.0cm}}c@{}}{$Joint$} & 17 $\pm$ 51 & 281 $\pm$ 406 & 50 $\pm$ 287 & -0.01 $\pm$ 0.10 & 22 $\pm$ 43 & 2 $\pm$ 6 & -12 $\pm$ 8 \\
\hline
\end{tabular}
\end{center}
\caption{Determination of $\fnl$ of some primordial and all galactic templates in the cleaned \texttt{SMICA}  CMB map at low resolution $n_{\mathrm{side}}=256$ with a 60 arcmin FWHM Gaussian beam from the 2015 Planck release. Because of the low $\ell_{\mathrm{max}}=300$, the analysis is performed using 39 bins.} 
\label{tab:cmb-low}
\end{table}

\subsection{Raw sky}
\label{sec:raw-sky}

\begin{figure}
  \centering
\includegraphics[width=0.66\linewidth]{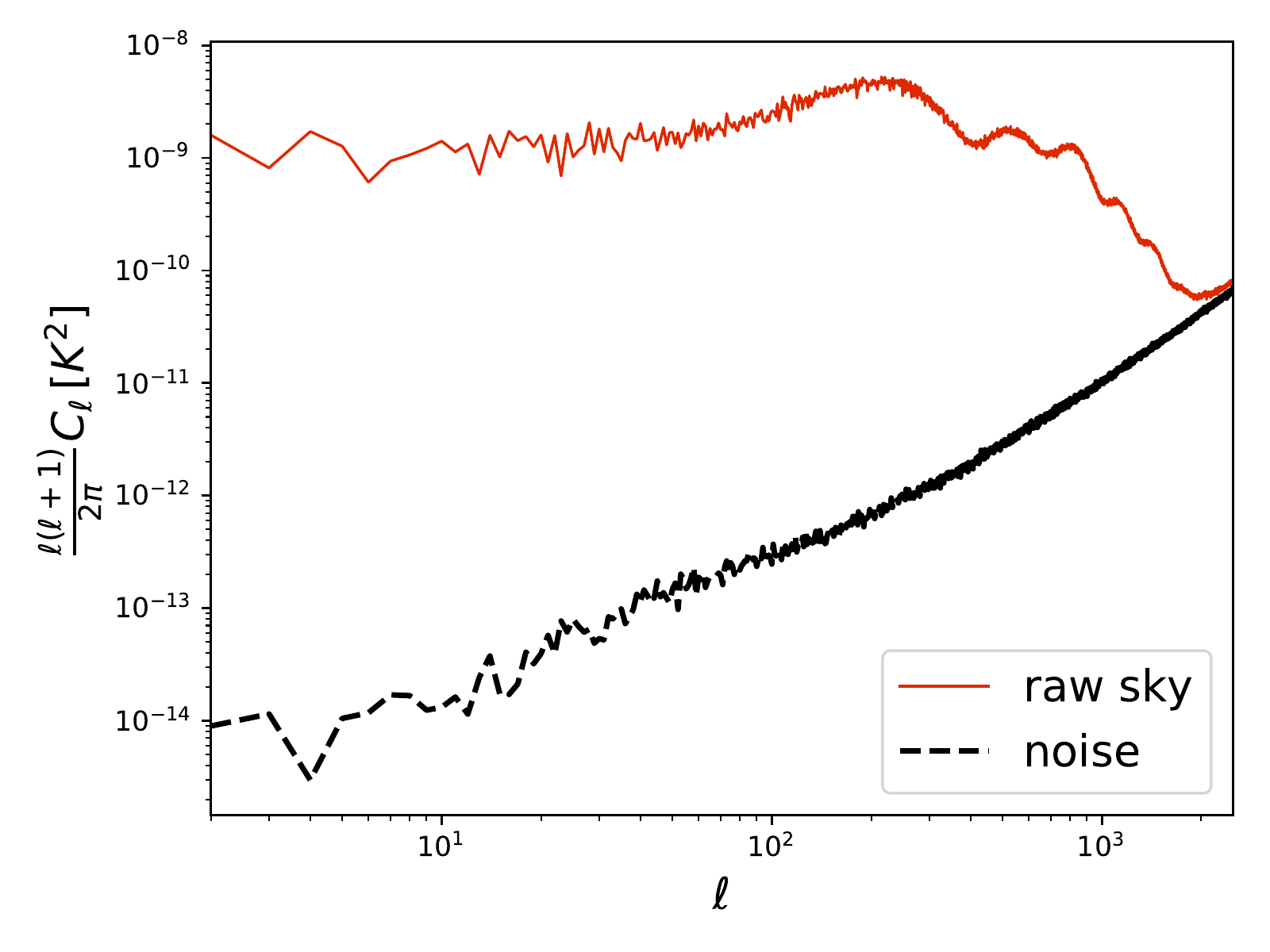}
  \caption{Power spectrum of the raw 143~GHz map as well as the estimated noise power spectrum.}
  \label{fig:raw-noise}
\end{figure}

After applying the foreground templates to the cleaned CMB map that is not supposed to contain any galactic foreground (which we confirmed), it is also interesting to study the raw 143~GHz Planck map. Again, we had to generate Gaussian simulations of this map to compute the linear correction. For this, we used the power spectrum of the map and we determined the noise power spectrum using the half-mission maps. The power spectra are shown in figure~\ref{fig:raw-noise}. The noise is modulated in pixel space using the hit-count map of figure~\ref{fig:hitcount-map} to make it anisotropic. The beam of this map can be approximated by a 7.3~arcmin FWHM Gaussian beam.

Results are given in tables \ref{tab:raw-high} and \ref{tab:raw-low}. We detected the expected amount of dust since $\fnl^{\mathrm{dust}}=1$ is within the $1\sigma$ error bars in both the independent and joint analyses. For the other foregrounds, the situation is similar to the previous section: the error bars are far too large for a detection (the synchrotron and AME shapes are not given here because the error bars are many orders of magnitude larger than the expected quantity in the map). To determine error bars, we did not have good simulations of the data map but we had of course Fisher forecasts. We made the simple but reasonable hypothesis that the factor between the real error bars and the Fisher ones due to the breakdown of the weak non-Gaussianity approximation is the same as for the anisotropic case in section~\ref{sec:gaussian-simulations}. Then it was easy to determine error bars for the primordial and the dust shapes that are a bit larger than in table~\ref{tab:gaussian-anisotropic}. However, for the CIB and the point sources, which are not the main object of study here, the situation is different because they were not present in the Gaussian simulations, so we could not apply this method. Hence we only give Fisher error bars for those two shapes, but because of the strong detection we know that they are underestimated. This is not an issue because they are uncorrelated to the local and the dust shapes. For the low-resolution case in table~\ref{tab:raw-low} we only have Fisher error bars for all shapes. In conclusion, the method also works correctly when applied to a raw sky map.

\begin{table}
\begin{center}
  \small
\begin{tabular}{lcccccc}
\hline
& Local & Equilateral & Orthogonal & P.S.$/10^{-29}$ & CIB$/10^{-27}$ & Dust \\
\hline
\multicolumn{1}{@{\hspace{0.7cm}}c@{\hspace{0.5cm}}}{$Indep$} & -61 $\pm$ 13 & 22 $\pm$ 71 & -12 $\pm$ 39 & 90 $\pm$ 4 & 28 $\pm$ 1 & 1.09 $\pm$ 0.25  \\
  \multicolumn{1}{@{\hspace{0.0cm}}c@{}}{$Joint$} & 13 $\pm$ 18 & -37 $\pm$ 81 & -81 $\pm$ 47 & 115 $\pm$ 9 & -11 $\pm$ 3 & 1.08 $\pm$ 0.32 \\
\hline
\end{tabular}
\end{center}
\caption{Determination of $\fnl$ in the raw 143~GHz map at high resolution $n_{\mathrm{side}}=2048$ from the 2015 Planck release. The analysis is performed with the usual choice of 57 bins. For the details on the error bars for the primordial and the dust shapes, see the main text. The CIB and point sources error bars are Fisher forecasts.} 
\label{tab:raw-high}
\end{table}

\begin{table}
\begin{center}
  \small
\begin{tabular}{lccccc}
\hline
& Local & Equilateral & Orthogonal & Dust & Free-free \\
\hline
\multicolumn{1}{@{\hspace{0.7cm}}c@{\hspace{0.5cm}}}{$Indep$} & -41 $\pm$ 54 & 232 $\pm$ 198 & 277 $\pm$ 177 & 0.88 $\pm$ 0.28 & 39 $\pm$ 50\\
  \multicolumn{1}{@{\hspace{0.0cm}}c@{}}{$Joint$} & -33 $\pm$ 94 & 543 $\pm$ 536 & 112 $\pm$ 388 & 0.85 $\pm$ 0.39 & -9 $\pm$ 69\\
\hline
\end{tabular}
\end{center}
\caption{Determination of $\fnl$ in the raw 143~GHz map at low resolution $n_{\mathrm{side}}=256$ from the 2015 Planck release. The analysis is performed with 39 bins. The error bars are all Fisher forecasts.} 
\label{tab:raw-low}
\end{table}

\section{Conclusion}
\label{sec:conclusion}

In this paper we used the binned bispectrum estimator to determine
the bispectra of different galactic foreground maps, as produced by
the \texttt{Commander} component separation method from Planck 2015 data
(rescaled to amplitudes representative for the 143~GHz Planck channel).
These galactic foreground
bispectra were then used as templates for other runs of the binned
bispectrum estimator applied to various types of maps: simulations,
raw sky maps, and cleaned CMB maps.

This paper serves different purposes. In the first place it is a proof
of concept. The possibility to determine the (binned) bispectrum of any
map is a clear advantage of the binned bispectrum estimator, and was
used in the official Planck releases to
present the bispectrum of the observed CMB. The fact that any provided
bispectrum, not only if an analytical template is known but also simply
any numerical bispectrum, can be used as template in the binned bispectrum
analysis pipeline, has also long been presented as an advantage of the
method. In fact the
possibility of combining these two advantages to do an analysis as
presented in this paper was already mentioned in the original
paper of the binned bispectrum estimator~\cite{Bucher:2009nm} and was
one of the motivations for developing it in the first place, but
had until now never been worked out explicitly. This paper proves that
this idea also works in practice.

Secondly this paper shows and discusses the bispectra of the various
galactic foregrounds, which is an interesting result in itself, even if
for the purposes of this paper it is only an intermediate step.
We found that the dust, the free-free and the anomalous microwave emissions
have very squeezed bispectra (similar to the local shape, but with an
opposite sign). The small-scale fluctuations of the dust radiation are
stronger in the large-scale dust clouds, so small-scale and large-scale
fluctuations are correlated (and a similar explanation is valid for
the other foregrounds). The synchrotron map as provided is different, as
its bispectrum is more similar to the equilateral shape, but we were able
to show that at least a large part of this effect is due to a residual
contamination by unresolved extra-galactic point sources. At 143 GHz (the
most important Planck frequency for CMB analysis) only the dust really 
contaminates the CMB signal, the other foregrounds giving contributions 
that are orders of magnitude smaller. An issue with the numerical
templates we determined is that they also depend on the mask applied
to the foregrounds and contain the characteristics of the experiment
like the beam and the noise. We showed that the choice of the mask is
very important because the foregrounds are localized in the galactic
plane close to the galactic mask, so a small change of mask
could give a large difference of bispectrum. This means the same mask
should be used for determining the template as for the final analysis.
It should be pointed out that for the purpose of studying the non-Gaussianity
of galactic foregrounds as goal in itself, the bispectrum is likely not the
best tool: a pixel-space based statistic to take into account the localized
nature of these foregrounds would seem more logical. However, our main
purpose is to investigate the impact these galactic foregrounds have on the
determination of primordial $f_\mathrm{NL}$ parameters in a bispectrum
analysis.

The third and final result of this paper is the $f_\mathrm{NL}$ analysis of real
sky maps, both raw and cleaned CMB, with these galactic bispectrum templates
to investigate in particular if any observable galactic residuals remain
in the cleaned CMB map and if a joint analysis of primordial and galactic
templates improves the determination of the primordial $f_\mathrm{NL}$.
But before doing that analysis we obviously
first tested and validated our methodology and our new analysis pipeline
on simulations. These tests were based on Gaussian realizations of the CMB
to which we added noise simulations and a known amount of dust. We showed
that both with isotropic and with anisotropic noise we managed to detect
the expected amount of dust in our maps. However, to do a joint analysis
with the primordial and the dust shapes, the usual choice of bins, while
acceptable, can be improved. With more bins at low $\ell$ one can better
discriminate between the templates that peak in the squeezed
configuration (local and dust especially). We also discussed the
effects of the (small) breakdown of the weak non-Gaussianity
approximation that occurs when we add the full dust map to the CMB
simulations (i.e.\ the expected amount of dust in raw sky
observations). The main consequence is that the real error bars become
larger than the Fisher forecasts.

The testing and validation having been successful, we then used the
numerical galactic templates on the cleaned \texttt{SMICA} CMB map of the 2015 Planck
release. Fortunately, we did not detect any residual of the dust. The error bars for the dust and
local shapes increase in a joint analysis with the usual
binning, again because of the choice of bins that is not optimal to
differentiate them. Finally, we applied the foreground templates to
the raw sky map at 143~GHz and the binned bispectrum estimator
succeeded in detecting the dust in it at the expected level (the
intensities of the other foregrounds at 143 GHz being too small 
to detect even if they were present in the map).

The work presented in this paper can be extended in several ways.
The additional functionality built into the binned bispectrum estimator
code to use numerical bispectra as templates opens new possibilities, and 
allows us to include the template of any component of which a map exists in
our bispectrum and $f_\mathrm{NL}$ analyses.
It would also be interesting to further study the galactic bispectra,
or their non-Gaussianity in general, together with an expert on galactic
astrophysics, to see if they can be physically understood. This could maybe
lead to building an analytical template for these bispectra that can also be
used by other bispectrum estimator codes.
Finally, the analysis of this paper will obviously have to be repeated on
the final 2018 Planck data. The hope is that the improved treatment of the
polarization maps in that release will make an extension to E-polarization
of this analysis viable as well.

\vspace{0.5cm}

{\bf Acknowledgements:} The authors would like to thank Ingunn Kathrine Wehus for providing the \texttt{Commander} foreground maps used in this paper and Ata Karakci for providing Planck Sky Model maps that were used in the preliminary phase of the paper. GJ and BvT thank the Institute of Theoretical Astrophysics of the University of Oslo, Norway, for a very pleasant and fruitful visit in June 2017. The authors thank Hans Kristian Eriksen, Ingunn Kathrine Wehus, Ata Karakci and Guillaume Patanchon for useful discussions. The binned bispectrum estimator code used and extended in this paper was originally written by Bartjan van Tent, Martin Bucher and Benjamin Racine. We gratefully acknowledge IN2P3 Computer Center (\url{https://cc.in2p3.fr}) for providing the computing resources and services needed for the analysis, as well as the use of the HEALPix (\url{https://healpix.sourceforge.io}) and CAMB (\url{https://camb.info}) computer codes.


\appendix
\section{Derivation of the variance of the bispectrum and the linear correction}
\label{sec:variance-appendix}

In this appendix, we recall the derivation of the variance of the bispectrum to explain the role of the linear correction. By definition, the variance is given by
\begin{equation}
  \label{eq:bispectrum-variance}
  \mathrm{Var}(B^\mathrm{obs}_{\ell_1 \ell_2 \ell_3}) = \langle (B^\mathrm{obs}_{\ell_1 \ell_2 \ell_3})^2 \rangle - \langle B^\mathrm{obs}_{\ell_1 \ell_2 \ell_3} \rangle^2
  \equiv V_{\ell_1 \ell_2 \ell_3}.
\end{equation}
In the weak non-Gaussianity regime, the average value of the bispectrum is negligible. This leaves us with computing the mean value of the product of two bispectra.

\subsection*{Isotropic case}

\begin{equation}
  \label{eq:variance-first-term}
\langle B_{\ell_1 \ell_2 \ell_3}(\hat\Omega) B_{\ell_4 \ell_5 \ell_6}(\hat\Omega') \rangle = \int_{S^2 \times S^2}\dif \hat\Omega \dif \hat\Omega'\, \langle M_{\ell_1}(\hat\Omega) M_{\ell_2}(\hat\Omega) M_{\ell_3}(\hat\Omega) M_{\ell_4}(\hat\Omega') M_{\ell_5}(\hat\Omega') M_{\ell_6}(\hat\Omega') \rangle.
 \end{equation}
One can use Wick's theorem for Gaussian fields to reduce the six-point correlation function to the sum of fifteen products of two-point correlation functions:
\begin{itemize}
\item 6 terms: each $\ell$ is paired with an element of the other triplet like in \\$\langle M_{\ell_1}(\hat\Omega) M_{\ell_4}(\hat\Omega')\rangle \langle M_{\ell_2}(\hat\Omega) M_{\ell_5}(\hat\Omega')\rangle \langle M_{\ell_3}(\hat\Omega) M_{\ell_6}(\hat\Omega') \rangle$;
 \item 9 terms: the rest (example: $\langle M_{\ell_1}(\hat\Omega) M_{\ell_2}(\hat\Omega)\rangle \langle M_{\ell_3}(\hat\Omega) M_{\ell_4}(\hat\Omega')\rangle \langle M_{\ell_5}(\hat\Omega') M_{\ell_6}(\hat\Omega') \rangle$).
 \end{itemize}

 We will explicitly compute the contribution of the 6 first terms below. But first, we show that in the isotropic case, the 9 last terms are zero. For this, we use the addition theorem,
 \begin{equation}
  \label{eq:ylm-addition-theorem}
  \sum\limits_{m=-\ell}^{\ell}\, Y_{\ell m}(\hat{\Omega})Y_{\ell m}^*(\hat{\Omega}') = \frac{2\ell+1}{4\pi}P_\ell(\hat{\Omega}\cdot\hat{\Omega}'),
\end{equation}
where $P_\ell$ is a Legendre polynomial, and the fact that the maps are real to compute the two-point correlation function
 \begin{equation}
   \label{eq:cancellation-variance}
   \begin{split}
   \langle M_{\ell}(\hat\Omega) M_{\ell'}(\hat\Omega')\rangle =
   \sum\limits_{m=-\ell}^{\ell}\sum\limits_{m'=-\ell'}^{\ell'} \langle a_{\ell m} a_{\ell' m'}^*\rangle Y_{\ell m}(\hat\Omega) Y_{\ell' m'}^*(\hat\Omega') 
   =C_\ell \delta_{\ell \ell'} \frac{2\ell+1}{4\pi}P_\ell(\hat\Omega\cdot\hat\Omega').   
   \end{split}
 \end{equation}
 Then we can perform the integration of our example term (and the eight others follow the same computation):
 \begin{equation}
   \label{eq:cancellation-variance-2}
   \begin{split}
     &\int_{S^2 \times S^2}\dif \hat\Omega \dif \hat\Omega'\,\langle M_{\ell_1}(\hat\Omega) M_{\ell_2}(\hat\Omega)\rangle \langle M_{\ell_3}(\hat\Omega) M_{\ell_4}(\hat\Omega')\rangle \langle M_{\ell_5}(\hat\Omega') M_{\ell_6}(\hat\Omega') \rangle\\
     &= \frac{(2\ell_1+1)(2\ell_3+1)(2\ell_5+1)}{(4\pi)^3} C_{\ell_1}C_{\ell_3}C_{\ell_5} \delta_{\ell_1 \ell_2}\delta_{\ell_3 \ell_4}\delta_{\ell_5 \ell_6}
     \int_{S^2 \times S^2}\dif \hat\Omega \dif \hat\Omega'\,
     P_{\ell_1}(1) P_{\ell_3}(\hat\Omega\cdot\hat\Omega') P_{\ell_5}(1)
   \end{split}
 \end{equation}
This integral can be solved using well-known properties of Legendre polynomials. First, we have $P_\ell(1) = 1$ and then we can use the integral
 \begin{equation}
   \label{eq:legendre-integral}
     \int_{S^2 \times S^2}\dif \hat\Omega \dif \hat\Omega'\,  P_\ell(\hat\Omega\cdot\hat\Omega')=0,
   \end{equation}
and we find the announced result that these terms vanish.   

Concerning the six first terms, we will also explicitly compute only the given example, but the correct permutations to obtain the five other terms will be in the final result. Substituting \eqref{eq:map} into the integral and using the fact that the maps are real, one obtains
 \begin{equation}
   \label{eq:remaining-variance}
   \begin{split}
     \int_{S^2 \times S^2}\dif \hat\Omega &\dif \hat\Omega'\,\langle M_{\ell_1}(\hat\Omega) M_{\ell_4}(\hat\Omega')\rangle \langle M_{\ell_2}(\hat\Omega) M_{\ell_5}(\hat\Omega')\rangle \langle M_{\ell_3}(\hat\Omega) M_{\ell_6}(\hat\Omega') \rangle\\
     &=
     C_{\ell_1}C_{\ell_2}C_{\ell_3} \delta_{\ell_1 \ell_4}\delta_{\ell_2 \ell_5}\delta_{\ell_3 \ell_6} \sum\limits_{m_1, m_2, m_3}\lh\int_{S^2}\dif \hat\Omega  Y_{\ell_1 m_1}(\hat\Omega)\, Y_{\ell_2 m_2}(\hat\Omega) Y_{\ell_3 m_3}(\hat\Omega)\rh\\ &\quad\times \lh\int_{S^2}\dif\hat\Omega'\, Y_{\ell_1 m_1}^*(\hat\Omega') Y_{\ell_2 m_2}^*(\hat\Omega') Y_{\ell_3 m_3}^*(\hat\Omega') \rh.     
   \end{split}
 \end{equation}
where one can recognize the Gaunt integral \eqref{eq:gaunt-integral}. Substituting it here and using the identity relation \begin{equation}
  \label{eq:3-j-identity}
  \sum\limits_{m_1 m_2 m_3}\begin{pmatrix}
     \ell_1 &  \ell_2 &  \ell_3\\
    m_1 & m_2 & m_3
  \end{pmatrix}^2=1,
\end{equation}
and the fact that the columns of Wigner 3$j$-symbols can be permuted when the parity condition is respected, one can find that the 6 terms give
 \begin{equation}
   \label{eq:variance-iso-1}
   \begin{split}
   \langle B_{\ell_1 \ell_2 \ell_3} B_{\ell_4 \ell_5 \ell_6} \rangle = h_{\ell_1 \ell_2 \ell_3}^2 C_{\ell_1}C_{\ell_2}C_{\ell_3} \biggl[&\delta_{\ell_1 \ell_4}\delta_{\ell_2 \ell_5}\delta_{\ell_3 \ell_6}+ \delta_{\ell_1 \ell_4}\delta_{\ell_2 \ell_6}\delta_{\ell_3 \ell_5} + \delta_{\ell_1 \ell_5}\delta_{\ell_2 \ell_4}\delta_{\ell_3 \ell_6}\\
   &+ \delta_{\ell_1 \ell_5}\delta_{\ell_2 \ell_6}\delta_{\ell_3 \ell_4}
   + \delta_{\ell_1 \ell_6}\delta_{\ell_2 \ell_4}\delta_{\ell_3 \ell_5}
   + \delta_{\ell_1 \ell_6}\delta_{\ell_2 \ell_5}\delta_{\ell_3 \ell_4}\biggr].
   \end{split}
 \end{equation}
 Hence the variance is
 \begin{equation}
   \label{eq:bispectrum-variance-iso}
   V_{\ell_1 \ell_2 \ell_3} = g_{\ell_1 \ell_2 \ell_3} h_{\ell_1 \ell_2 \ell_3}^2 C_{\ell_1}C_{\ell_2}C_{\ell_3}.
 \end{equation}

\subsection*{Anisotropic case}

As explained in section \ref{sec:variance}, with observational data from an actual experiment we cannot use the isotropy assumption. This would lead to a large increase of the variance \eqref{eq:bispectrum-variance-iso}, because the nine terms described just before are no longer zero. However, it is possible to show that adding the simple linear correction given in \eqref{eq:linear-correction} to the cubic term of the angle-averaged bispectrum solves this issue. To verify this, we will derive the variance similarly to the previous section, the main difference being that integrations are performed on $S^2 \setminus \mathcal{M}$ instead of $S^2$. Again, in the weak non-Gaussianity regime, we only have to compute the average of the product of two bispectra $\langle B_{\ell_1 \ell_2 \ell_3}^{\mathrm{obs}}(\hat\Omega) B_{\ell_4 \ell_5 \ell_6}^{\mathrm{obs}}(\hat\Omega')\rangle$ and there are three types of terms:
\begin{itemize}
  \item 1 term: product of the two cubic terms: $\langle M_{\ell_1} M_{\ell_2} M_{\ell_3} M_{\ell_4} M_{\ell_5} M_{\ell_6} \rangle$ (this is the only term present in the isotropic case);
  \item 6 terms: product of a linear term with a cubic term, e.g.\ $\langle - M_{\ell_1} M_{\ell_2} M_{\ell_3} M_{\ell_4}  \rangle\langle M_{\ell_5} M_{\ell_6}\rangle$;
  \item 9 terms: product of two linear terms, e.g.\ $\langle M_{\ell_1} M_{\ell_4} \rangle\langle M_{\ell_2} M_{\ell_3}\rangle\langle M_{\ell_5} M_{\ell_6}\rangle$.
\end{itemize}
We have seen how the first term gives 15 contributions if we use Wick's theorem to transform the six-point correlation function into a combination of products of three two-point correlation functions. We have also seen that in the isotropic case, only the six terms where each multipole among $(\ell_1,~\ell_2,~\ell_3)$ is coupled with an element of the other triplet $(\ell_4,~\ell_5,~\ell_6)$ are non-zero. The same can be done for the four-point correlation function and each combination of a linear with a cubic term will give three terms, hence a total of eighteen terms. Note also that each term derived from the linear correction contains necessarily two $\ell$'s of the same triplet that are coupled (it is in the definition of the linear term). Hence, it means that they cannot cancel the six terms of the isotropic case. The new terms (i.e.\ the terms that are not present in the isotropic case) are nine from the six-point correlation function, nine from the product of two linear terms, and eighteen from terms with the four-point correlation functions (with a minus sign) and it is then easy to check that they exactly cancel each other. So finally
\begin{equation}
  \begin{split}
  \langle B^{\mathrm{obs}}_{\ell_1 \ell_2 \ell_3} B^{\mathrm{obs}}_{\ell_4 \ell_5 \ell_6} \rangle =
  &\int \dif \hat\Omega \dif \hat\Omega'\, \left[ \langle  M_{\ell_1} M_{\ell_4}\rangle\langle  M_{\ell_2} M_{\ell_5}  \rangle\langle M_{\ell_3} M_{\ell_6}\rangle + (14)(26)(35) \right. \\
  &\qquad\left.+ (15)(24)(36) + (15)(26)(34) + (16)(24)(35) + (16)(25)(34)\right],
  \end{split}
\end{equation}
where we use an obvious shorthand notation to indicate the other permutations of filtered maps. It is important to note that we recover the same variance as in the isotropic case without a linear term (except for the integration interval). This proves that the estimator with this linear correction is optimal, with the assumption that $C_{\ell m, \ell' m'}$ is diagonal (thus equation \eqref{eq:cancellation-variance} is valid here). However, the integration interval for the two integrals is $S^2 \setminus \mathcal{M}$. In the
$f_\mathrm{sky}$ approximation, which many tests have shown to be a good approximation, we calculate the
integrals as if the interval were the full sky $S^2$, and then add appropriate factors of $f_\mathrm{sky}$
at the end to compensate for the partial sky.
Then, performing the same last steps of the calculation as before, the variance is given by
\begin{equation}
  \label{eq:variance-appendix}
  V_{\ell_1 \ell_2 \ell_3} =
  g_{\ell_1 \ell_2 \ell_3} \frac{h_{\ell_1 \ell_2 \ell_3}^2}{f_\mathrm{sky}}
  (b_{\ell_1}^2 C_{\ell_1} + N_{\ell_1}) (b_{\ell_2}^2 C_{\ell_2} + N_{\ell_2}) (b_{\ell_3}^2 C_{\ell_3} + N_{\ell_3}),
\end{equation}
when including the effect of the beam and the noise and has a form similar to the isotropic case
\eqref{eq:bispectrum-variance-iso}. The factor of $1/f_\mathrm{sky}$ can easily be understood given that
the variance of a quantity determined from $N$ data points scales as $1/N$ and here the number of data
points roughly corresponds to the number of observed pixels on the sky.

\section{Weights of bispectral shapes}
\label{ap:weights-bisp-shap}

In this appendix, we give another representation of the different bispectra discussed in this paper (primordial shapes, ISW-lensing, extra-galactic and galactic foregrounds) well-suited to understand the correlation coefficients given in e.g.~tables~\ref{tab:corr_coeff_2015}, \ref{tab:corr_coeff_dust} and \ref{tab:corr_coeff_others}.

\begin{figure}
  \centering 
  \includegraphics[width=0.48\linewidth]{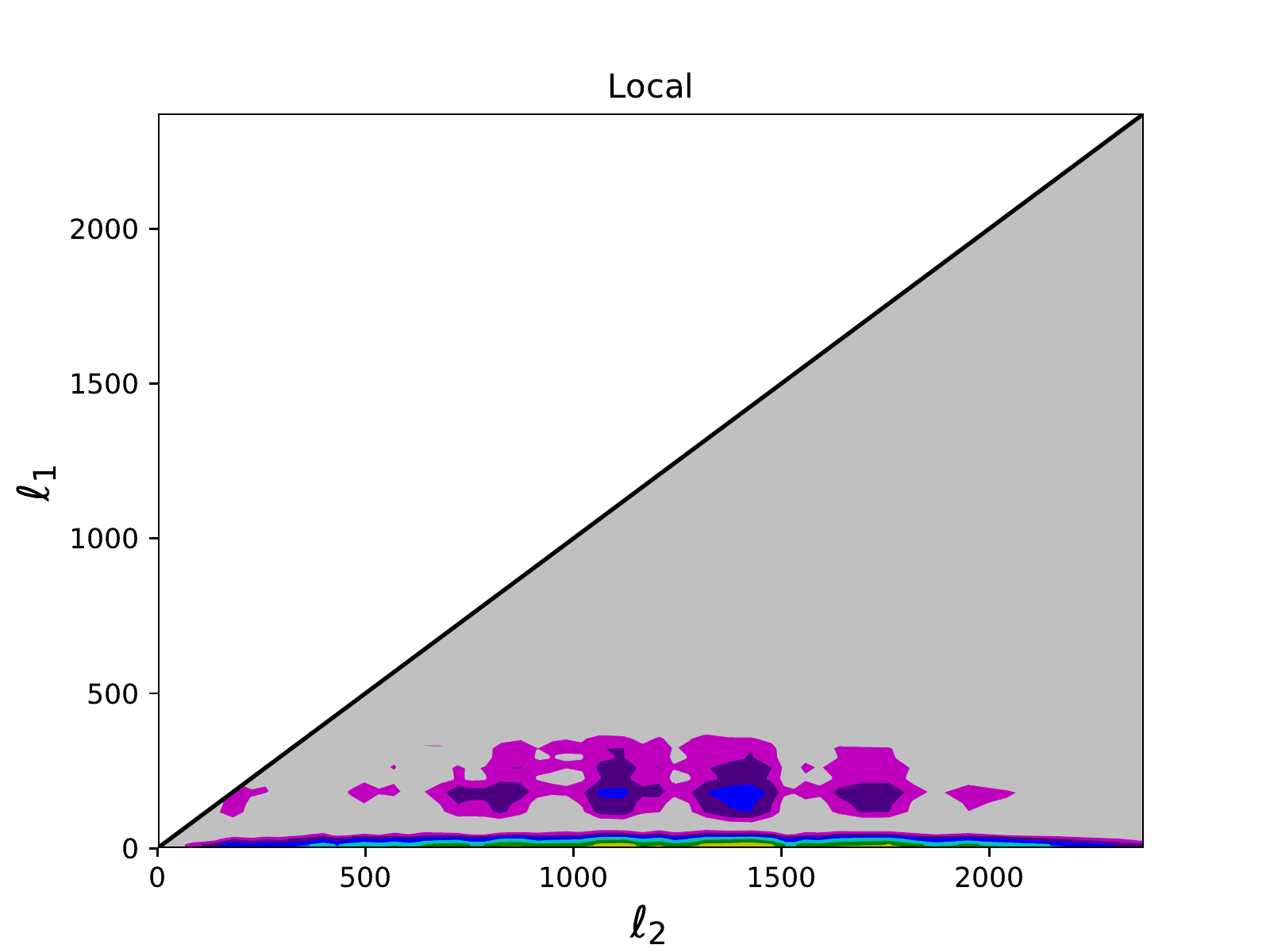} 
  \includegraphics[width=0.48\linewidth]{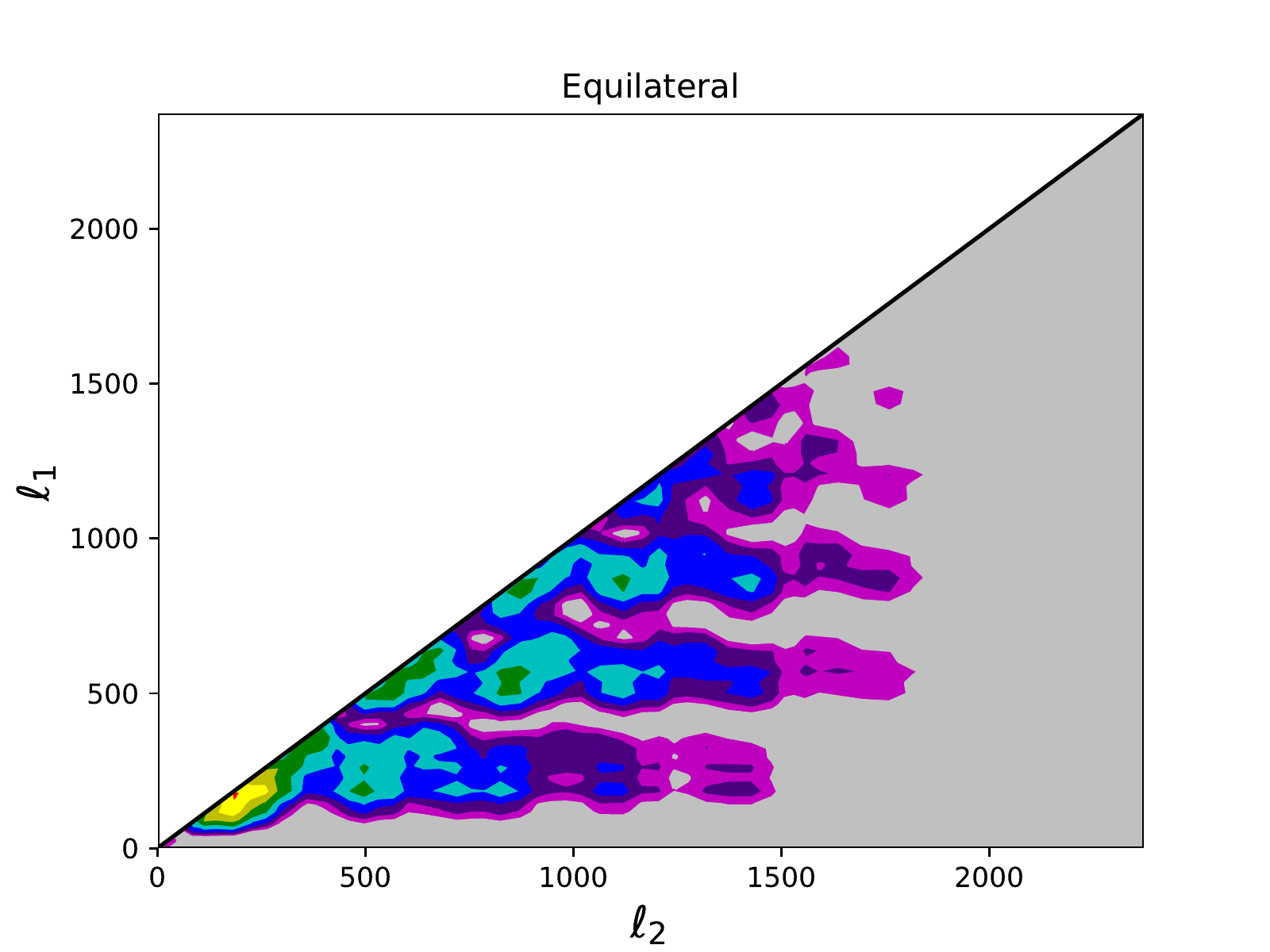}
  \includegraphics[width=0.48\linewidth]{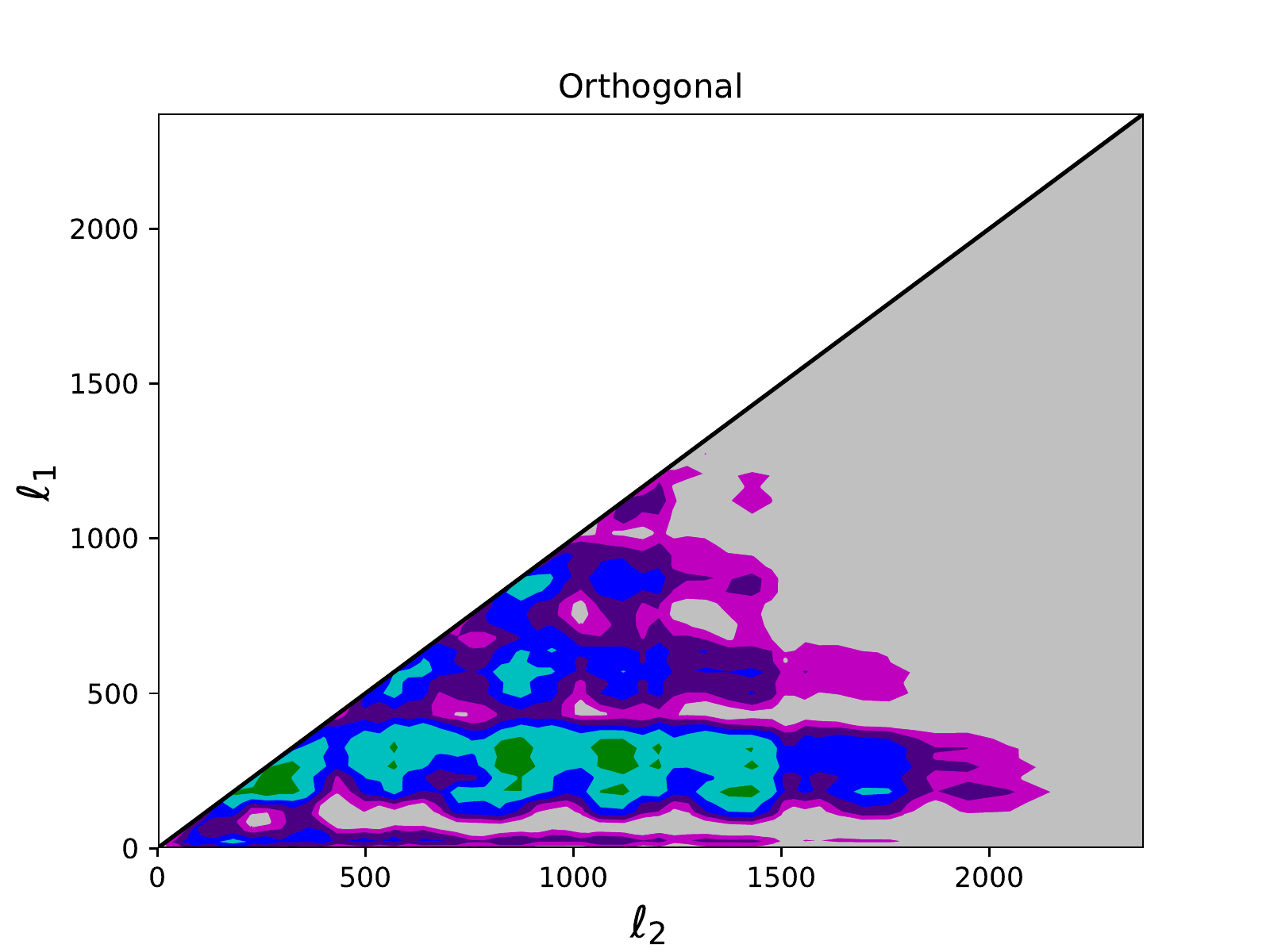}
  \includegraphics[width=0.48\linewidth]{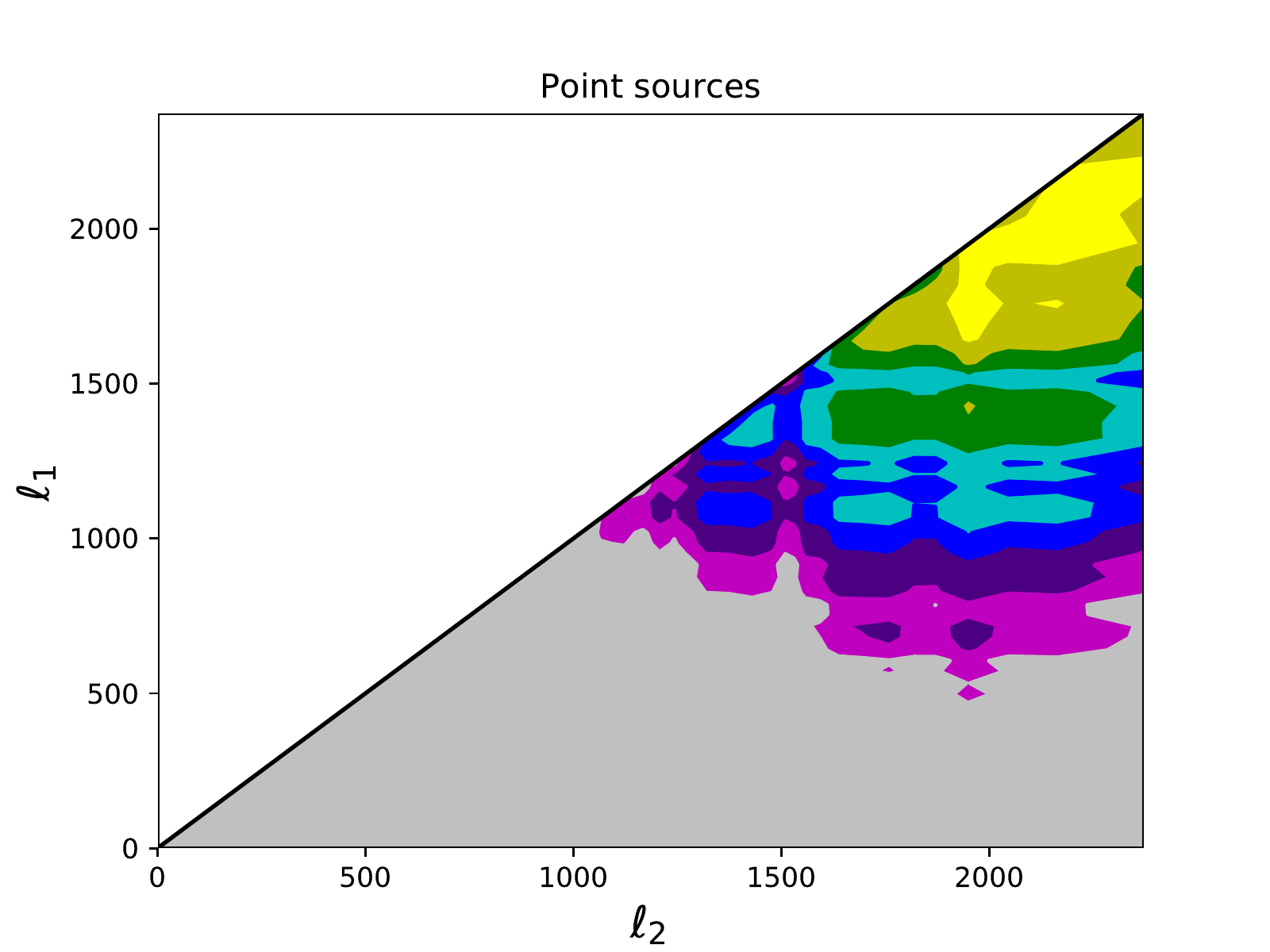}
  \includegraphics[width=0.48\linewidth]{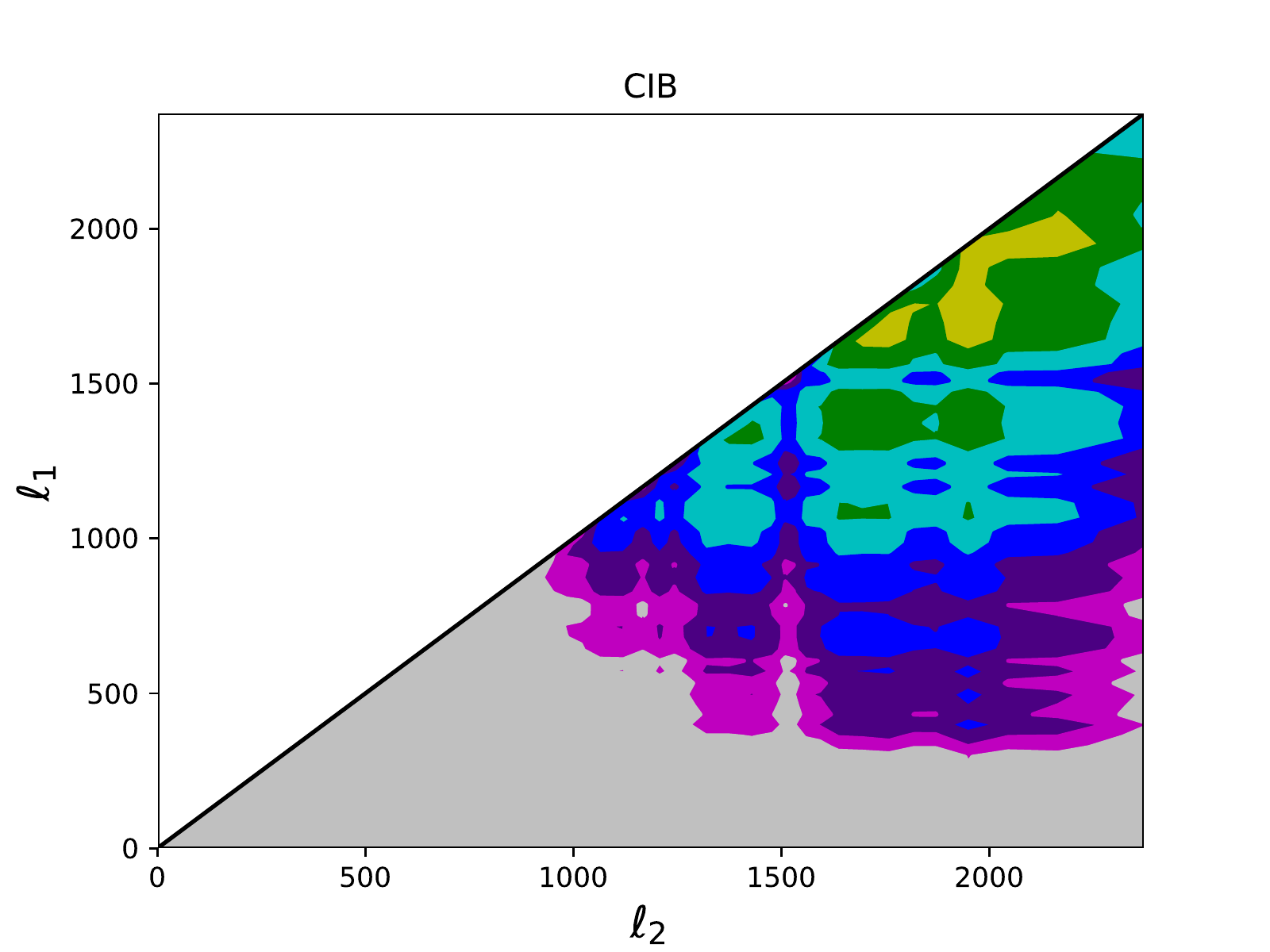} 
  \includegraphics[width=0.48\linewidth]{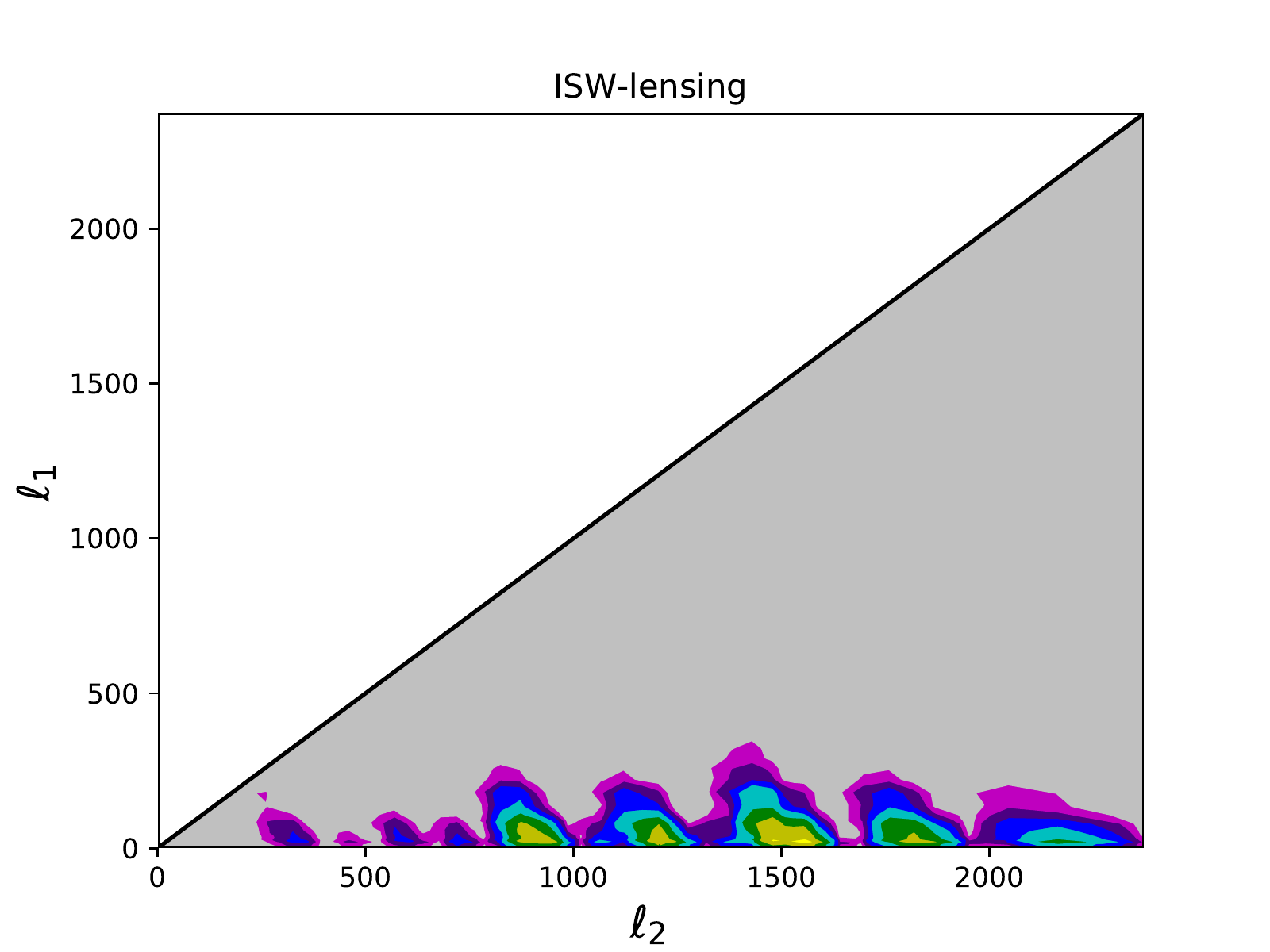}
  \includegraphics[align=c,width=0.48\linewidth]{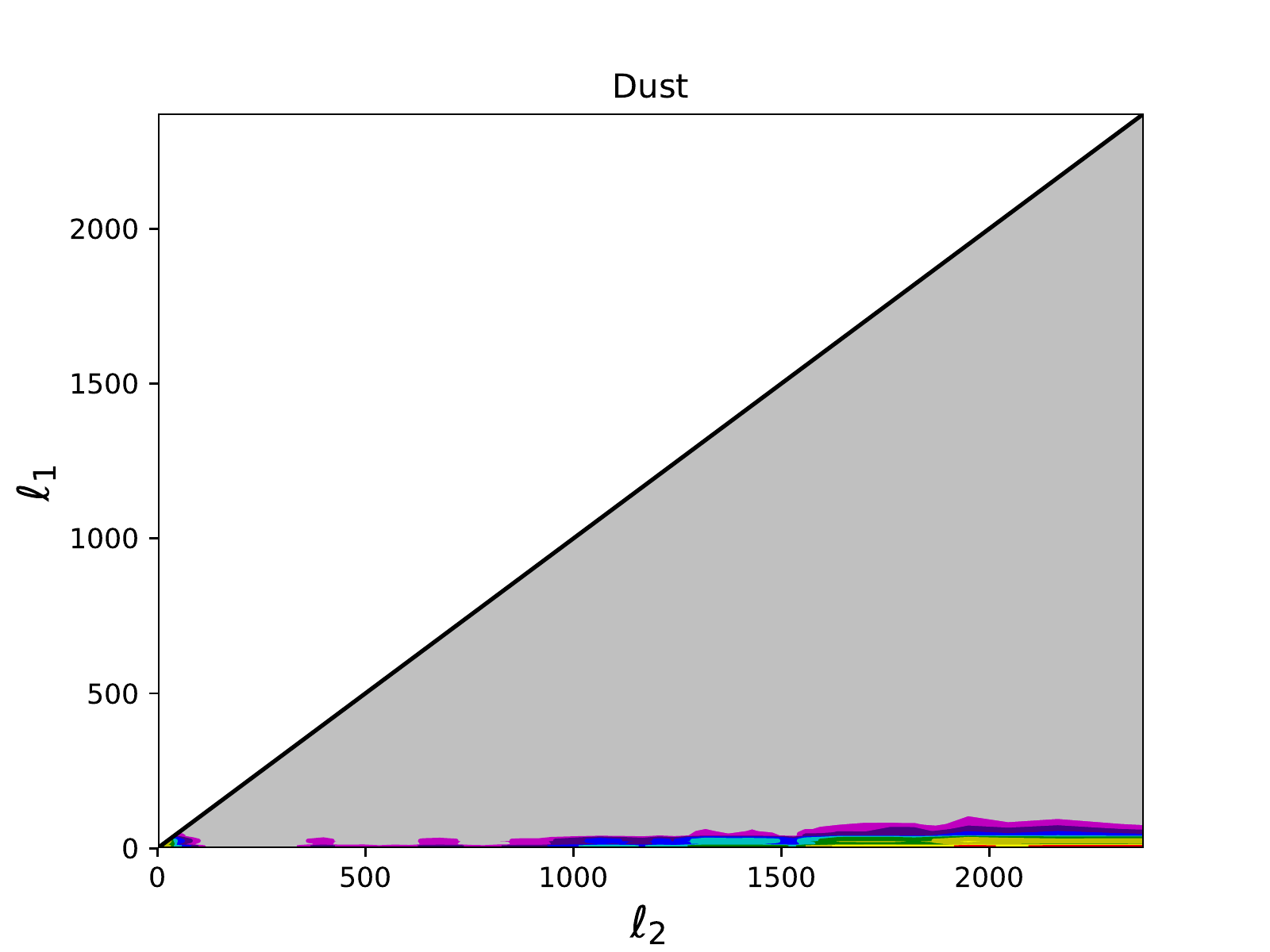}    
  \includegraphics[align=c,width=0.50\linewidth]{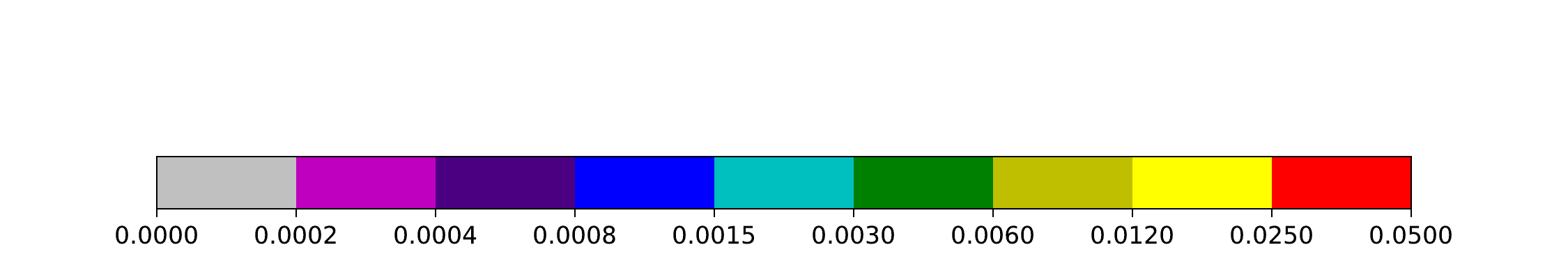}
  \caption{Weights of the bispectral shapes discussed in this paper at high resolution. Note that the colour scale is logarithmic.}
  \label{fig:weights-highres}
\end{figure}

\begin{figure}
  \centering    
  \includegraphics[width=0.48\linewidth]{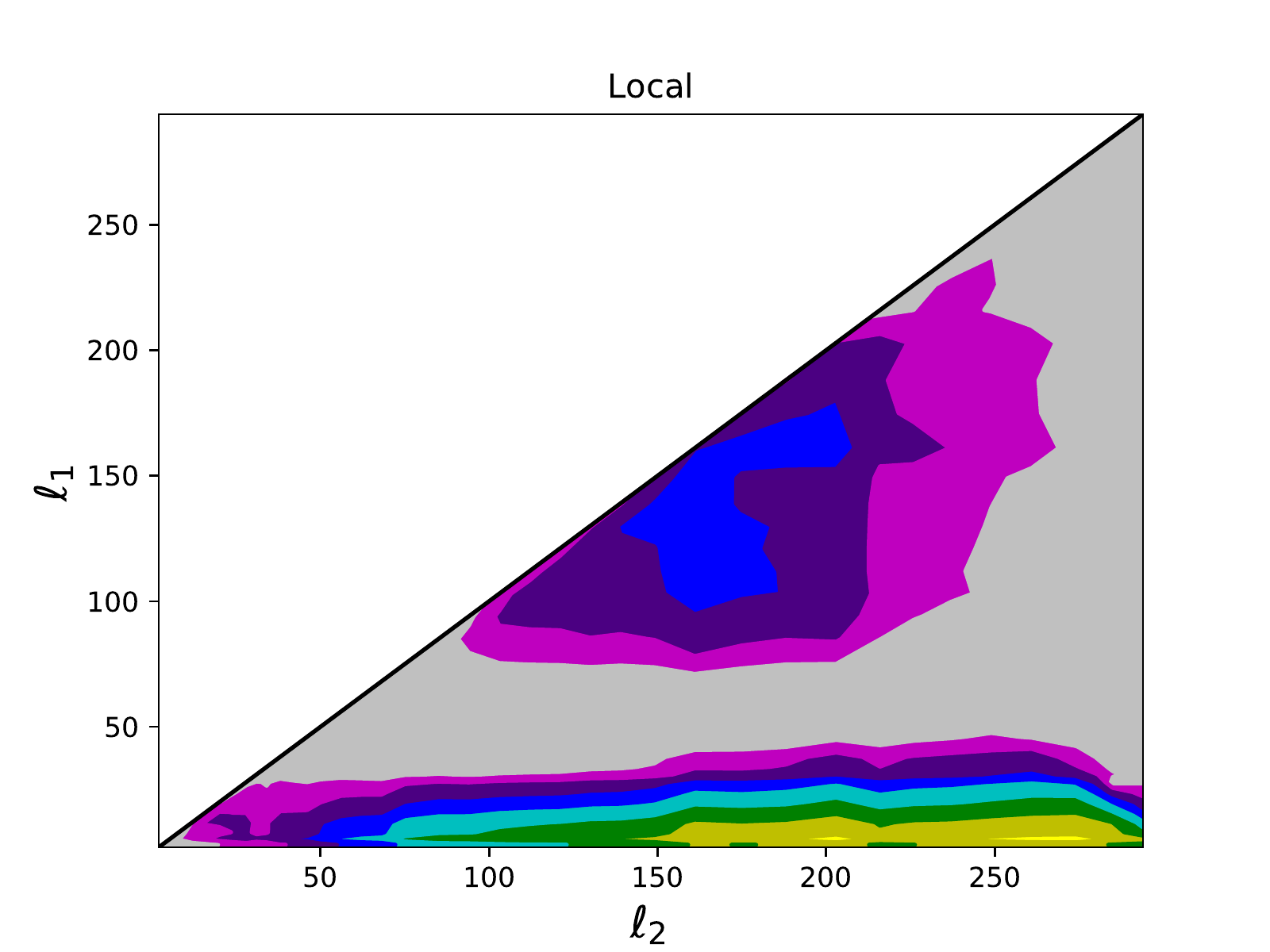}
  \includegraphics[width=0.48\linewidth]{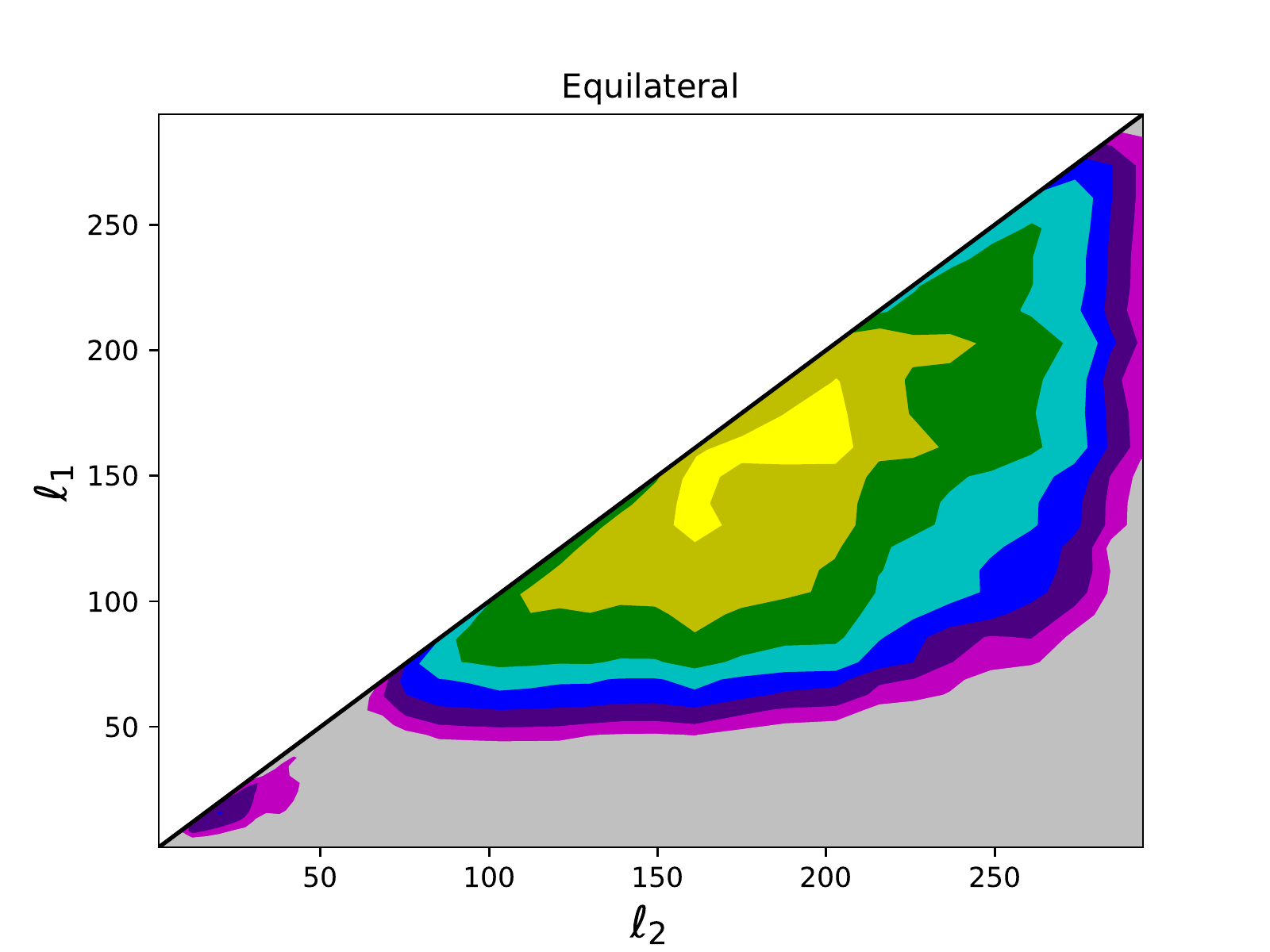}
  \includegraphics[width=0.48\linewidth]{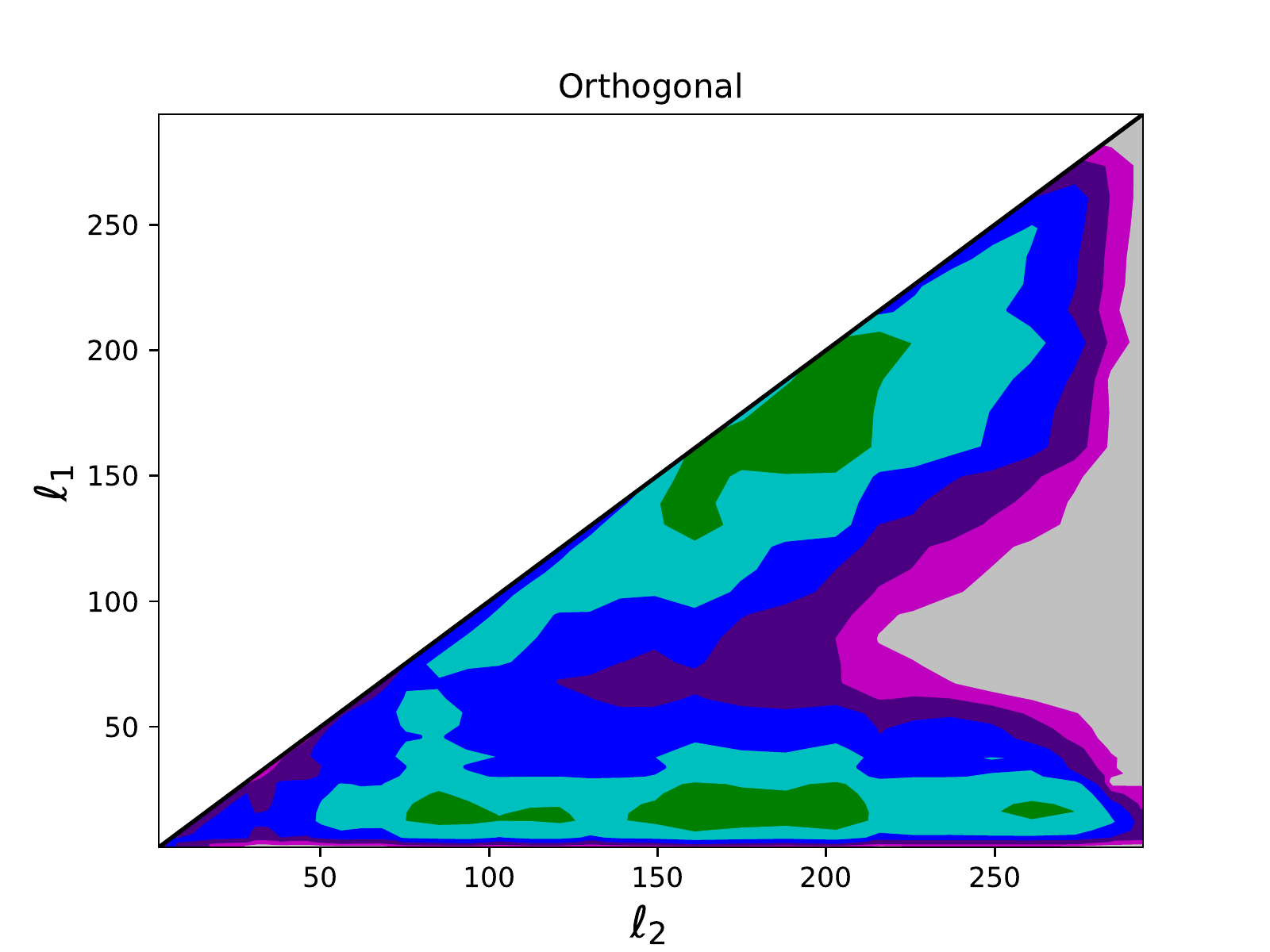} 
  \includegraphics[width=0.48\linewidth]{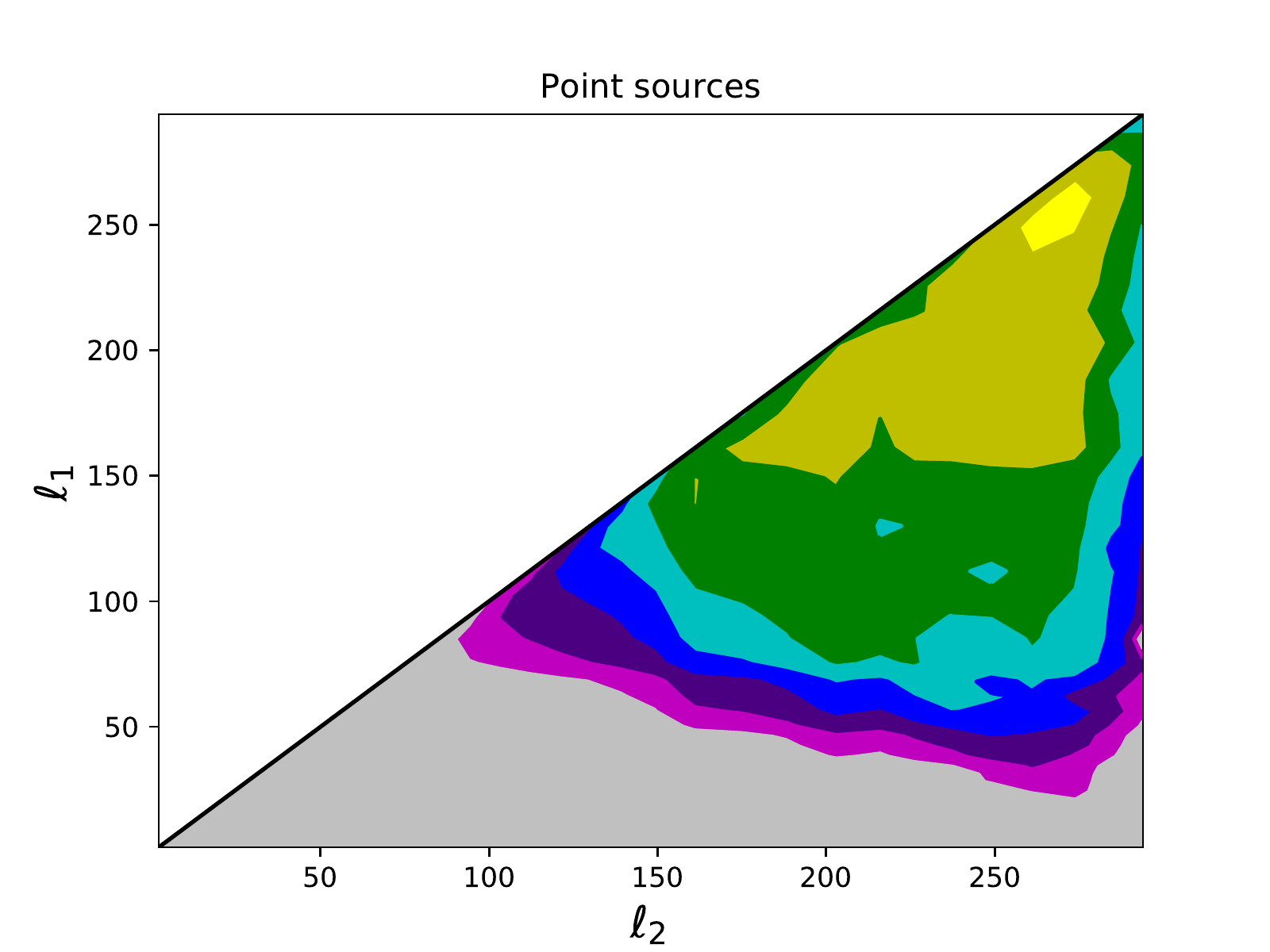} 
  \includegraphics[width=0.48\linewidth]{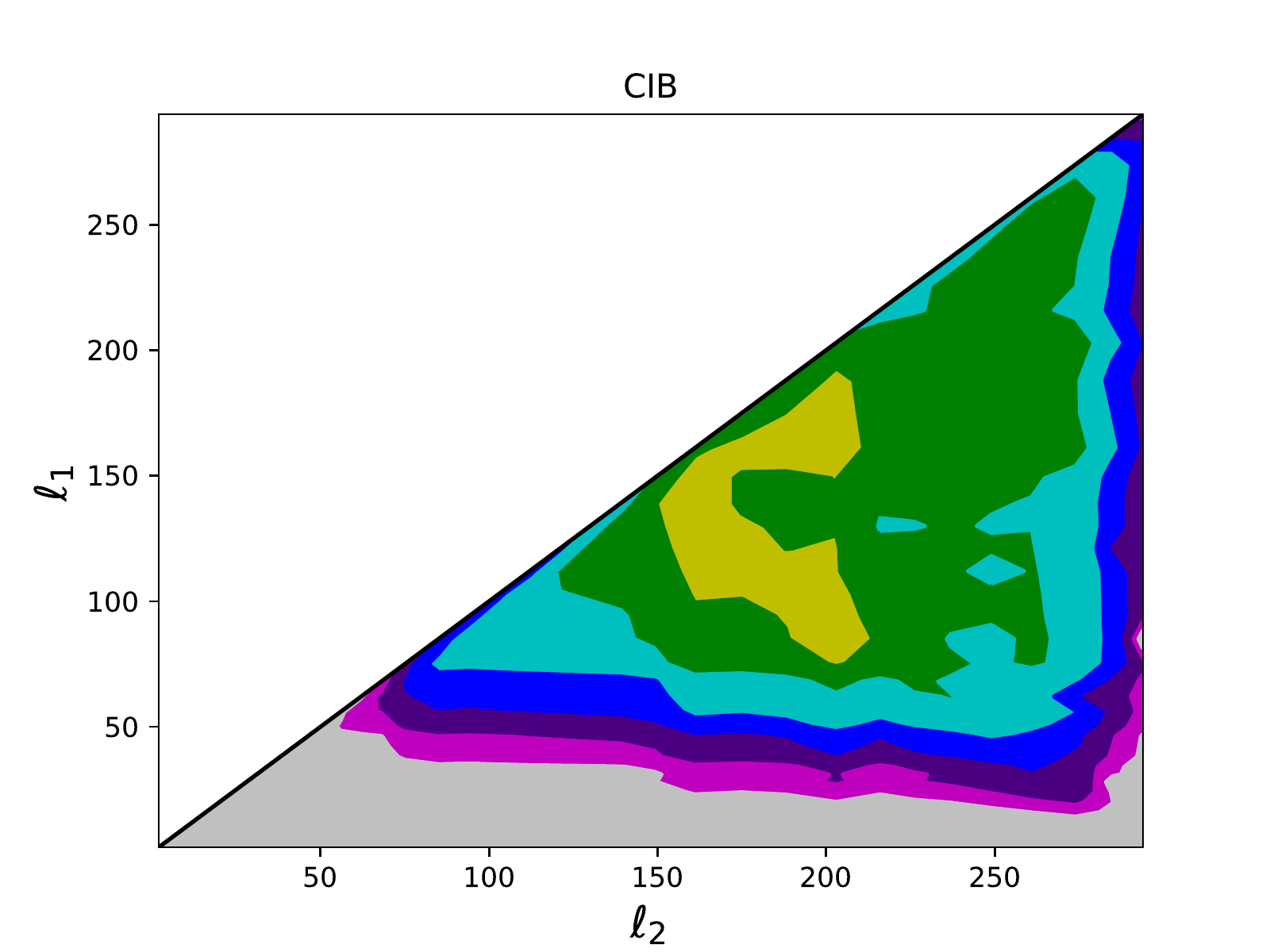} 
  \includegraphics[width=0.48\linewidth]{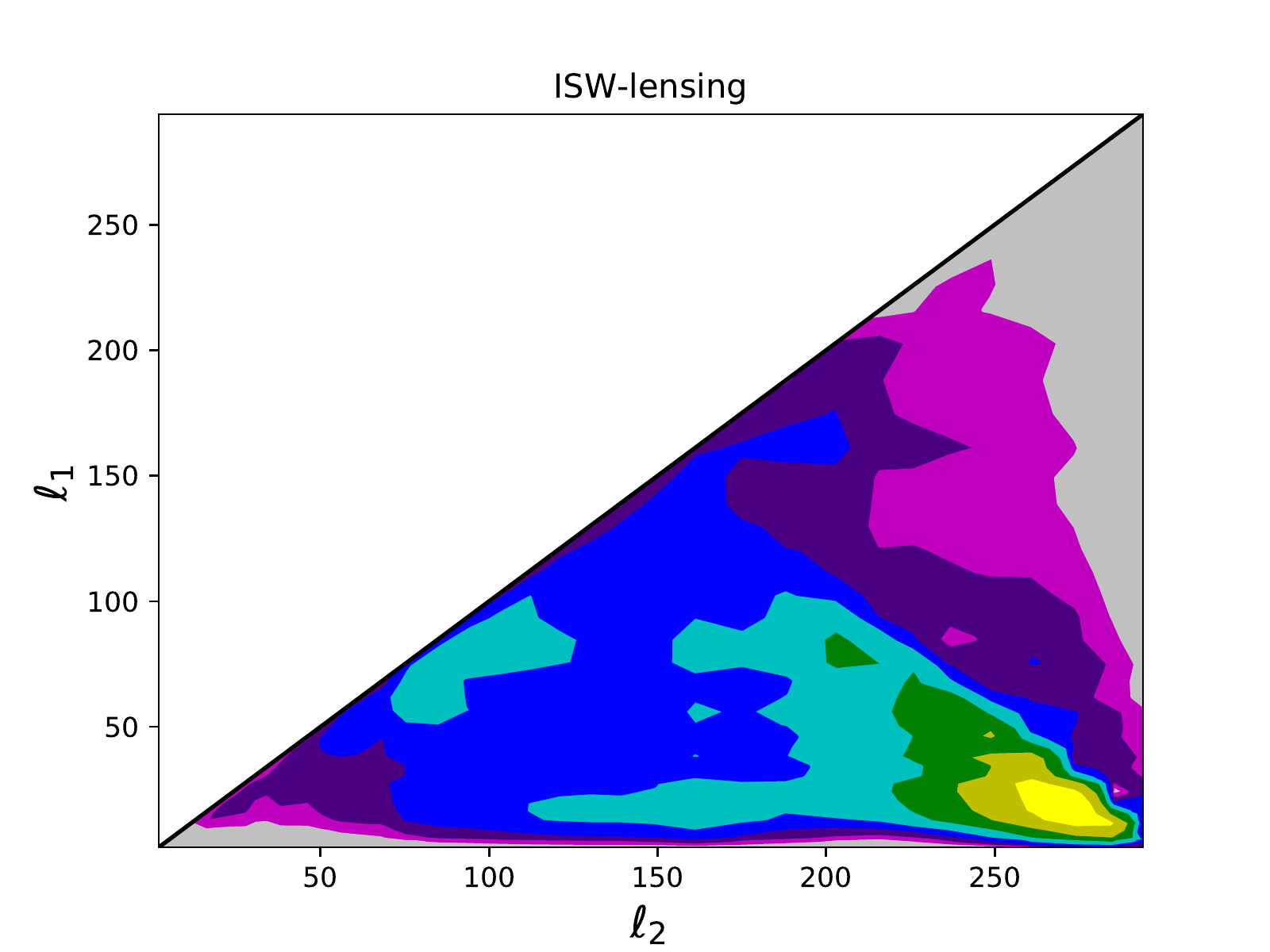}
  \includegraphics[align=c, width=0.49\linewidth]{plots/appendices/colorbar.pdf}
  \caption{Weights of the standard bispectral shapes at low resolution. Note the difference on the axes with the previous figure. The colour scale is the same, but the weights are normalized to one here over a much reduced region of multipole space with $\ell<300$.}
  \label{fig:weights-lowres_1}
\end{figure}

\begin{figure}
  \centering
  \includegraphics[width=0.48\linewidth]{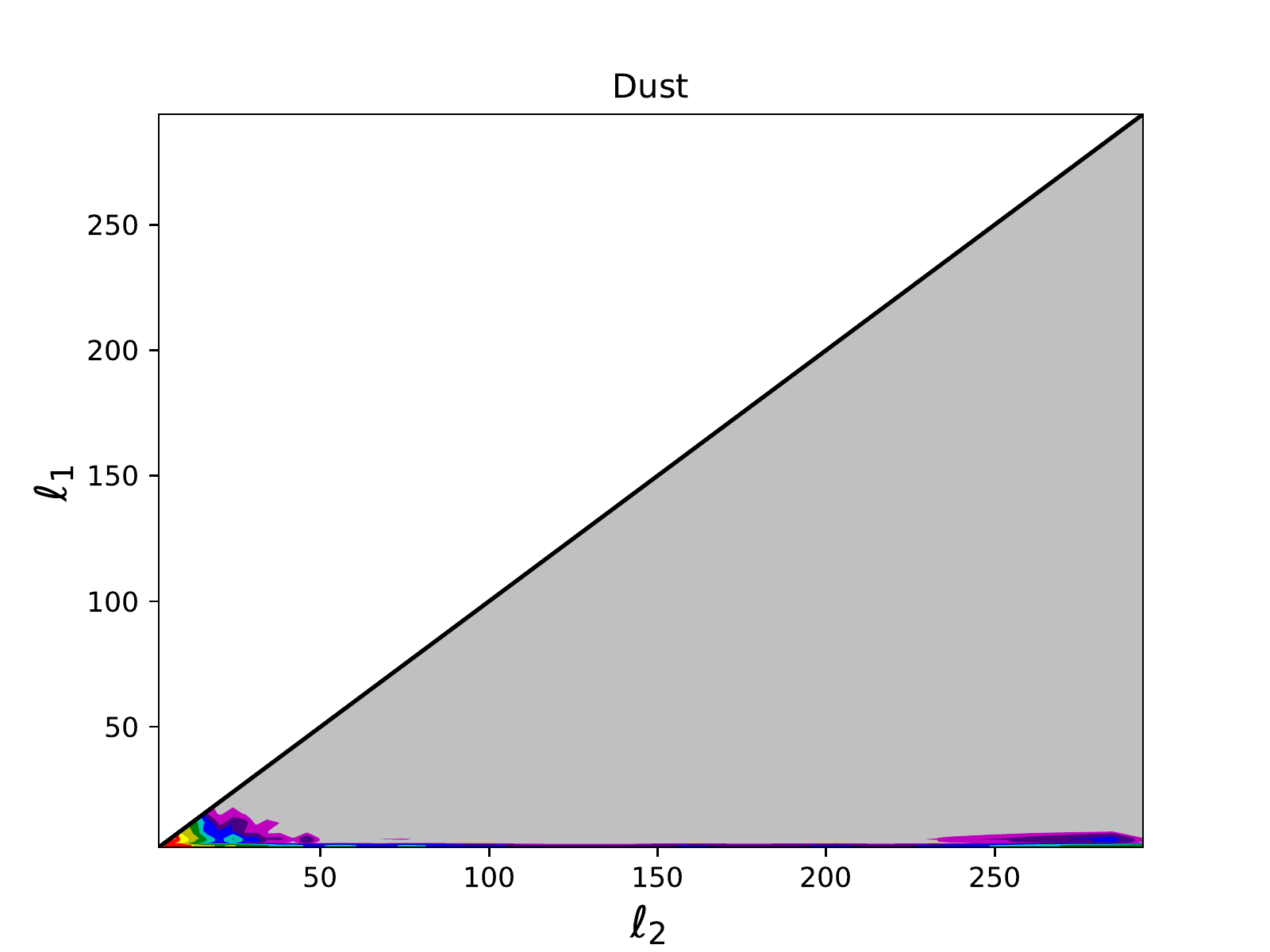}
  \includegraphics[width=0.48\linewidth]{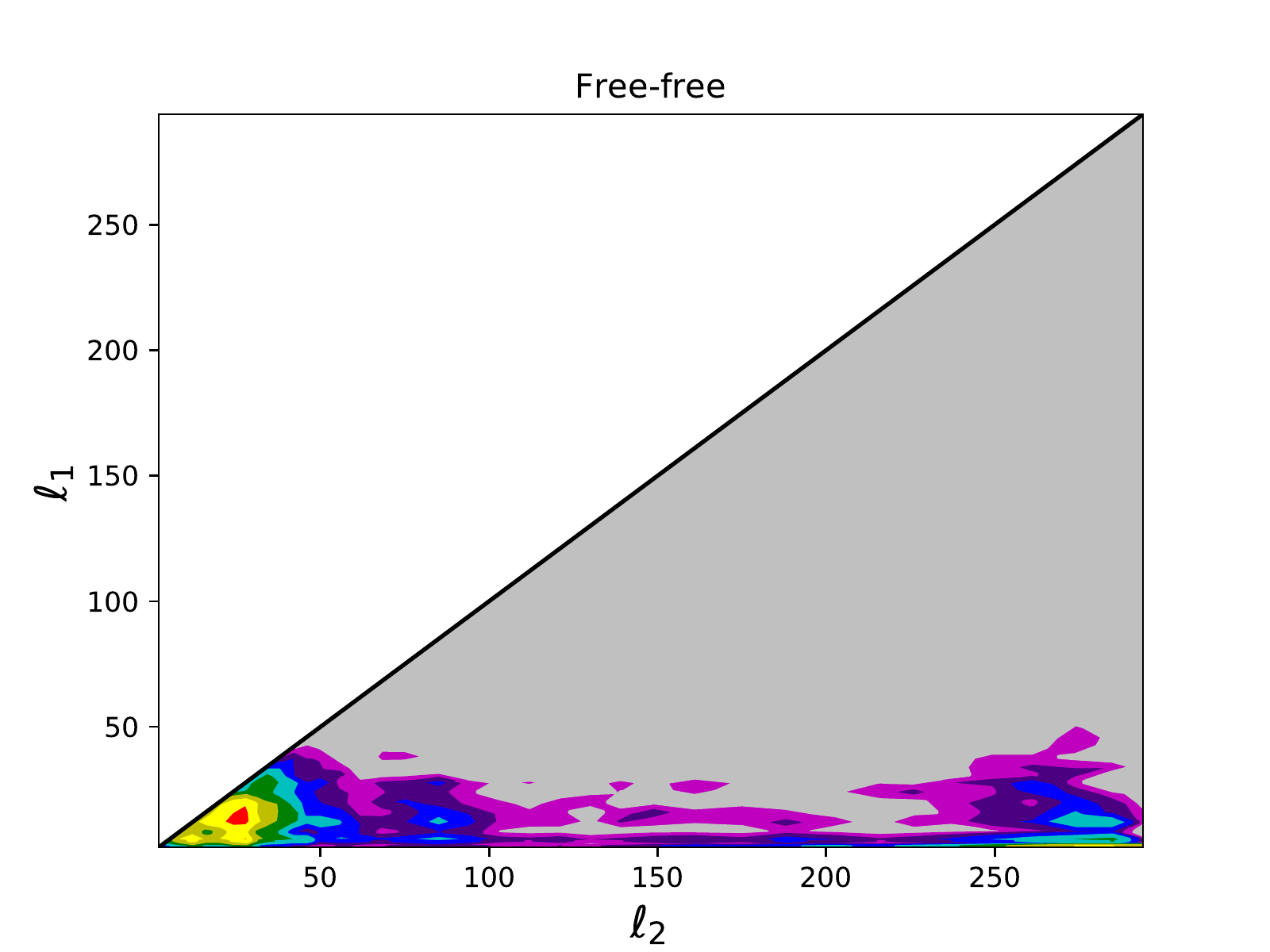}
  \includegraphics[width=0.49\linewidth]{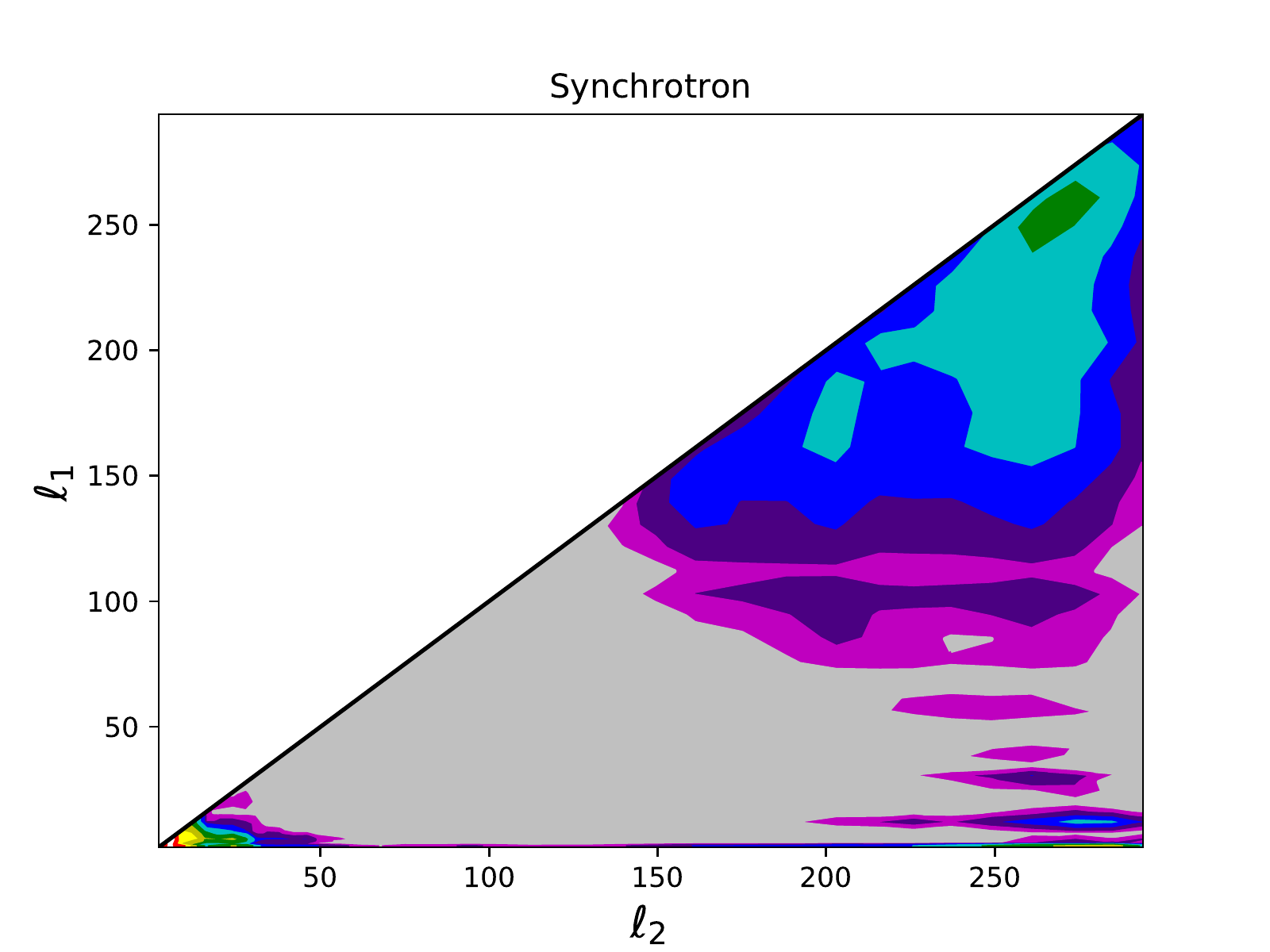}
  \includegraphics[width=0.49\linewidth]{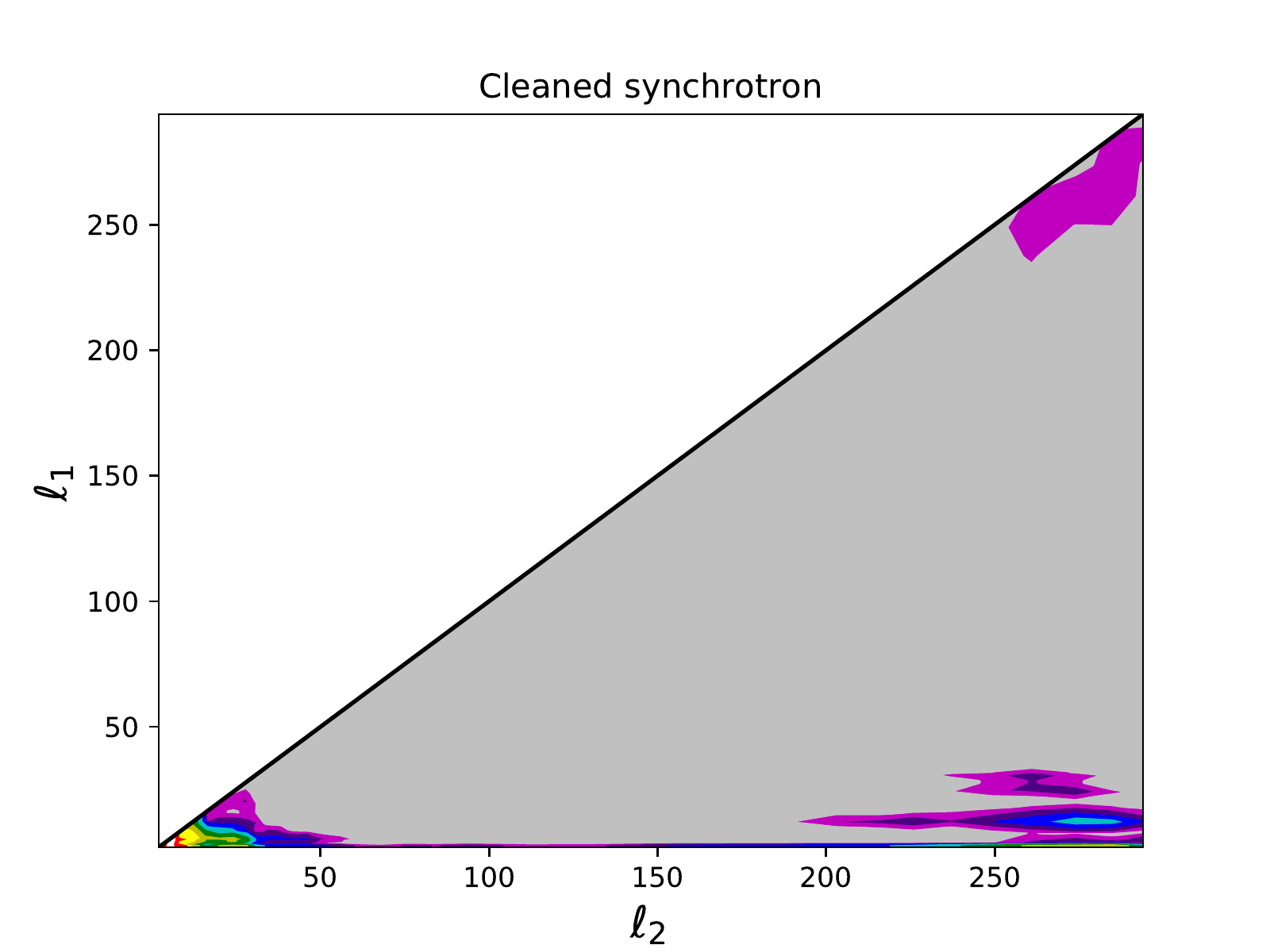}
  \includegraphics[align=c, width=0.49\linewidth]{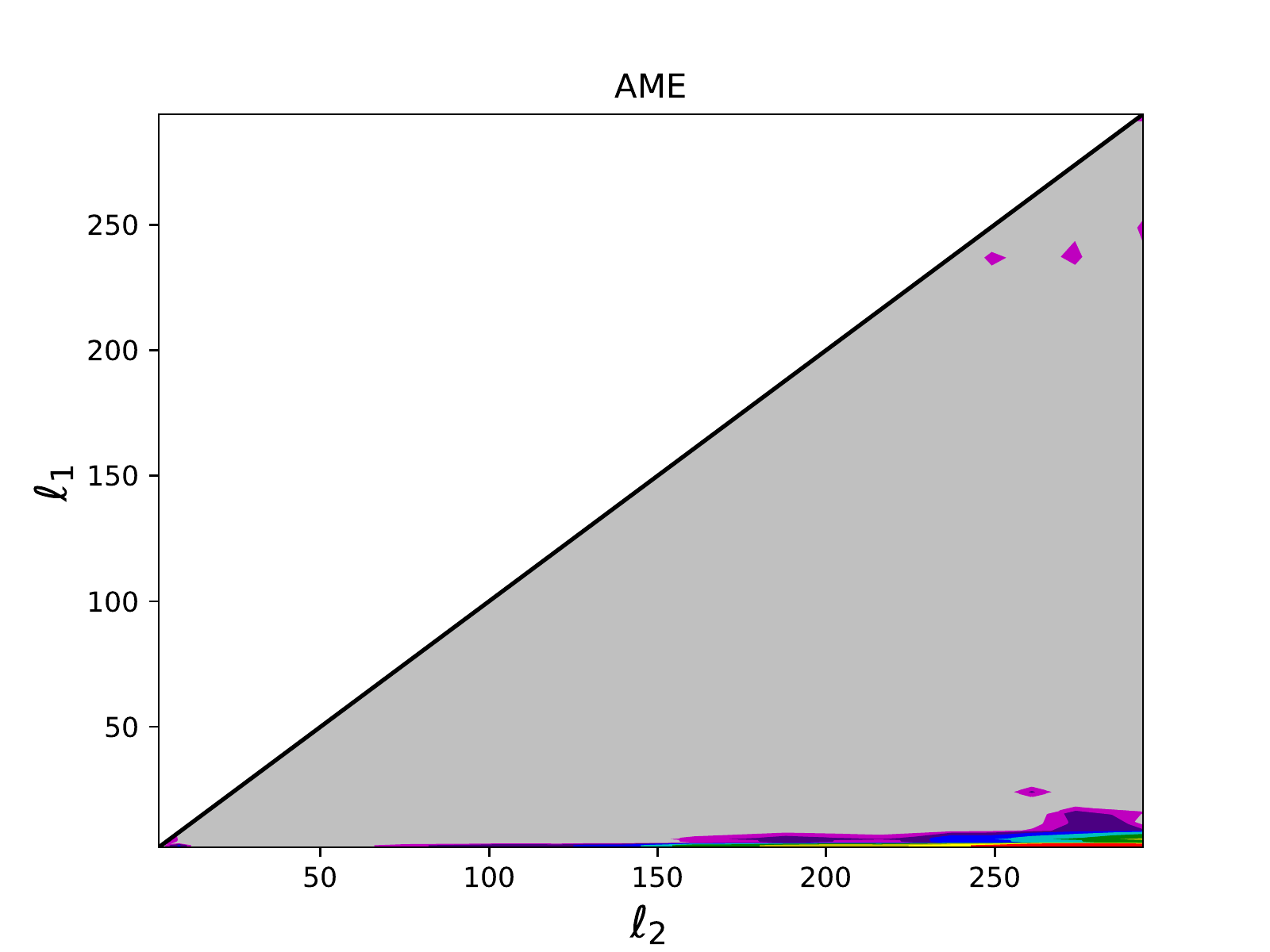}
  \includegraphics[align=c, width=0.49\linewidth]{plots/appendices/colorbar.pdf}
  \caption{Weights of the foreground bispectral shapes discussed in this paper at low resolution.}
  \label{fig:weights-lowres}
\end{figure}

The weight of a single multipole configuration ($\ell_1, \ell_2, \ell_3$) of a bispectral shape $B_{\ell_1 \ell_2 \ell_3}$ is defined by \cite{Bucher:2009nm}
\begin{equation}
  \label{eq:weight}
  w_{\ell_1 \ell_2 \ell_3}= \frac{1}{\langle B, B\rangle}\frac{(B_{\ell_1 \ell_2 \ell_3})^2}{V_{\ell_1 \ell_2 \ell_3}}.
\end{equation}
It is the inverse of the variance of the ratio of the observed and theoretical bispectra divided by $\langle B, B\rangle$ which is the denominator of the estimator for $\fnl$ and normalizes the sum of the weights to one. In other words,
\begin{equation}
  \label{eq:estimator-weight}
  \hat{f}_\mathrm{NL} = \sum\limits_{\ell_1 \ell_2 \ell_3}w_{\ell_1 \ell_2 \ell_3} \frac{B_{\ell_1 \ell_2 \ell_3}^{\mathrm{obs}}}{B_{\ell_1 \ell_2 \ell_3}},
\end{equation}
where $B^\mathrm{obs}/B$ can be viewed as an $\fnl$ estimator based on just a single $\ell$-triplet. These equations are the same for bin-triplets ($i_1, i_2, i_3$). Figures~\ref{fig:weights-highres}, \ref{fig:weights-lowres_1}, and~\ref{fig:weights-lowres} show the weights of the different theoretical and numerical shapes discussed in this paper at both high and low resolution, with the usual choice of 57 bins. Instead of using a few slices of $\ell_3$ like in section \ref{sec:foregrounds}, we summed over $\ell_3$. It has the advantage that now the whole bispectrum is used in one figure, but of course we lose the information about the variation of the bispectrum as a function of $\ell_3$. A larger weight means that the region of multipole space is more important for the template. Conversely, a large observed non-Gaussianity in that region of multipole space means that it is more likely to be that particular shape.

In this kind of plot, shapes that peak in squeezed configurations will have a colored band/line at the bottom of the figure (low $\ell_1$). As expected it is present for the different foregrounds for very low $\ell_1$ ($< \mathcal{O}(20)$), including synchrotron (which was not visible in figure~\ref{fig:other-templates}). As expected, the characteristic line of a squeezed bispectrum can be seen for the local and the ISW-lensing shapes, but also for the orthogonal shape (which explains why it is somewhat correlated to the foregrounds).

Shapes that peak in equilateral configurations have a large weight along the diagonal black line of these plots, when the three $\ell$'s are of the same order. The primordial equilateral shape is the strongest for three low $\ell$'s, while the point sources and the CIB are more non-Gaussian at higher multipoles. It is easy to see the correlation between the point sources and the synchrotron bispectra which peak when the three $\ell$'s are over 150. An additional remark is necessary about the orthogonal shape. Indeed by definition it is orthogonal to the equilateral shape (uncorrelated), which is not visible in these figures because they both have similar acoustic peaks. It is an effect of the sum over $\ell_3$ which hides the differences of these bispectra.


\bibliographystyle{JHEP}
\bibliography{biblio}

\end{document}